\documentclass[11pt]{article}
\usepackage{a4wide}
\usepackage{amsmath,amssymb,amsfonts}
\usepackage{graphicx,subfigure,epsfig}
\usepackage{xcolor}
\usepackage[T1]{fontenc}
\usepackage{upquote,multirow,slashed}
\usepackage{cite,appendix}
\usepackage{empheq}
\usepackage{lipsum}

\definecolor{ForestGreen}{rgb}{0.15,0.70,0.15}
\usepackage[colorlinks=true]{hyperref}
\hypersetup{
 linktocpage=true,
 citecolor=ForestGreen,
 linkcolor=blue,
 urlcolor=blue}
\date{\vspace{-5ex}}

\numberwithin{equation}{section}

\begin{document}

\title{\vspace{-2cm}
{\small \hfill MPP-2019-39 \\} 
\vspace{2cm}
\textbf{A prescription for projectors to compute helicity amplitudes in D dimensions}}

\author{Long Chen\footnote{\textit{E-mail}: longchen@physik.rwth-aachen.de} \\ [.5cm]
Institut f\"ur Theoretische Teilchenphysik und Kosmologie, \\
    RWTH Aachen University,  52056 Aachen, Germany}

\maketitle

\noindent\rule{\textwidth}{.5pt}
\begin{abstract}
This article discusses a prescription to compute polarized dimensionally regularized amplitudes, providing a recipe for constructing simple and general polarized amplitude projectors in D dimensions that avoids conventional Lorentz tensor decomposition and avoids also dimensional splitting.
Because of the latter, commutation between Lorentz index contraction and loop integration is preserved within this prescription, which entails certain technical advantages.  
The usage of these D-dimensional polarized amplitude projectors results in helicity amplitudes that can be expressed solely in terms of external momenta, but different from those defined in the existing dimensional regularization schemes.
Furthermore, we argue that despite being different from the conventional dimensional regularization scheme (CDR), owing to the amplitude-level factorization of ultraviolet and infrared singularities, our prescription can be used, within an infrared subtraction framework, in a hybrid way without re-calculating the (process-independent) integrated subtraction coefficients, many of which are available in CDR. 
This hybrid CDR-compatible prescription is shown to be unitary. 
We include two examples to demonstrate this explicitly and also to illustrate its usage in practice.
\end{abstract}
\noindent\rule{\textwidth}{.5pt}

\thispagestyle{empty}

\clearpage

\noindent\rule{\textwidth}{.5pt}
{
\hypersetup{linkcolor=blue}
\tableofcontents
}
\noindent\rule{\textwidth}{.5pt}

\thispagestyle{empty}
\vspace{1cm}

\allowdisplaybreaks


\section{Introduction}
\label{SEC:introduction}

Helicity scattering amplitudes in Quantum Field Theory (QFT) encode the full dependence
on the spin degrees of freedom of the particles involved in the scattering, and are the building blocks 
for computing various kinds of physical observables through which we try to understand 
the interactions among particles observed in nature.
The incorporation of spin degrees of freedom, or polarization effects, in terms of  
spin- respectively polarization-dependent physical observables, leads to a richer phenomenology.
Such observables offer valuable means to discriminate different dynamical models, 
in particular for discovering potential Beyond-Standard-Model effects.
For a review of the role of particle polarizations in testing the Standard Model and searching for new physics, 
we refer to refs.~\cite{Lampe:1998eu,Leader:2001gr,Accomando:1997wt,MoortgatPick:2005cw} 
and references therein.

Unlike physical observables, individual scattering amplitudes in QFT generally possess 
infrared\footnote{We use the term ``infrared'' (IR) to denote both soft and collinear divergences.} 
(IR) and ultraviolet (UV) divergences, and thus a regularization scheme (RS) for handling these
intermediate divergences needs to be introduced. 
Dimensional regularization~\cite{tHooft:1972tcz,Bollini:1972ui} is by far the most convenient one 
to use in gauge theories as it respects gauge and Lorentz invariance\footnote{The treatment of
$\gamma_5$ in dimensional regularization requires special attention.}, renders all loop integrals
invariant under arbitrary loop momentum shifts, and allows one to handle both UV and IR divergences in the same manner.
The key ingredient of dimensional regularization is the analytic continuation of 
loop momenta to $D=4-2\epsilon$ spacetime dimensions with indefinite $\epsilon$. 
Having done this, one is still left with some freedom regarding the dimensionality of the momenta
of the external particles, of algebraic objects like the spacetime-metric tensor and Dirac matrices, 
as well as the number of polarizations of both external and internal particles. 
This gives rise to different dimensional regularization variants (for a review see e.g. 
ref.~\cite{Gnendiger:2017pys} and references therein), which in general leads to different 
expressions for singular amplitudes. 
Apparently the RS dependence is intimately connected to the singularity structures of amplitudes, 
which fortunately obey a nice factorization form at the amplitude level~\cite{Sen:1982bt,Collins:1989bt,Catani:1998bh,Sterman:2002qn,Aybat:2006wq,Dixon:2008gr,Gardi:2009qi,Gardi:2009zv,Becher:2009cu,Becher:2009kw,Becher:2009qa,Feige:2014wja}.
The result for a physical quantity, such as a physical cross section which is free of any such divergence, 
must not depend on the RS that has been used. 
However, in practice, such a result is obtained as a sum of several partial contributions, 
which usually are individually divergent and computed separately before being combined. 
Therefore, these intermediate results can depend on the RS, and have to be computed consistently 
to ensure the cancellation of the spurious RS-dependence.

The conventional dimensional regularization (CDR)\footnote{By the acronym ``CDR'' 
we refer in this article to the usual CDR~\cite{Collins:1984xc} where, 
in addition, $\gamma_5$ is treated by Larin's prescription~\cite{Larin:1991tj,Larin:1993tq}.} 
scheme~\cite{Collins:1984xc} is a very popular RS, where all vector bosons are treated as D-dimensional objects. 
It is conceptually the simplest one and does guarantee a consistent treatment.
It is typically employed in calculating (unpolarized) amplitude interferences 
where the sum over the polarizations of an external particle is conveniently made
by using the respective unpolarized Landau density matrix. 
For computing helicity amplitudes at the loop level, the two commonly used RS 
are the \textquotesingle t Hooft-Veltman (HV) scheme~\cite{tHooft:1972tcz} 
and the Four-Dimensional-Helicity (FDH) scheme~\cite{Bern:1991aq,Bern:2002zk}. 
In the FDH, the usage of spinor-helicity representations~\cite{DeCausmaecker:1981wzb,DeCausmaecker:1981jtq,Gunion:1985vca,Kleiss:1985yh,Xu:1986xb,Kleiss:1986qc,Dittmaier:1998nn,Schwinn:2005pi,Arkani-Hamed:2017jhn}
and unitarity-cut based methods~\cite{Bern:1994zx,Bern:1994cg,Bern:1997sc,Britto:2004nc,Bern:2007dw} lead to compact expressions 
for helicity amplitudes, which are computationally very advantageous, 
while the proper renormalization procedure for non-supersymmetric theories beyond one loop order 
requires some expertise~\cite{Kilgore:2011ta,Kilgore:2012tb,Gnendiger:2016cpg,Gnendiger:2017rfh}.
Another widely used dimensional regularization variant, the Dimensional-Reduction (DRED) scheme~\cite{Siegel:1979wq}, 
was initially devised for application to supersymmetric theories and was later shown to be applicable also to non-supersymmetric theories~\cite{Capper:1979ns,Jack:1993ws}. 
The DRED and FDH have much in common, while there are also subtle differences 
between the two~\cite{Bern:2002zk,Kilgore:2012tb,Broggio:2015dga,Gnendiger:2017pys}.

For computing D-dimensional helicity amplitudes, especially for amplitudes at the loop level, one typically uses the projection method, see, e.g.,~refs.\cite{Karplus:1950zza,Passarino:1978jh,Kniehl:1990iva,Binoth:2002xg}, which is based on Lorentz covariant tensor decomposition of scattering amplitudes (with external state vectors being stripped off).
The entire dependence of loop amplitudes on loop integrals is encoded in the Lorentz invariant decomposition coefficients which multiply the relevant Lorentz tensor structures. 
Lorentz tensor decomposition is commonly employed in QFT, exploiting its symmetry under the Lorentz group, for instance, in the study 
of hadron structure functions 
that describe deep-inelastic lepton-hadron scattering, in the Passarino-Veltman reduction procedure~\cite{Passarino:1978jh},
and also in the systematic constructions of dimensionally regularized QCD helicity amplitudes~\cite{Abreu:2018jgq,Boels:2018nrr}.

Despite being very generic, versatile, and widely used in many high-order perturbative calculations, 
there are a few aspects of the Lorentz tensor decomposition approach that makes the traditional projection method 
not so easy to be carried out in certain cases, as will be discussed in detail in the next section.
For example, besides facing complexities in deriving D-dimensional projectors for tensor 
decomposition coefficients in some multiple-parton, multiple-scale scattering processes, 
evanescent Lorentz structures\footnote{The evanescent Lorentz structures appearing in a Lorentz tensor decomposition should not be confused with \textit{operator mixings} in the renormalization of composite operators in effective field theories \cite{Collins:1984xc,Buras:1989xd}, nor with evanescent terms in the DRED or FDH regularized Lagrangian~\cite{vanDamme:1984ig,Jack:1993ws,Jack:1994bn,Stockinger:2005gx,Harlander:2006rj,Bern:2002zk,Kilgore:2011ta,Kilgore:2012tb,Gnendiger:2016cpg,Gnendiger:2017rfh}.} can appear in the D-dimensional basis for the loop amplitudes in question. 
Their presence can lead to intermediate spurious poles in the resulting D-dimensional projectors~\cite{Binoth:2002xg,Glover:2004si,Gehrmann:2015ora}. 
Furthermore, when there are several external fermions involved in the scattering~\cite{Glover:2004si,Abreu:2018jgq}, 
the complete and linearly independent set of basis structures in D dimensions will generally increase with the perturbative order at which the virtual amplitude is computed (as the Dirac algebra is formally infinite-dimensional in non-integer D dimensions).

As is well known, when computing polarized amplitudes using spinor-helicity representations, such as in ref.~\cite{Bern:2001dg} for four photon scattering amplitudes in FDH, Lorentz tensor decomposition is typically not used.
While given the impressive long list of high-order QCD calculations of important phenomenological consequences 
done in CDR and, moreover, having in mind the aforementioned critical features of D-dimensional Lorentz tensor decomposition, 
it should be justified to think of possible add-ons in order to facilitate the computations of polarized amplitudes 
in a way fully compatible with CDR. 
In this article we propose an alternative regularization prescription of external states 
(for both bosons and fermions) in order to avoid Lorentz tensor decomposition in the conventional projection method 
for extracting helicity amplitudes. The prescription outlined below is devised to be fully compatible with CDR 
so that certain results known in CDR can be directly recycled.

As will become clear in following sections, the idea is based on the following simple observation. 
In 4 dimensions, there are only four linearly independent Lorentz 4-vectors, 
and hence any Lorentz 4-vector can be expressed linearly using just three linearly independent 
Lorentz 4-vectors with the aid of the Levi-Civita tensor.
Therefore all polarization vectors can be built up by just using three linearly 
independent external momenta in a Lorentz covariant way, provided that there are 
enough linearly independent momenta involved in the process.  
This basic mathematical fact is of course well known, and without surprise it was already 
exploited about forty years ago in calculating (tree-level) multiple photon bremsstrahlung processes in massless QED~\cite{DeCausmaecker:1981wzb,DeCausmaecker:1981jtq}.
It was initially used for simplifying the massless QED vertex by rewriting the 
slashed photon polarization vector in terms of the slashed momenta of external charged fermions
(from which the photon was radiated), a trick that preluded the introduction of the   
4-dimensional massless spinor-helicity formalism~\cite{Gunion:1985vca,Kleiss:1985yh,Xu:1986xb,Kleiss:1986qc}.
In this article, instead of seeking simplifications of the gauge interaction vertices of fermions in 
4-dimensional massless theories, this mathematical fact is employed for finding a CDR-compatible 
way to directly project out polarized loop amplitudes, circumventing Lorentz tensor decomposition. 
Furthermore, despite being different from CDR, we would like to argue that thanks to the amplitude-level factorization of IR singularities in the UV renormalized amplitudes~\cite{Sen:1982bt,Collins:1989bt,Catani:1998bh,Sterman:2002qn,Dixon:2008gr,Aybat:2006wq,Gardi:2009qi,Gardi:2009zv,Becher:2009cu,Becher:2009kw,Becher:2009qa,Feige:2014wja}, such a prescription can be used in a hybrid way together with results known in CDR to obtain RS-independent finite remainders of loop amplitudes, without the need to recalculate the integrated subtraction coefficients involved in an IR subtraction framework. 
In other words, we will show that such a hybrid CDR-compatible prescription is unitary
in the sense defined in refs.~\cite{vanDamme:1984ig,Catani:1996pk}.~\\

The article is organized as follows. 
In the next section, the conventional projection method for computing polarized
amplitudes is reviewed with comments on a few aspects which motivated the work
presented in this article.
In section~\ref{SEC:Prescription} the proposed prescription to obtain polarized 
dimensionally regularized scattering amplitudes is presented in detail.
Section~\ref{SEC:unitarity} is devoted to the discussion of the unitarity of the 
hybrid regularization prescription of section~\ref{SEC:Prescription}.
In particular we show that pole-subtracted RS-independent finite remainders
are always obtained and furthermore demonstrate this feature in the context of an IR subtraction method.
In section~\ref{SEC:examples}, we provide two examples of calculating finite remainders of virtual amplitudes 
in order to illustrate the usage of the prescription and to comment on a few practical points worthy of attention. 
We conclude in section~\ref{SEC:conclusion}.

\section{A Recap of the Projection Method}
\label{SEC:projectionmethod}

In this section, we review the projection method for computing polarized amplitudes, and discuss a few aspects that motivated the work in this article.

Lorentz covariant tensor decomposition is commonly employed in theoretical physics, exploiting the fact that the QFT is invariant under the Lorentz group.
In particular, the projection method, (see, e.g.,~refs.\cite{Karplus:1950zza,Passarino:1978jh,Kniehl:1990iva,Binoth:2002xg},) based on Lorentz covariant tensor decomposition, can be used to obtain helicity amplitudes for a generic scattering process at any loop order. 
The entire dependence of scattering amplitudes on loop integrals is encoded in their Lorentz-invariant decomposition coefficients that multiply the corresponding Lorentz tensor structures and are independent of the external particles' polarization vectors.
These Lorentz-invariant decomposition coefficients are sometimes called \textit{form factors} 
of the amplitudes, a relativistic generalization of the concept of charge distributions. 
In order to extract these form factors containing dimensionally regularized loop integrals, 
projectors defined in D dimensions should be constructed and subsequently applied directly 
to the amplitude, which can proceed diagram by diagram.

\subsection{Gram matrix and projectors}
\label{SEC:projectionmethod:recap}

Scattering amplitudes in QFT with Poincar\'e symmetry are multi-linear in the 
state vectors of the external particles, i.e., proportional to the tensor product 
of all external polarization vectors, to all loop orders in perturbative calculations, 
as manifestly shown by the Feynman diagram representations. 
The color structure of QCD amplitudes can be conveniently described using
the color-decomposition~\cite{Berends:1987cv,Mangano:1987xk,Mangano:1987kp,Mangano:1988kk,Bern:1990ux} 
or the color-space formalism of ref.~\cite{Catani:1996vz}. 
QCD amplitudes are thus viewed as abstract vectors in the color space of external colored particles.
Since projecting QCD amplitudes onto the factorized color space and spin (Lorentz) structures 
can be done independently of each other, we suppress for ease of notation possible color indices of scattering amplitudes in the following discussions.

As nicely summarized and exploited in~\cite{Boels:2017gyc,Boels:2018nrr},   
every scattering amplitude in Lorentz-invariant QFT is a vector in a linear space spanned by a finite set of Lorentz 
covariant structures, in dimensional regularization at any given perturbative order.  
These structures are constrained by physical requirements such as on-shell kinematics 
and symmetries of the dynamics. Scattering amplitudes can thus be written as a linear combination of 
a set of chosen Lorentz basis structures, where the decomposition coefficients are functions 
of Lorentz invariants of external kinematics. All non-rational dependence
of the decomposition coefficients on external kinematics appear via loop integrals.   
This implies the following linear ansatz for a scattering amplitude 
$\hat{\mathcal{M}}$ at a fixed perturbative order, 
\begin{eqnarray} \label{EQ:ampFFsprimitive}
\hat{\mathcal{M}} = \sum_{n=1}^{N_P} c_n ~\hat{T}_n \, , 
\end{eqnarray}
where each form factor $c_n$ is a function of Lorentz invariants of external momenta, 
and each Lorentz structure $\hat{T}_n$ is multi-linear in the external polarization 
state vectors. 
$N_P$ denotes the total number of Lorentz structures involved in the Lorentz tensor decomposition.
In general, $\hat{T}_n$ contains contractions of external gauge bosons' 
polarization state vectors with either the spacetime-metric tensor connecting two different 
polarizations or with external momenta, and contains also products of Dirac matrices sandwiched 
between external on-shell spinors. 
The Levi-Civita tensor can also occur if the scattering process involves parity-violating interactions. 
The complete and linearly independent set of Lorentz structures 
for $\hat{\mathcal{M}}$ at any given perturbative order depends on its 
symmetry properties as well as the Lorentz and Dirac algebra in use.

Note that, as discussed in detail for the four-quark scattering amplitude 
$q \bar{q} \rightarrow Q\bar{Q}$ in~\cite{Glover:2004si,Abreu:2018jgq}, 
the complete and linearly independent set of D-dimensional basis structures must in general 
be enlarged according to the perturbative order at which $q \bar{q} \rightarrow Q\bar{Q}$ 
is computed, because the Dirac algebra is infinite-dimensional for non-integer dimensions. 
At each perturbative order only a finite number of linearly independent Lorentz structures 
can appear in an amplitude, as is evident from inspecting the corresponding Feynman diagrams
which is a set of finite elements.~\\

To be specific, we consider in the following the Lorentz tensor decomposition of scattering amplitudes 
in CDR at fixed order in perturbation theory.
In the discussion of the projection method below, we investigate also how to uncover linear dependent relations 
among a set of (preliminarily chosen) Lorentz tensor structures arising from on-shell constraints, 
without making explicit reference to the origin of these linear dependencies.

Let us assume that by construction the set of the $N_P$  Lorentz structures 
$\hat{T}_n$ in eq.~(\ref{EQ:ampFFsprimitive}), denoted by $\mathbf{T}_{P} \equiv \{
\hat{T}_1, \cdots, \hat{T}_{N_P}\}$, is linearly complete for the $\hat{\mathcal{M}}$ in question,
but the  $\hat{T}_n$ may not be linearly independent of each other. 
For an analogy we recall the representation of QCD amplitudes in terms of a set of 
color structures in color space without demanding linear independence of these color structures.
Let us thus call eq.~(\ref{EQ:ampFFsprimitive}) a \textit{primitive} Lorentz covariant decomposition 
of $\hat{\mathcal{M}}$. 
Possible linear relations among the $N_P$ Lorentz structures $\hat{T}_n$ due to Lorentz and/or Dirac algebra 
and also on-shell constraints, such as equations of motion as well as transversality satisfied 
by external state vectors, can be uncovered by computing their $N_P$$\times$$N_P$ Gram matrix 
$\hat{\mathrm{\mathbf{G}}}$, whose matrix elements are defined by  
\begin{eqnarray} \label{EQ:grammatrix}
\hat{\mathrm{\mathbf{G}}}_{ij} = \langle \hat{T}^{\dagger}_i , \hat{T}_j \rangle\,.
\end{eqnarray} 
The symbol $\langle \hat{T}^{\dagger}_i, \hat{T}_j \rangle$ denotes the Lorentz-invariant inner product 
between these two linear Lorentz structures. 
It is typically defined as the trace of the matrix product of $\hat{T}_i$'s hermitian conjugate, 
i.e.~$\hat{T}^{\dagger}_i$, and $\hat{T}_j$ with tensor products of external state vectors (spinors) being
substituted by the corresponding unpolarized Landau density matrices.  
In other words, this Lorentz-invariant quantity can be viewed as the interference
between two linear Lorentz structures $\hat{T}_i$ and $\hat{T}_j$ summed over all helicity states of 
external particles in accordance with certain polarization sum rules 
(encoded in the unpolarized Landau density matrices).

This $N_P$$\times$$N_P$ Gram matrix $\hat{\mathrm{\mathbf{G}}}$ in eq.~(\ref{EQ:grammatrix}) can be 
used to determine the linearly independent subset of $\mathbf{T}_{P}$ spanning the 
vector space where the considered amplitude $\hat{\mathcal{M}}$ lives. 
If the determinant of $\hat{\mathrm{\mathbf{G}}}$ is not identically zero, then the set $\mathbf{T}_{P}$
is both  complete and linearly independent, and thus qualifies as a basis of the vector space 
where $\hat{\mathcal{M}}$ lives.   
Otherwise, $\hat{\mathrm{\mathbf{G}}}$ is not a full-rank matrix, and its matrix rank 
$N_R \equiv \mathrm{R}[\hat{\mathrm{\mathbf{G}}}]$ tells us the number of 
linearly independent members of $\mathbf{T}_{P}$. 
Since $\mathbf{T}_{P}$ is assumed to be linearly complete w.r.t. $\hat{\mathcal{M}}$ by 
construction, $N_R$ is thus the number of basis elements of a linear basis of the vector space that 
contains $\hat{\mathcal{M}}$.

The number $N_P$$-$$N_R$ of linear dependent relations in $\mathbf{T}_{P}$ can be extracted from 
the \textit{null-space} of this Gram matrix $\hat{\mathrm{\mathbf{G}}}$. 
Technically, the null-space of a matrix $\mathrm{M}$ (not necessarily a square matrix) is 
the solution space of the homogeneous system of linear algebraic equations
defined by taking this matrix $\mathrm{M}$ as the system's coefficient matrix. 
The null-space of $\hat{\mathrm{\mathbf{G}}}$ can be conveniently represented as 
a list of linearly independent $N_P$-dimensional basis vectors of the 
solution space of the homogeneous linear algebraic system defined by $\hat{\mathrm{\mathbf{G}}}$. 
The length of this list of basis vectors is equal to the dimension of 
$\hat{\mathrm{\mathbf{G}}}$ minus its matrix rank, i.e.,~$N_P$$-$$N_R$. 
For the information we would like to extract\footnote{To just
identify the linearly dependent columns and/or rows of the multivariate Gram matrix, numerical samples of this matrix 
at a few test points are usually enough.}, this null-space provides 
the complete set of linear combination coefficients (being rational in the external kinematics) 
of the column vectors of $\hat{\mathrm{\mathbf{G}}}$ that lead to vanishing $N_P$-dimensional vectors. 
After having removed those linearly dependent columns (and their corresponding transposed rows), 
we end up with a reduced full-rank Gram matrix among the thus-selected linearly independent set of Lorentz structures, 
denoted by $\mathbf{T}_{R}$. 
The set $\mathbf{T}_{R}$ can then be directly taken as the basis of the vector space of $\hat{\mathcal{M}}$.

Elimination of redundancies in the set $\mathbf{T}_{P}$ for $\hat{\mathcal{M}}$ involving external gauge bosons, 
e.g.,~due to Ward identities of local gauge interactions, can be effectively accounted for by choosing physical 
polarization sum rules for those external gauge bosons 
(with their reference vectors chosen as momenta of other external particles).
This point can be easily seen once we realize that any unphysical structure,  
which may happen to be just one specific $\hat{T}_n$ or a linear combination of some of them
(with rational coefficients in external kinematics), 
gets nullified by the physical polarization sum rules of external gauge bosons.
Notice, however, reduction in the number of linearly independent basis structures 
of $\hat{\mathcal{M}}$ due to additional process-specific symmetries such as 
charge, parity, and/or Bose symmetry is not achieved by analyzing $\hat{\mathrm{\mathbf{G}}}$ in this way. 
Instead they have to be accounted for from the outset when determining the primitive set $\mathbf{T}_P$ in eq.~(\ref{EQ:ampFFsprimitive}). ~\\

In terms of the thus-determined basis $\mathbf{T}_{R}$, the linear decomposition of $\hat{\mathcal{M}}$ 
can be recast into 
\begin{eqnarray} \label{EQ:ampFFs}
\hat{\mathcal{M}} = \sum_{n=1}^{N_R} \tilde{c}_n ~\hat{T}_n ~,~
\end{eqnarray}
and the Gram matrix $\hat{\mathrm{\mathbf{G}}}_R$ of $\mathbf{T}_{R}$ with matrix elements defined 
similarly as eq.~(\ref{EQ:grammatrix}) is now an invertible $N_R$$\times$$N_R$ matrix.

Now we are ready to discuss projectors $\hat{P}_n$ for the Lorentz decomposition coefficients (or form factors) $\tilde{c}_n$ of $\hat{T}_n$ in eq.~(\ref{EQ:ampFFs}).
They are defined by
\begin{eqnarray} \label{EQ:projectordef}
\tilde{c}_n = \langle \hat{P}^{\dagger}_n , \hat{\mathcal{M}} \rangle~ ~~ 
\text{for any $n ~\in~ \{1, \cdots, N_R\}$} \, ,
\end{eqnarray} 
where the same Lorentz-invariant inner product operation as in eq.~(\ref{EQ:grammatrix}) is 
used in the above projection. 
The defining equation (\ref{EQ:projectordef}) of $\hat{P}_n$ holds for any linear object 
 from the vector space spanned by the basis $\mathbf{T}_{R}$, rather than just for a particular 
scattering amplitude $\hat{\mathcal{M}}$.
Inserting eq.~(\ref{EQ:ampFFs}) into 
eq.~(\ref{EQ:projectordef}) then, taking the aforementioned property into account,  the defining equation  
 for the  projectors translates into 
\begin{eqnarray} \label{EQ:projectordefequvi}
\langle \hat{P}^{\dagger}_n , \hat{T}_m \rangle = \delta_{nm} \quad \text{for any $n,m ~\in~ \{1, \cdots, N_R\}$} \,. 
\end{eqnarray}

Each projector $\hat{P}^{\dagger}_n$ can be expressed in terms of a linear combination 
of hermitian conjugate members of $\mathbf{T}_{R}$ that span also a  vector space.
We thus write 
\begin{eqnarray} \label{EQ:projectoransatz}
\hat{P}^{\dagger}_n = \sum_{k=1}^{N_R} \hat{\mathrm{\mathbf{H}}}_{nk} \, \hat{T}^{\dagger}_k \, ,
\end{eqnarray}
where the elements $\hat{\mathrm{\mathbf{H}}}_{nk}$ are to be determined. 
Inserting eq.~(\ref{EQ:projectoransatz}) into eq.~(\ref{EQ:projectordefequvi}) and using the definition of Gram matrix elements we get
\begin{eqnarray}
\sum_{k=1}^{N_R} \hat{\mathrm{\mathbf{H}}}_{nk} \, \big(\hat{\mathrm{\mathbf{G}}}_R\big)_{km} = \delta_{nm}~, 
 \quad \text{i.e.,} \; \hat{\mathrm{\mathbf{H}}} \, \hat{\mathrm{\mathbf{G}}}_R = \hat{1}. 
\end{eqnarray}
Recall that $\hat{\mathrm{\mathbf{G}}}_R$ is invertible by the aforementioned trimming procedure.  
This then answers the question of how to construct, in general, the projectors $\hat{P}^{\dagger}_n$ 
from linear combinations of the hermitian conjugates of $\mathbf{T}_{R}$ in a systematic algorithmic manner.
In the special and ideal case of a norm-orthogonal basis $\mathbf{T}_{R}$, 
its Gram matrix $\hat{\mathrm{\mathbf{G}}}_R$ is equal to the identity matrix of dimension $N_R$ 
and hence $\hat{\mathrm{\mathbf{H}}}  = \hat{\mathrm{\mathbf{G}}}_R^{-1} = \hat{1} $.
Subsequently, we have $\hat{P}^{\dagger}_n = \hat{T}^{\dagger}_n$, as is well known
for a norm-orthogonal basis.~\\

By taking the Dirac traces and keeping all Lorentz indices in D dimensions in the projection, 
these Lorentz-invariant tensor decomposition coefficients, or form factors, 
are evaluated in D dimensions. These form factors are independent of the external polarization vectors, 
and all their non-rational dependence on external momenta is confined to loop integrals. 
Scalar loop integrals appearing in these form factors can be reduced to a finite 
set of master integrals with the aid of the linear integration-by-parts (IBP) identities~\cite{Tkachov:1981wb,Chetyrkin:1981qh}. 
Once these dimensionally regularized form factors have been determined, 
external particles' state vectors can be conveniently chosen in 4 dimensions, 
leading to helicity amplitudes in accordance with the HV scheme. 
In fact, once the (renormalized) virtual amplitudes are available at hand in such a 
D-dimensional tensor-decomposed form (with all Lorentz-invariant form factors computed 
in D dimensions), then changing the regularization convention for the external particles' states 
consistently in both the virtual amplitude and the corresponding IR-subtraction terms, 
should not alter the finite remainder that is left after subtracting all poles, 
although the individual singular pieces do change accordingly.

\subsection{Comments on the D-dimensional projection}
\label{SEC:projectionmethod:comments}

We now discuss a few delicate aspects of the Lorentz tensor decomposition in D dimensions 
that motivated the work presented in this article.

In general, the Gram matrix $\hat{\mathrm{\mathbf{G}}}$ or $\hat{\mathrm{\mathbf{G}}}_R$ 
computed using Lorentz and Dirac algebra in CDR depends on the spacetime dimension D. 
 We can examine its 4-dimensional limit by inserting 
$D=4-2\epsilon$ and check whether its determinant, power-expanded in $\epsilon$, 
is zero or not in the limit $\epsilon = 0$. 
A determinant vanishing at $\epsilon = 0$ implies the presence of Lorentz structures in the D-dimensional 
linearly independent basis set $\mathbf{T}_{R}$ that are redundant in 4 dimensions.

To be more specific, we can compute the matrix rank of $\hat{\mathrm{\mathbf{G}}}_R$ 
at $D =4$, denoted by $\mathrm{R}[\hat{\mathrm{\mathbf{G}}}_R^{D=4}]$, and  
the difference $N_R$$-$$\mathrm{R}[\hat{\mathrm{\mathbf{G}}}_R^{D=4}]$
tells us the number of Lorentz structures  
appearing in $\mathbf{T}_R$ that are redundant in D=4. 
Furthermore, if we compute the null-space of the 4-dimensional limit of 
$\hat{\mathrm{\mathbf{G}}}$, then we can explicitly uncover all these special linear 
relations among $\hat{T}_n$ due to the constraint of integer dimensionality\footnote{Any potential non-linear 
relation among the $\hat{T}_n$ is irrelevant here as we use a linear basis.} 
in a similar way as one identifies $\mathbf{T}_{R}$ out of $\mathbf{T}_{P}$.
These special linear relations can be used to construct exactly the number 
$N_R$$-$$\mathrm{R}[\hat{\mathrm{\mathbf{G}}}_R^{D=4}]$ of evanescent Lorentz structures out of $\mathbf{T}_{R}$ that are non-vanishing in D dimensions but vanishing in 4 dimensions\footnote{Alternatively, one could achieve this by employing the Gram-Schmidt orthogonalization procedure to $N_R$$-$$\mathrm{R}[\hat{\mathrm{\mathbf{G}}}_R^{D=4}]$ number of the (4-dimensional) redundant structures in $\mathbf{T}_{R}$.}. 
In this way, the original basis set $\mathbf{T}_R$ can be re-cast into a union of two subsets: 
one is linearly independent and complete in 4 dimensions, 
and the other one only consists of $N_R$$-$$\mathrm{R}[\hat{\mathrm{\mathbf{G}}}_R^{D=4}]$ evanescent Lorentz structures. 
Such a reformulation of the Lorentz tensor decomposition basis in D dimensions can thus be very useful in exhibiting the additional non-four-dimensional structures involved in the virtual amplitude.~\\

In case the number of structures in $\mathbf{T}_{R}$ is not very small 
(say, not less than 10) and if there are several kinematic variables involved, 
algebraically inverting $\hat{\mathrm{\mathbf{G}}}_R$ can be computationally 
quite cumbersome~\cite{Boels:2018nrr}. 
Moreover, the resulting projectors constructed in the above fashion may be hardly 
usable if the amplitudes themselves are already quite complicated. 
This situation occurs naturally in multiple-parton multiple-scale scattering processes.
Possible simplifications may be obtained by suitably recombining the linear basis structures 
in $\mathbf{T}_{R}$ classified into several groups, such that they are mutually orthogonal 
or decoupled from each other~\cite{Boels:2018nrr}. 
For example, we could divide the set of tensor structures into symmetric and anti-symmetric
sectors, and also choose the anti-symmetrized product basis for strings of Dirac matrices~\cite{Dugan:1990df,Gehrmann:2011aa}. 
This amounts to choosing the basis structures in $\mathbf{T}_{R}$ such that a partial triangularization 
of the corresponding Gram matrix $\hat{\mathrm{\mathbf{G}}}_R$ is achieved already by construction. 
This will facilitate the subsequent inversion operation, and also make the results simpler. 
In addition, in case the set of tensor structures all observe factorized forms in terms of products 
of a smaller set of lower rank tensor structures, then this factorization can also be exploited 
to greatly facilitate the construction of projectors~\cite{Binoth:2002xg}. 
Alternatively, it is also a good practice to ``compactify'' the vector space as much as possible, 
before the aforementioned construction procedure is applied, by employing all possible physical constraints 
and symmetries, such as parity and/or charge symmetry of the amplitudes in question, 
and also by fixing the gauge of the external gauge bosons~\cite{Gehrmann:2013vga,vonManteuffel:2015msa,Boels:2017gyc,Boels:2018nrr}.

Other than the aforementioned technical complexity in inverting the Gram matrix, there is another delicate point about the Lorentz tensor decomposition approach in D dimensions, as already briefly mentioned above. 
In cases where the external state consists only of bosons, a list of fixed number of Lorentz tensor structures is indeed linearly complete in D dimensions to all orders in perturbation theory~\cite{Binoth:2002xg,Gehrmann:2013vga}. 
However, if external fermions are involved in the scattering, the complete and linearly independent set of basis structures will generally increase with the perturbative order at which the scattering amplitude is computed, because the Dirac algebra is formally infinite dimensional in non-integer D dimensions, as discussed for the four-quark scattering amplitude $q \bar{q} \rightarrow Q\bar{Q}$ in~\cite{Glover:2004si,Abreu:2018jgq}.
Of course, at each given perturbative order only a finite number of linearly independent Lorentz structures can appear in an amplitude, because the corresponding Feynman diagrams are just a set of finite elements. 
These additional D-dimensional Lorentz structures are either evanescent by themselves or will lead to additional evanescent structures of the same number computed by the procedure discussed above.

The last comment we would like to make about the projection method in D dimensions is the possible appearance of intermediate spurious poles in these projectors~\cite{Binoth:2002xg,Glover:2004si,Gehrmann:2015ora}, which are closely related to the presence of the aforementioned evanescent Lorentz structures in the D-dimensional linearly independent basis. 
Since the presence of evanescent Lorentz structures in the D-dimensional basis implies a Gram matrix that vanishes in 4 dimensions, one expects that projectors resulting from its inverse can contain poles in 4$-$D=$2\epsilon$, for instance in~\cite{Binoth:2002xg} for four-photon scattering.
Of course, all intermediate spurious poles generated this way in the individual form factors projected out should cancel in the physical amplitudes composed out of them, such as helicity amplitudes or linearly polarized amplitudes.~\\

All these sometimes cumbersome issues discussed above\footnote{We remark that after the initial release of this work, refs.~\cite{Ahmed:2019udm,Peraro:2019cjj,Peraro:2020sfm} managed to address some of the issues related to the conventional form factor decomposition, highlighting the advantage of removing evanescent tensor structures.} motivated the work that will be presented in the following: the construction of simple and general polarized amplitude projectors in D dimensions that avoids conventional Lorentz tensor decomposition, yet is still fully compatible with CDR.

\section{The Prescription}
\label{SEC:Prescription}

The idea behind the proposed prescription to obtain polarized dimensionally
regularized scattering amplitudes can be briefly outlined as follows, 
with details to be exposed in the subsequent subsections.

For external gauge bosons of a scattering amplitude, massless and/or massive, 
we decompose each external polarization  vector 
in terms of external momenta. 
We then keep the form of Lorentz covariant decomposition fixed while formally promote 
all its open Lorentz indices, which are now all carried by external momenta, 
from 4 dimensions to D dimensions, like every Lorentz vector in CDR. 
If external fermions are present in the scattering amplitude,  
strings of Dirac matrices sandwiched between external on-shell spinors will show up. 
For each open fermion line, we first rewrite this quantity as a trace of products of 
Dirac matrices with the aid of external spinors' Landau density matrices, 
up to an overall Lorentz-invariant normalization factor.
The space-like polarization vectors of a massive spinor can also be represented 
in terms of external momenta.  Again, once such a momentum basis representation 
is established in 4 dimensions, the Lorentz covariant form will be kept \textit{fixed} 
while all open Lorentz and Dirac indices, carried by external momenta and/or 
Dirac matrices as well as the spacetime-metric tensor, will be respectively promoted in accordance with CDR.

As scattering amplitudes are multi-linear in the state vectors of the external particles 
to all loop orders in perturbation theory, the tensor products of momentum basis representations 
of all external gauge bosons and all properly re-written external spinor products, 
with their open indices promoted accordingly as in CDR, 
will be taken as the \textit{external} projectors for polarized amplitudes. 
Helicity amplitude projectors of a generic scattering process defined in this way
naturally obey a simple factorized pattern as the tensor product of 
the respective polarization projector of each external gauge boson and open fermion line. 
Features and subtleties worthy of attention during these rewriting procedures  
will be discussed and explained below.

\subsection{Momentum basis representations of polarization vectors}
\label{SEC:prescription:MBR}

Let us start with the cases where all external states are bosons.
We recall that the polarization vector $\varepsilon^{\mu}_{\lambda}(p)$ of a physical vector-boson state 
of momentum $p^\mu$ has to satisfy $\varepsilon^{\mu}_{\lambda}(p)~ p_{\mu} = 0$. 
Here the subscript $\lambda$ labels the number of physical spin degrees of freedom, 
i.e., $\lambda=1,2,3$ in D=4 dimensions.
By convention the physical polarization state vectors are orthogonal and normalized by 
$\varepsilon^{*}_{\lambda}(p) \cdot \varepsilon_{\lambda'}(p) = -\delta_{\lambda \lambda'}$.
The polarization vectors of a massless gauge boson obey an additional condition 
in order to encode the correct number of physical spin degrees of freedom.
In practice, this additional condition is usually implemented by introducing an auxiliary 
reference vector $\hat{r}_{\mu}$ that is not aligned with the boson's momentum 
but otherwise arbitrary, to which the physical polarization vectors have to be orthogonal, 
$\varepsilon^{\mu}_{\lambda}(p)~ \hat{r}_{\mu} = 0$.
Thus, the reference vector $\hat{r}_{\mu}$ and the boson's momentum $p_{\mu}$ define 
a plane to which the massless gauge boson's physical polarization vectors are orthogonal.
We also recall that in CDR the number of physical polarizations of a massless gauge
boson in D dimension is taken to be D$-$2. This is in contrast to our prescription, 
where the number of physical polarizations remains two in D dimensions, see below.

\subsubsection{The $2\to 2$ scatterings among massless gauge bosons}
\label{SEC:prescription:MBR:2to2}

Let us first consider a prototype  $2\to 2$ scattering among 4 external 
massless gauge bosons:
\begin{align} \label{momentaassignment}
g_1(p_1) + g_2(p_2) \rightarrow g_3(p_3) + g_4(p_4), 
\end{align}
with on-shell conditions $p_j^2 = 0,~ j=1,...,4.$
The  Mandelstam variables  associated with \eqref{momentaassignment}
\begin{eqnarray} \label{EQ:kinematicinvariants}
 s \equiv \left(p_1 + p_2 \right)^2 = \left(p_3 + p_4 \right)^2 ~, \qquad
 t \equiv \left(p_2 - p_3 \right)^2 = \left(p_1 - p_4 \right)^2
\end{eqnarray} 
encode the independent external kinematic invariants.

The representation of the gauge bosons' polarization state vectors 
in terms of three linearly independent external momenta, $p_1, p_2, p_3$ 
 can be determined in the following way. 
We first write down a Lorentz covariant parameterization ansatz for the linear representation 
and then solve the aforementioned orthogonality and normalization conditions 
for the linear decomposition coefficients. 
Once we have established a definite Lorentz covariant decomposition form initially in 4 dimensions 
solely in terms of external momenta and kinematic invariants, 
this covariant form will be kept and used as the \textit{definition} 
of the corresponding polarization state vector in D dimensions.

\begin{figure}[tbh!]
\centering
\includegraphics[width=12cm,height=6cm]{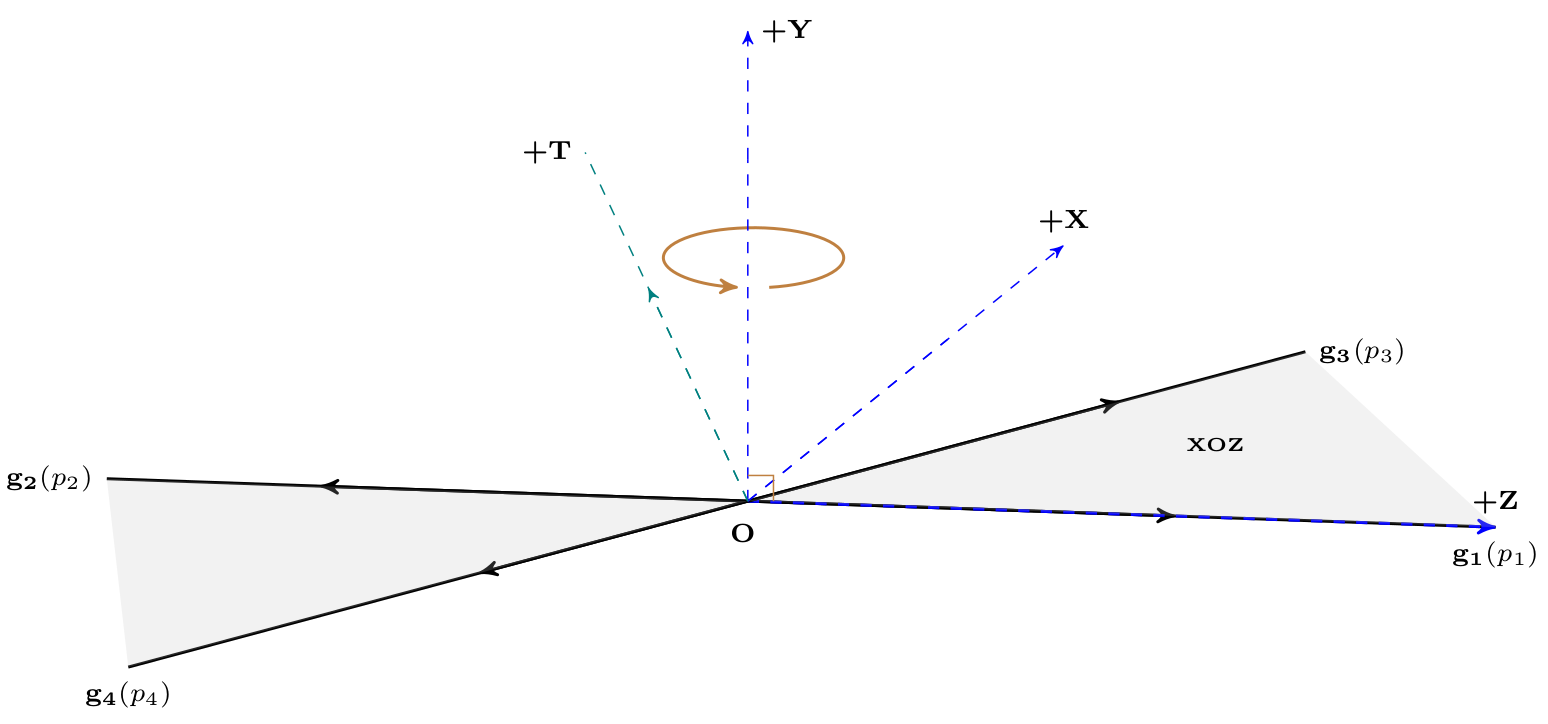}
\caption{The chosen coordinate system in the center-of-mass reference
frame of the two incoming particles.}
\label{FIG:coordinatesystem}
\end{figure}

While the decomposition of polarization state vectors in terms of external momenta 
is Lorentz covariant, it is very helpful to have in mind 
a particular reference frame where a clear geometric picture can be established to
illustrate the choices of and constraints on polarization state vectors.  
To this end, we consider in the following discussion the center-of-mass frame of 
the two incoming particles, as illustrated in figure~\ref{FIG:coordinatesystem}, 
where the beam axis is taken as the Z-axis with its positive direction chosen along $p_1$.
Furthermore, the scattering plane determined by $p_1$ and $p_3$ is chosen as the X-O-Z plane 
with $p_3$ having a non-negative X-component by definition. 
The positive direction of the Y-axis of the coordinate system will be determined 
according to the \textit{right-hand} rule. The reference frame is shown 
in figure~\ref{FIG:coordinatesystem}.~\\

Let's now come to the momentum basis representations of the polarization vectors 
in this reference frame. 
There are two common basis choices regarding the transverse polarization states, 
the linear  and the circular polarization basis, 
the latter represents  helicity eigenstates of gauge bosons.
These two bases can be related via a ${\pi}/{2}$ rotation in the complex plane. 
In the following, we will first establish a Lorentz covariant decomposition of 
a set of elementary linear polarization state vectors in terms of external 4-momenta 
and then compose circular polarization states of all external gauge bosons, 
i.e., their helicity eigenstates, out of these elementary ones.

For the two initial-state massless gauge bosons $g_1(p_1)$ and $g_2(p_2)$, 
whose momenta are taken as the reference momenta for each other, 
we first introduce a common linear polarization state vector $\varepsilon_{X}^{\mu}$ 
along the X-axis direction, i.e., transverse to the beam axis but within the X-O-Z plane. 
The  set of equations that determines $\varepsilon_{X}^{\mu}$  reads: 
\begin{eqnarray} \label{EQ:XpolEQs}
&&\varepsilon^{\mu}_{X}  = c^{X}_1~ p^{\mu}_1 + c^{X}_2~ p^{\mu}_2 + c^{X}_3~ p^{\mu}_3  \, ,\nonumber\\ 
&&\varepsilon_{X} \cdot p_1 = 0 \, ,\nonumber\\
&&\varepsilon_{X} \cdot p_2 = 0 \, ,\nonumber\\
&&\varepsilon_{X} \cdot \varepsilon_{X} = -1 \, .
\end{eqnarray}
Solving eq.~(\ref{EQ:XpolEQs}) for the coefficients $c^{X}_1,~ c^{X}_2,~ c^{X}_3$,
and subsequently inserting the solution back to the first line of eq.~(\ref{EQ:XpolEQs}),  
we obtain the following momentum basis representation for $\varepsilon_{X}$: 
\begin{eqnarray} \label{EQ:XpolMBR}
\varepsilon^{\mu}_{X}  = \mathcal{N}_{X} 
\Big(t~ p^{\mu}_1 + (-s-t)~ p^{\mu}_2 + s~ p^{\mu}_3 \Big) \, ,
\end{eqnarray}
where $\mathcal{N}^{-2}_{X} = -t s (s + t)$.
Notice that one may choose to include the overall Lorentz-invariant normalization factor $\mathcal{N}_{X}$ in the very last step of the computation of polarized loop amplitudes, for instance after UV renormalization and IR subtraction if an IR subtraction method is employed. 
In this way, we never have to deal with $\mathcal{N}_{X}$, i.e.,~with a square root explicitly 
in the intermediate stages. 
If we choose to incorporate the overall normalization factors only at the level of squared amplitudes (or interferences), then  square roots of kinematic invariants never appear. 
Furthermore, we can always by convenience define this overall normalization factor such that the coefficients exhibited in eq.~(\ref{EQ:XpolMBR}) are polynomials in the external kinematic invariants (rather than rational functions). 
This can be helpful as computer algebra systems are typically more efficient when dealing with polynomials only.

Concerning the two final-state massless gauge bosons $g_3(p_3)$ and $g_4(p_4)$, whose momenta are also taken as reference momenta for each other, we can introduce a common linear polarization state vector $\varepsilon_{T}^{\mu}$
defined to be transverse to $p_3$ and $p_4$ but still lying within the X-O-Z plane,
in analogy to $\varepsilon_{X}^{\mu}$. 
The definition of $\varepsilon_{T}^{\mu}$ then translates into the following set of equations:
\begin{eqnarray} \label{EQ:TpolEQs}
&&\varepsilon^{\mu}_{T}  = c^{T}_1~ p^{\mu}_1 + c^{T}_2~ p^{\mu}_2 + c^{T}_3~ p^{\mu}_3  \, , \nonumber\\ 
&&\varepsilon_{T} \cdot p_3 = 0 \, , \nonumber\\
&&\varepsilon_{T} \cdot p_4 = 0 \, , \nonumber\\
&&\varepsilon_{T} \cdot \varepsilon_{T} = -1 \, .
\end{eqnarray}
Solving eq.~(\ref{EQ:TpolEQs}) for the coefficients $c^{T}_1,~ c^{T}_2,~ c^{T}_3$, 
 one  obtains
\begin{eqnarray} \label{EQ:TpolMBR}
\varepsilon^{\mu}_{T}  = \mathcal{N}_{T} 
\Big(t~ p^{\mu}_1 + (s+t)~ p^{\mu}_2 + (-s-2 t)~ p^{\mu}_3 \Big) \, ,
\end{eqnarray}
where $\mathcal{N}_{T}^{~-2} = -t s (s + t)$.
The comments given above on $\varepsilon_{X}^{\mu}$ apply here as well.

The last elementary polarization state vector needed for constructing
 helicity eigenstates of all four external massless gauge bosons is the one 
orthogonal to $p_1$, $p_2$, and $p_3$, denoted by $\varepsilon_{Y}$,
which is thus perpendicular to the X-O-Z plane. 
In 4 dimensions, we obtain it using the Levi-Civita tensor:
\begin{eqnarray} \label{EQ:YpolMBR}
\varepsilon^{\mu}_{Y}  = \mathcal{N}_{Y}~ 2\epsilon^{\nu\rho\sigma\mu} p_{1\nu} p_{2 \rho} p_{3 \sigma}  
= \mathcal{N}_{Y}~ 2\epsilon^{\mu}_{p_1 p_2 p_3}  \, ,
\end{eqnarray}
where $\mathcal{N}_{Y}^{~-2} = - t s (s + t)$, and in the last line we introduced the short-hand notation $\epsilon^{\mu}_{p_1 p_2 p_3} \equiv \epsilon^{\nu\rho\sigma\mu} p_{1\nu} p_{2 \rho} p_{3 \sigma}$\footnote{We use the convention $\epsilon^{0123} = +1$ and $\epsilon_{\mu\nu\rho\sigma} = -\epsilon^{\mu\nu\rho\sigma}$.}.
We have introduced a factor 2 in eq.~(\ref{EQ:YpolMBR}) so that $\mathcal{N}^{~2}_{X} = \mathcal{N}^{~2}_{Y} =\mathcal{N}^{~2}_{T} = 1/(-t s (s + t))$.

A comment concerning $\epsilon^{\mu\nu\rho\sigma}$ is appropriate here.
The above polarization state vectors will be eventually used in D-dimensional calculations. 
To this end, following~\cite{Larin:1991tj,Larin:1993tq,Moch:2015usa}, 
we will treat $\epsilon^{\mu\nu\rho\sigma}$ merely as a symbol denoting an object whose 
algebraic manipulation rules consist of the following two statements.
\begin{itemize}
\item 
Antisymmetry: it is completely anti-symmetric regarding any odd permutation of its arguments.
\item 
Contraction Rule\footnote{There is a subtle point concerning this when there are 
multiple Levi-Civita tensors in the contraction, related to the choice of pairing, 
as will be briefly commented on in section \ref{SEC:examples:eeQQ}.}: 
the product of two $\epsilon^{\mu\nu\rho\sigma}$ is replaced by a combination 
of products of spacetime-metric tensors $g^{\mu\nu}$ of the same tensor rank according to 
the following fixed pattern:
\begin{eqnarray} \label{EQ:LeviCivitaContRule}
\epsilon^{\mu\nu\rho\sigma} \epsilon^{\mu'\nu'\rho'\sigma'} 
= \mathrm{Det}\Big[g^{\alpha \alpha'} \Big]~, 
\text{~  with $\alpha = \mu,~\nu,~\rho,~\sigma$ and $\alpha' = \mu',~\nu',~\rho',~\sigma'$,} 
\end{eqnarray}
which agrees with the well-known mathematical identity for Levi-Civita tensors in 4 dimensions.
\end{itemize}

Using eq.~(\ref{EQ:LeviCivitaContRule}) with the D-dimensional spacetime-metric tensor 
in determining $\mathcal{N}_{Y}$ in eq.~(\ref{EQ:YpolMBR}), 
one gets ${\mathcal{N}_{Y}^{-2}}=(3-D) s t (s + t)$ with $D = 4 - 2 \epsilon$. 
 Because $\mathcal{N}_{Y}$ is an overall normalization factor 
which must be used consistently in computing both the (singular) 
virtual loop amplitudes, the UV-renormalization counter-terms, 
as well as potential IR subtraction terms, 
 it is merely a normalization convention 
whether the explicit $D$ appearing in $\mathcal{N}_{Y}$ is set to 4 or to $4 - 2 \epsilon$, 
on which the final 4-dimensional finite remainder should not depend
(albeit the individual singular objects do of course differ).
This point can be made even more transparent if one chooses to incorporate this overall normalization 
factor in the last stage of the consistent computation of finite remainders 
where the 4-dimensional limit has already been explicitly taken. 
~\\

The circular polarization state vectors of all four external massless gauge bosons,
namely their helicity eigenstates, can be easily constructed from the three linear polarization states given above by a suitable ${\pi}/{2}$ rotation in the complex plane. 
Here we take a convention where the two helicity eigenstates of each gauge boson are given by\footnote{Alternatively one could choose the more systematically formulated Jacob-Wick phase convention as introduced in~ref.\cite{Jacob:1959at}, although physics in the end should not depend on artificial phase conventions for quantum states as long as it is always the same convention adopted consistently throughout the computation (see section~\ref{SEC:prescription:comments} for more discussions).}  
\begin{eqnarray} \label{EQ:LP2HLmassless}
\varepsilon_{\pm}(p_1;p_2) &=& \frac{1}{\sqrt{2}} \left( \varepsilon_{X} \pm 
i \, \varepsilon_{Y} \right) \, , \nonumber\\
\varepsilon_{\pm}(p_2;p_1) &=& \frac{1}{\sqrt{2}} \left( \varepsilon_{X} \mp 
i \, \varepsilon_{Y} \right)  \, ,\nonumber\\
\varepsilon_{\pm}(p_3;p_4) &=& \frac{1}{\sqrt{2}} \left( \varepsilon_{T} \pm 
i \, \varepsilon_{Y} \right)  \, ,\nonumber\\
\varepsilon_{\pm}(p_4;p_3) &=& \frac{1}{\sqrt{2}} \left( \varepsilon_{T} \mp 
i \, \varepsilon_{Y} \right) \, ,
\end{eqnarray}
where the first argument of $\varepsilon_{\pm}(p;r)$ is the particle's momentum while the second 
shows the reference momentum. Eq.~(\ref{EQ:LP2HLmassless}) shows that the helicity flips 
once the particle's 3-momentum gets reversed or if the polarization vector is subject to complex conjugation. 
Furthermore, owing to the Ward identities fulfilled by the gauge amplitudes, the representations
of helicity state vectors in eq.~(\ref{EQ:LP2HLmassless}) can be further reduced respectively 
for each gauge boson by removing the component proportional to the gauge boson's own 4-momentum. 
For instance, for the gauge boson $g_1$ with 4-momentum $p_1$, the component of $\varepsilon_{X}$ 
proportional to $p_1^{\mu}$ in eq.~(\ref{EQ:XpolMBR}) can be safely dropped when 
constructing $\varepsilon_{\pm}(p_1;p_2)$, and similar reductions hold also for the 
other gauge bosons. 
As will become clear in the following discussions, when loop integrals in amplitudes are kept in their unreduced symbolic form, it is beneficial to first perform the projection in the linear polarization basis, and then have the helicity amplitudes composed while reducing the results~(see section~\ref{SEC:prescription:comments} for more comments on this).
Since these elementary linear polarization state vectors will be used to construct helicity states 
of several scattered particles, we should keep their complete momentum basis representation forms 
as given by eqs. \eqref{EQ:XpolMBR}, \eqref{EQ:TpolMBR}, and \eqref{EQ:YpolMBR}.~\\

We emphasize again that in our prescription the number of physical polarizations in D dimensions 
of a massless gauge boson remains two, see eq.~\eqref{EQ:LP2HLmassless}. 
In order to illustrate the resulting differences to CDR, let us do a simple exercise about polarization sums. 
In CDR the sum over the D-dimensional polarization states of a massless gauge boson $g_1$ with 4-momentum 
$p_1^{\mu}$ and gauge reference vector $r^{\mu}=p^{\mu}_2$ (cf. eq.~\eqref{momentaassignment}) is 
\begin{eqnarray}\label{EQ:polsumCDRPhys}
\sum_{\bar{\lambda} = \pm,~ D-4} \bar{\varepsilon}_{\bar{\lambda}}^{~\mu}(p_1;p_2)\bar{\varepsilon}_{\bar{\lambda}}^{\,*\,\nu}(p_1;p_2)
&=&  -g^{\mu\nu} + \frac{p_1^{\mu} p_2^{\nu} + p_2^{\mu} p_1^{\nu}}{p_1 \cdot p_2} \nonumber\\
&=&  -g^{\mu\nu} + \frac{2}{s} \left(p_1^{\mu} p_2^{\nu} + p_2^{\mu} p_1^{\nu} \right)
\end{eqnarray}
which is also the unpolarized Landau density matrix of the polarization states of $g_1$.
All Lorentz indices in \eqref{EQ:polsumCDRPhys} are D dimensional and the  symbol $\bar{\lambda}$ labels
the $D$$-$$2$ numbers of polarization states $\bar{\varepsilon}_{\bar{\lambda}}^{~\mu}(p_1;p_2)$ in D dimensions.
On the other hand, in our prescription we sum over just the two transverse polarization states of 
$g_1$ that are \textit{defined} by their respective momentum basis representations in 
eqs.~(\ref{EQ:XpolMBR}), \eqref{EQ:TpolMBR}.
We get 
\begin{eqnarray}\label{EQ:polsumMBR}
\sum_{\lambda = X,Y} \varepsilon^{~\mu}_{\lambda}(p_1;p_2)\varepsilon^{\,*\, \nu}_{\lambda}(p_1;p_2) 
&=& \frac{1}{D-3} \left(-g^{\mu\nu} + \frac{D-2}{s} \left(p_1^{\mu} p_2^{\nu} + p_2^{\mu} p_1^{\nu} \right) \right) \nonumber\\
&+&\frac{4-D}{D-3} 
\Big(\frac{t}{s (s+t)} p_1^{\mu} p_1^{\nu} 
+\frac{s+t}{s t} p_2^{\mu} p_2^{\nu}
+\frac{s}{t (s+t)} p_3^{\mu} p_3^{\nu} 
\nonumber\\&&~~~
+\frac{1}{s+t} \left(p_1^{\mu} p_3^{\nu} + p_3^{\mu} p_1^{\nu} \right)
-\frac{1}{t} \left(p_2^{\mu} p_3^{\nu} + p_3^{\mu} p_2^{\nu} \right)\, ,
\Big)
\end{eqnarray}
where, as part of the definition of this expression, we have rewritten the 
product of two Levi-Civita tensors in $\varepsilon^{\mu}_{Y}(p_1;p_2)\,\varepsilon^{\,*\,\nu}_{Y}(p_1;p_2)$ 
in terms of spacetime-metric tensors.  
Apparently eq.~(\ref{EQ:polsumMBR}) is not identical to eq.~(\ref{EQ:polsumCDRPhys}),\footnote{Note that 
with our prescription unpolarized squared amplitudes are supposed to be computed by incoherently summing over squared helicity amplitudes, and not by using polarization sums like \eqref{EQ:polsumMBR}.} but the two 
expressions agree of course in D=4 dimensions.~\\

Before we move on to establish explicit momentum basis representations of longitudinal polarization
vectors for massive vector bosons and also for massive fermions, let us emphasize that 
by construction these momentum basis representations of polarization state vectors 
fulfill all the defining physical constraints, i.e., orthogonality to momenta and 
reference vectors, which are assured even if the open Lorentz indices (carried
by either the external momenta or the Levi-Civita symbol) are taken to be  D-dimensional.

Mathematically, the procedure of determining norm-orthogonal polarization vectors
eqs.~\eqref{EQ:XpolEQs}, \eqref{EQ:TpolEQs} from a given set of linearly independent momenta 
in 4 dimensions resembles the Gram-Schmidt orthogonalization procedure. 
Our key insight here is that we establish these Lorentz covariant decomposition 
representations initially in 4 dimensions in a form that facilitates the subsequent promotion 
of their open Lorentz indices from 4 to D, resulting in expressions which will be 
taken as their \textit{definitions} in D dimensions.

\subsubsection{Massive particles in the final state}

Next we consider the scattering process eq.~(\ref{momentaassignment}) but 
with massive final-state vector bosons, for instance W or Z bosons, 
with on-shell conditions 
\begin{eqnarray} \label{EQ:onshellmassive}
p_1^2 = p_2^2 = 0~,~~ p_3^2 = p_4^2 = m^2 \, . 
\end{eqnarray}

Concerning the three elementary physical polarization state vectors, 
$\varepsilon^{\mu}_{X}~,~ \varepsilon^{\mu}_{T}~,~ \varepsilon^{\mu}_{Y}$, 
the above constructions can be repeated but with slightly different 
kinematics.
It is straightforward to arrive at the following explicit representations:
\begin{eqnarray} \label{EQ:TranspolMBRmassive}
\varepsilon^{\mu}_{X} &=& \mathcal{N}_{X} \, 
\big((t-m^2)~ p^{\mu}_1 + (-s-t+m^2)~ p^{\mu}_2 + s~ p^{\mu}_3 \big) \, ,\nonumber\\
\varepsilon^{\mu}_{T} &=&  \mathcal{N}_{T} 
\big((t+m^2)~ p^{\mu}_1 + (s+t-3m^2)~ p^{\mu}_2 + (-s-2 t+2m^2)~ p^{\mu}_3 \big) \, ,\nonumber\\
\varepsilon^{\mu}_{Y} &=& \mathcal{N}_{Y}~ 2\epsilon^{\mu}_{p_1 p_2 p_3} \, ,
\end{eqnarray}
with the normalization factors 
\begin{eqnarray} \label{EQ:norms_massive}
\mathcal{N}_{X}^{~-2} &=& s \left(-t (s + t) + 2 m^2 t - m^4 \right) \, ,\nonumber\\
\mathcal{N}_{T}^{~-2} &=& - s t (s + t) + 2 m^2 t (3 s + 2 t) - m^4 (s + 8 t) + 4 m^6 \, , \nonumber\\
\mathcal{N}_{Y}^{~-2} &=& s \left(-t (s + t) + 2 m^2 t - m^4\right) \, ,
\end{eqnarray}
which, as already emphasized above, could be conveniently chosen to be 
incorporated only at the last stage of the computation.~\\

Compared to the massless case, the helicity eigenstates of massive gauge bosons are 
reference-frame dependent and their helicities are not Lorentz-invariant.  
Helicity eigenstates constructed from the above elementary linear polarization state 
vectors are defined in the center-of-mass reference frame of the two colliding particles.
The third physical polarization state of a massive gauge boson is described by 
the longitudinal polarization vector (defined in the same reference frame), 
which has its spatial part aligned with the momentum of the boson.
For the massive particle $g_3(p_3)$ these conditions translate into 
the following set of equations for its longitudinal polarization vector $\varepsilon_{L3}^{\mu}$:
\begin{eqnarray} \label{EQ:L3polEQs}
&&\varepsilon_{L3}^{\mu}  = c^{L3}_1 \left( p^{\mu}_1 + p^{\mu}_2 - p^{\mu}_3 \right) 
+ c^{L3}_2 p^{\mu}_3  \, , \nonumber\\ 
&&\varepsilon_{L3}^{\mu} \cdot p_3 = 0 \, ,\nonumber\\
&&\varepsilon_{L3} \cdot \varepsilon_{L3} = -1 \, .
\end{eqnarray}
Solving eq.~(\ref{EQ:L3polEQs}) for $c^{L3}_1,~ c^{L3}_2$, one obtains 
\begin{eqnarray} \label{EQ:L3polMBR}
\varepsilon^{\mu}_{L3}  = \mathcal{N}_{L3} 
\big(-2 m^2~ \left( p^{\mu}_1 + p^{\mu}_2 \right) + s~ p^{\mu}_3 \big) \, ,
\end{eqnarray}
where $\mathcal{N}_{L3}^{~-2} = s m^2 (s-4 m^2)$. 
For the massive vector boson $g_4(p_4)$ one gets for its longitudinal polarization vector $\varepsilon_{L4}^{\mu}$: 
\begin{eqnarray} \label{EQ:L4polMBR}
\varepsilon^{\mu}_{L4}  = \mathcal{N}_{L4} 
\big((s -2 m^2)~ \left( p^{\mu}_1 + p^{\mu}_2 \right) - s~ p^{\mu}_3 \big) \, ,
\end{eqnarray}
where $\mathcal{N}_{L4} = \mathcal{N}_{L3}$.
 By construction the defining physical properties, such as orthogonality to the momenta, 
are fulfilled by these momentum basis representations, even if 
their open Lorentz indices are taken to be  D-dimensional. 
We emaphasize that in our prescription the number of physical polarizations of a massive vector boson 
remains three in D dimensions.~\\

There are also polarization vectors associated with massive fermions. 
The helicity eigenstate of a massive fermion with 4-momentum $k$ can be described by 
a Dirac spinor, e.g.~$u(k,S_k)$, characterized by the normalized space-like polarization 
vector $S_k^{\mu}$. Its components are 
\begin{eqnarray} 
S_k^{\mu} = \Big(\frac{|\vec{k}|}{m}, \frac{k^0}{m} \frac{\vec{k}}{|\vec{k}|}\Big) \, ,
\end{eqnarray} 
where $k^0$ and $m$ are, respectively, the energy and mass of the massive fermion,
while $\vec{k}$ represents its 3-momentum. 
Interestingly, this polarization vector has the same momentum basis decomposition form 
as the longitudinal polarization vector of a massive vector boson 
(of the same momentum), provided the same external kinematic configuration applies. 
By identifying $p_3^{\mu} = k^{\mu}$, eq.~(\ref{EQ:L3polMBR}) can be viewed as the momentum 
basis representation of $S_k^{\mu}$ for the same external kinematic configuration as above. 
Namely, 
\begin{eqnarray} \label{EQ:PLpolMBR}
S_k^{\mu} = \mathcal{N}_{S_k} 
\Big(-2 m^2~\left( p^{\mu}_1 + p^{\mu}_2 \right) + s~ k^{\mu} \Big)
\end{eqnarray}  
with $\mathcal{N}_{S_k}^{~-2} = s m^2 (s-4 m^2)$.
This is because the set of norm-orthogonal conditions that $S^{\mu}$ has to fulfill, 
namely $k \cdot S = 0~,~ S \cdot S = -1~,~ \vec{S} \parallel \vec{k}$, 
which are sufficient to determine it up to an overall phase factor, 
are exactly the same as those that the longitudinal polarization vector in eq.~(\ref{EQ:L3polMBR}) 
has to fulfill.

\subsection{Normalized tensor products of external spinors}
\label{SEC:prescription:NTS}

In cases where external fermions are involved in scattering amplitudes, 
strings of Dirac matrices sandwiched between external on-shell spinors 
will show up. In order to evaluate each  open fermion line using 
\textit{trace} techniques, we employ the standard trick of multiplying and dividing 
this quantity by appropriate auxiliary Lorentz-invariant spinor inner products,
which can be traced back to~ref.~\cite{Bjorken:1966kh}.
Pulling out the chosen overall Lorentz-invariant normalization factor, 
the rest can be cast into a trace of products of Dirac matrices with the aid 
of Landau density matrices of external spinors.
The momentum basis representations of space-like  (massive) fermion polarization vectors, such as eq.~(\ref{EQ:PLpolMBR}), can be used 
 in these density matrices.
For massless fermions, the spin density matrices are reduced to 
left- respectively right-chirality projectors, which thus spares us from introducing 
any explicit polarization vector in this case. 
This is because helicity states of massless fermions coincide with chiral spinors. ~\\

From a single open fermion line in a Feynman diagram, we get a contribution 
which can be generically written as $\langle \psi_A|\,\hat{\mathrm{M}}\,|\psi_B \rangle$.
The symbol $\hat{\mathrm{M}}$ denotes a product of Dirac matrices with their Lorentz indices 
either contracted or left open, and $|\psi_A\rangle,~|\psi_B \rangle$ stand for 
the two external on-shell Dirac spinors, either of $u$-type or $v$-type, of this open fermion line. 
Viewed as a spinor inner product, $\langle \psi_A|\,\hat{\mathrm{M}}\,|\psi_B \rangle$  
can always be rewritten as a trace of a product of Dirac-matrices 
in the Dirac-spinor space:
\begin{eqnarray} \label{EQ:SFLtrace1}
\langle \psi_A|\,\hat{\mathrm{M}}\,|\psi_B \rangle = \mathrm{Tr} \Big{[} 
|\psi_B \rangle  \langle \psi_A|\,\hat{\mathrm{M}} \Big{]}.
\end{eqnarray}
This formal rewriting is not really useful unless we can further 
exploit the matrix structure of the external spinors' tensor product $|\psi_B \rangle \langle \psi_A|$ 
in the spinor space (explicitly in terms of elementary Dirac matrices), so as to apply trace techniques. 
To this end, we rewrite $|\psi_B \rangle \langle \psi_A|$ 
by introducing an auxiliary spinor inner product along the following line:
\begin{eqnarray} \label{EQ:TPextsps1}
|\psi_B \rangle \langle \psi_A| 
&=& 
\frac{\langle \psi_B|\,\hat{\mathrm{N}}\,|\psi_A \rangle }{\langle \psi_B|\,\hat{\mathrm{N}}\,|\psi_A \rangle }
|\psi_B \rangle  \langle \psi_A| \nonumber\\
&=& \frac{1}{\langle \psi_B|\,\hat{\mathrm{N}}\,|\psi_A \rangle}
|\psi_B \rangle \langle \psi_B|\,\hat{\mathrm{N}}\,|\psi_A \rangle \langle \psi_A| \nonumber\\
&=& \mathcal{N}_{AB}~
|\psi_B \rangle  \langle \psi_B|\, \hat{\mathrm{N}} \,|\psi_A \rangle \langle \psi_A| \, , 
\end{eqnarray}
where $\mathcal{N}_{AB} \equiv ({\langle \psi_B|\,\hat{\mathrm{N}}\,|\psi_A \rangle})^{-1}$.
The auxiliary matrix $\hat{\mathrm{N}}$ is only required to have a non-vanishing
matrix element $\langle \psi_B|\hat{\mathrm{N}}|\psi_A \rangle$, and otherwise can be chosen 
to be as simple as desired.
For instance, for massive external spinors of some particular helicity configurations,
 $\hat{\mathrm{N}}$ may be chosen to be the identity matrix in spinor space, provided that 
the spinor inner product between those helicity spinors is not vanishing. 
A generally valid and simple choice is $\hat{\mathrm{N}} = \gamma_{\mu} p^{\mu}$ 
with a 4-momentum $p^{\mu}$ that is not linearly dependent on the on-shell momenta $p_A$ and $p_B$
of $\langle \psi_A|$ and $|\psi_B \rangle$, respectively.

We manipulate eq.~(\ref{EQ:TPextsps1}) further by first substituting the Landau density matrices for 
$|\psi_A \rangle \langle \psi_A|$ and $|\psi_B \rangle \langle \psi_B|$, conventionally given by 
\begin{eqnarray} \label{EQ:LDMofDSP}
u(p,S_p) \otimes \bar{u}(p,S_p) &=& \left(\slashed{p} + m \right) \frac{1+\gamma_5 \slashed{S}_p}{2} \, , \nonumber\\
v(p,S_p) \otimes \bar{v}(p,S_p) &=& \left(\slashed{p} - m \right) \frac{1+\gamma_5 \slashed{S}_p}{2} \, .
\end{eqnarray} 
Then we simplify the resulting composite Dirac matrix object 
before finally obtaining a form that is suitable for being unambiguously used 
in eq.~(\ref{EQ:SFLtrace1}) with the trace to be done in D dimensions.

There are several equivalent forms of these on-shell Dirac-spinors' projectors 
in 4 dimensions. In particular, one may commute the on-shell projection 
operator $\slashed{p} \pm m$ and the polarization projection operator 
$({1+\gamma_5 \slashed{S}_p})/{2}$ using $p \cdot S_p = 0$ and the anticommutativity of $\gamma_5$.
However, it is well known that a fully anticommuting $\gamma_5$ can not be thoroughly implemented in dimensional regularization in an algebraically consistent way (see e.g.~\cite{Collins:1984xc}), 
if we still want this object to coincide with the usual $\gamma_5$ in 4 dimensions. 
In this article, we adopt a particular variant of a non-anticommuting $\gamma_5$ prescription 
 formulated in ref.~\cite{Larin:1991tj,Larin:1993tq}, 
conventionally known as Larin's scheme, whose equivalent but more efficient implementations in high-order 
perturbative calculations are discussed in ref.~\cite{Moch:2015usa}.  
In principle, one could distinguish the $\gamma_5$ appearing in the external projectors and inside the amplitude. 
The prescription for the \textit{external} projectors proposed here is not tied to applying a non-anticommuting $\gamma_5$ prescription to the axial currents or other $\gamma_5$-related objects \textit{inside} the amplitudes. 
Any appropriate $\gamma_5$ prescription can of course be used as long as its application to the amplitudes in question is justified.
For the sake of clarity, the appearances of the symbol $\gamma_5$ in the following, in particular in the 
computations presented in the examples of this article, should be regarded just for bookkeeping purposes and their interpretations is based on~\cite{Larin:1991tj,Larin:1993tq,Moch:2015usa}. 
As a consequence of this prescription, $\gamma_5$ no longer anticommutes with all 
Dirac $\gamma$ matrices, and 4-dimensional equivalent forms of eq.~(\ref{EQ:LDMofDSP}) 
are no longer necessarily algebraically equivalent in D dimensions.

In order to eliminate potential ambiguities --- after having  simplified eqs.~(\ref{EQ:SFLtrace1},\ref{EQ:TPextsps1},\ref{EQ:LDMofDSP}) using 4-dimensional Lorentz and Dirac algebra as much as possible ---, we should agree on one definite fixed form of eqs.~(\ref{EQ:SFLtrace1},\ref{EQ:TPextsps1},\ref{EQ:LDMofDSP}), solely in terms of a string of Dirac $\gamma$ matrices with fixed product ordering, the Levi-Civita tensor, and external momenta. 
We may call these their canonical forms in 4 dimensions. 
This allows an unambiguous interpretation\footnote{This is up to a potential subtlety 
related to the contraction order among multiple Levi-Civita tensors~\cite{Moch:2015usa}, 
as will be commented on in section~\ref{SEC:examples:eeQQ}.} 
of the expression in D dimensions where it will be manipulated according to the $D$-dimensional algebra 
after being inserted back into eq.~(\ref{EQ:SFLtrace1}).

Let us now be more specific about this by working out a representative case, 
a single open fermion line with two massive external $u$-type spinors, $u(p_A,S_A)$ and $u(p_B,S_B)$.
We choose $\hat{\mathrm{N}} = \slashed{q}$ where $q^{\mu}$ is a 4-momentum 
that is linearly independent of $p_A$ and $p_B$. 
Pulling out the normalization factor $\mathcal{N}_{AB} = \left({\bar{u}(p_A,S_A)\, \slashed{q}\, u(p_B,S_B)}\right)^{-1}$, 
eq.~(\ref{EQ:TPextsps1}) reads in this case:
\begin{eqnarray}\label{EQ:TPextspsaux}
\frac{1}{\mathcal{N}_{AB}}~ u(p_B,S_B) \otimes \bar{u}(p_A,S_A) 
&=& 
\left(\slashed{p}_B + m \right) 
\frac{1+\gamma_5 \slashed{S}_B}{2}
\slashed{q}
\frac{1+\gamma_5 \slashed{S}_A}{2} 
\left(\slashed{p}_A + m \right) \, ,
\end{eqnarray}
which can be brought into the form
\begin{eqnarray} \label{EQ:TPextsps2}
\frac{1}{\mathcal{N}_{AB}}~ u(p_B,S_B) \otimes \bar{u}(p_A,S_A) 
&=& \left(\slashed{p}_B + m \right) 
\frac{1}{4}\slashed{q}
\left(\slashed{p}_A + m \right) \nonumber\\
&+& \left(\slashed{p}_B + m \right) 
\frac{1}{4} 
\left(\frac{-i}{3!} \epsilon_{\gamma \gamma \gamma S_B} \right) \slashed{q}
\left(\slashed{p}_A + m \right) \nonumber\\
&+& \left(\slashed{p}_B + m \right) 
\frac{1}{4}\slashed{q} 
\left(\frac{-i}{3!} \epsilon_{\gamma \gamma \gamma S_A} \right)
\left(\slashed{p}_A + m \right) \nonumber\\
&+& \left(\slashed{p}_B + m \right) 
\frac{1}{4} \slashed{S}_B \slashed{q} \slashed{S}_A 
\left(\slashed{p}_A + m \right) \, .
\end{eqnarray} 
Strictly speaking, eq.~(\ref{EQ:TPextsps2}) is identical to \eqref{EQ:TPextspsaux} only in 4 dimensions. 
The unambiguous eq.~(\ref{EQ:TPextsps2}), which no longer contains any explicit $\gamma_5$, will be taken as the definition of  $u(p_B,S_B) \otimes \bar{u}(p_A,S_A)$ when it is inserted into eq.~(\ref{EQ:SFLtrace1}) and manipulated in accordance with the D-dimensional algebra.

Notice that in eq.~\eqref{EQ:TPextspsaux} the auxiliary matrix 
$\slashed{q}$ and the polarization projection operators were 
placed inside the on-shell projection operators $\slashed{p}_{I} +  m$ $(I=A,B)$,
a point which will be explained and become clear in section~\ref{SEC:unitarity:PSA}.  
We emphasize again that the momentum basis representations of the
 helicity polarization vectors $S^{\mu}_A$ and $S^{\mu}_B$ of massive fermions
will be eventually inserted, whose open Lorentz indices are carried 
by external momenta that are assumed to be D-dimensional.
Similar rewritings and definitions like eq.~(\ref{EQ:TPextsps2}) can be made 
also for fermion lines with external $v$-type  spinors, whose Landau density 
matrices are given in eq.~(\ref{EQ:LDMofDSP}).

In practice, it is very convenient to keep projections associated with each of 
the four terms in eq.~(\ref{EQ:TPextsps2}) separate from each other, 
at least in the initial stage with unreduced amplitudes, for two reasons.
First, this organization is in accordance with the power of the Levi-Civita tensor 
appearing in the terms, which is advantageous especially when the contributing Feynman diagrams 
are also split into terms with even and odd products of $\gamma_5$ (arising from, e.g.,~axial vertices).  
Second, for a fermion with fixed momentum, its polarization vector, e.g.~$S_A$ or $S_B$ in eq.(\ref{EQ:TPextsps2}), 
changes just by an overall minus sign when its helicity is flipped.
Therefore, the expressions of eq.~(\ref{EQ:SFLtrace1}) for the four different helicity configurations
can all be obtained by suitably combining the traces in eq.~(\ref{EQ:SFLtrace1}) of the product 
of $\hat{\mathrm{M}}$ and each of the four terms in eq.~(\ref{EQ:TPextsps2}).
Using \eqref{EQ:TPextsps2} we need to project these four individual projections separately just once, 
out of which all four different helicity configurations can be obtained. 
 Notice that in general the normalization factor $\mathcal{N}_{AB}$ in eq.~(\ref{EQ:SFLtrace1}) 
depends on the helicities of the external fermions A and B, 
as will be explicitly shown in the example given in section~\ref{SEC:examples:eeQQ}.

Once a definite unambiguous form of the right-hand side of eq.~(\ref{EQ:TPextsps2}) has been established 
in 4 dimensions, it will be kept fixed while all open Lorentz and Dirac indices will be promoted 
in accordance with computations in CDR. 
Additionally, just like the aforementioned normalization factors associated with the gauge boson's polarization vectors, 
the factor $\mathcal{N}_{AB}$ in eq.~(\ref{EQ:SFLtrace1}) is an overall normalization factor 
which must be adopted consistently in computing all amplitudes involved in the calculations of finite remainders. 
If one chooses to incorporate this overall normalization factor in the very last stage of calculating 
finite remainders where the four-dimensional limit can already be taken, 
it is then evident that we can evaluate these Lorentz invariant factors in 4 dimensions.~\\

As already mentioned above, in the massless limit the spin density matrices in eq.~(\ref{EQ:LDMofDSP}) 
are reduced to left- or right-chiral projectors. Thus no polarization vectors are needed.
For instance, the massless limit of eq.~(\ref{EQ:TPextsps2}) with $++$ helicity configuration reads: 
\begin{eqnarray} \label{EQ:TPextsps2massless}
\frac{1}{\mathcal{N}_{AB}}~ u(p_B,+) \otimes \bar{u}(p_A,+) 
&=& 
\slashed{p}_B  
\frac{1-\gamma_5 }{2}
\slashed{q}
\frac{1+\gamma_5}{2} 
\slashed{p}_A \nonumber\\
&=& \frac{1}{2}  
\Big(
\slashed{p}_B  
\slashed{q} 
\slashed{p}_A 
- \slashed{p}_B  
\Big(\frac{-i}{3!} \epsilon_{\gamma \gamma \gamma q} \Big)
\slashed{p}_A \Big). 
\end{eqnarray}
The remarks below eq.~\eqref{EQ:TPextsps2} concerning the use of this equation in D-dimensional calculations
apply also to \eqref{EQ:TPextsps2massless}.
The above reformulations of tensor products of two external helicity spinors, 
such as eq.~(\ref{EQ:TPextsps2}) and eq.~(\ref{EQ:TPextsps2massless}), 
can be applied to each single open fermion line, besides using for each external boson
the momentum basis representation of its polarization vector.~\\

To summarize, the tensor product of momentum basis representations of all external gauge bosons' polarization vectors 
and all properly re-written external spinor products, such as those given by 
eqs.~\eqref{EQ:XpolMBR} - \eqref{EQ:YpolMBR}, and eq.~(\ref{EQ:TPextsps2}), (\ref{EQ:TPextsps2massless}), 
with their open indices promoted in accordance with CDR, will be taken as the external projectors for polarized amplitudes. 
Polarized amplitudes, at least in their unreduced form,  are thus first projected in the linear polarization basis for external gauge bosons 
and the basis indicated by eq.~(\ref{EQ:TPextsps2}), (\ref{EQ:TPextsps2massless}) 
for each open fermion line, before being subsequently combined to form helicity amplitudes. 
It is a good practice to first combine Levi-Civita tensors that appear in external projectors in order to 
reach an unambiguous canonical form that is homogeneous in the Levi-Civita tensor whose power is at most one,
at least for scattering processes with less than 5 external particles.
(See the next section for discussions of $2\to 3$ processes.) 
As a consequence of this operation, for some projectors new D-dimensional non-factorized tensors may arise that are
 different  from the original bookkeeping forms.  
Helicity amplitudes can be subsequently obtained from these  polarized amplitudes by
linear combinations, such as those implied in eq.~(\ref{EQ:LP2HLmassless}) --  
although it may be case-dependent when it is beneficial to perform this combination.
(See section~\ref{SEC:prescription:comments} for more discussions.) 
The transformation matrix of polarized scattering amplitudes among four massless gauge bosons from the linear to the circular polarization basis is a $16\times16$ constant matrix that can be extracted from eq.~(\ref{EQ:LP2HLmassless}). 
Likewise, constant transformation matrices can also be extracted from eq.~(\ref{EQ:TPextsps2}) 
for massive fermion lines and from eq.~(\ref{EQ:TPextsps2massless}) for massless fermion lines.~\\

Eventually every helicity amplitude composed in this way is manifestly given as a function of Lorentz invariant 
variables solely dependent on external momenta and  the space-time dimension D. 
This is owing to the fact that the momentum basis representations of polarization vectors 
allow us to find a Lorentz covariant representation of the tensor product of external particle states 
solely in terms of external momenta and algebraic constants, 
such as the space-time metric tensor, the Levi-Civita tensor and Dirac matrices, 
which permits a formal D-dimensional extension. 
Subsequently, this makes it feasible to directly take these objects as the external polarization projectors.

From the point of view of the projection method as outlined in section~\ref{SEC:projectionmethod:recap},
the set of external polarization projectors described above might be loosely viewed as 
a special choice of Lorentz decomposition basis which by construction are orthogonal among each other.
Consequently, the corresponding Gram matrix is diagonal and its inversion is trivial. 
Furthermore, each structure that arises from such a decomposition is directly related to a physical quantity, 
and therefore its (explicit and/or implicit) singularity structure is protected by physical constraints obeyed by these physical quantities. 
In addition, the transformations from these primary projections to the helicity amplitudes are constants that can be easily extracted. 
In this way the issues related to the conventional form-factor decomposition as discussed in section~\ref{SEC:projectionmethod:comments} are circumvented, similar to how this is achieved in the computation of polarized amplitudes using spinor-helicity representations but now in a manifestly CDR-compatible way.

\subsection{Projectors for $2\to 3$ scattering processes}
\label{SEC:prescription:beyond4P}

In the preceding sections we have discussed a prototype $2 \to 2$ scattering process where there are only three linearly independent external momenta, and consequently 
the Lorentz-invariant projected amplitude cannot contain a term composed of one Levi-Civita tensor fully contracted with external momenta.
This fact can lead to a reduction of terms that are to be included in external projectors. 
For instance, if the $2\to 2$ scattering process is parity-invariant, then all terms in the external projectors that are linear in Levi-Civita tensor can be dropped from the outset. 
This simplification no longer occurs if there are more than four particles involved in the scattering, e.g.,~in a $2\to 3$ process.
However, having at hand a complete set of linearly independent 4-momenta in the 4-dimensional Minkowski spacetime, offers an opportunity to eliminate the explicit appearance of the Levi-Civita tensor $\epsilon^{\mu\nu\rho\sigma}$ from the external projectors by applying the trick used in defining the van Neerven-Vermaseren basis~\cite{vanNeerven:1983vr} (see, e.g.,~eq.~(\ref{EQ:NVtrick}) below).
The same trick can be applied to any other Lorentz tensors as well, including the spacetime-metric tensor in 4 dimensions. 
Below we discuss a few technical aspects of applying the proposed projection prescription to the scattering process with 5 external particles. 
In particular, we mainly focus on the cases with 5 (massless) gauge bosons, which has the highest rank as a Lorentz tensor, 
while the presence of fermions can be dealt with by combining with the discussion in section~\ref{SEC:prescription:NTS}.\footnote{See ref.~\cite{Chawdhry:2020for} for how this prescription is applied in the computation of the helicity amplitudes for the process $q\bar{q} \rightarrow \gamma \gamma \gamma$ at 2-loop order.}
~\\

The construction procedure devised in section~\ref{SEC:prescription:MBR} works for a vector boson, massless or massive, with an arbitrary choice of reference vectors in any scattering process.
In particular, the reference vector choice made in section~\ref{SEC:prescription:MBR:2to2} amounts to taking the ``beam-axis''
vector $p_1 + p_2$ as the reference vector for all 4 external gauge bosons, because shifting a reference vector $r^{\mu} \rightarrow r^{\mu} + \alpha \, p^{\mu}$, where $p^{\mu}$ is the momentum of the gauge boson, does not change any physical amplitude. (See Appendix~\ref{append:ffvp} for an extension to more general cases)
Here, we would like to present a compact explicit formula for linear polarization states of a (massless) gauge boson 
that can be conveniently used in any multiple-parton scattering process in massless QCD, e.g.~5-gluon scattering amplitudes.

Let us denote the light-like momentum of the gauge boson by $p^{\mu}$ and its gauge-reference vector $r^{\mu}$ is chosen 
to be light-like as well, just for the sake of simplicity, where $r \cdot p \neq 0$.  
In addition, we assume that there exists another auxiliary Lorentz vector, denoted by $q^{\mu}$, which is required 
to be linearly independent of $p^{\mu}$ and $r^{\mu}$. 
Under this condition, one can repeat the procedure of section~\ref{SEC:prescription:MBR} and 
 arrive at the following formula for the two physical linear polarization states of a gauge boson:
\begin{eqnarray} \label{EQ:TranspolMBRGeneric}
\varepsilon^{\mu}_{X} &=& \mathcal{N}_{X} 
\Big(p \cdot r \, q^{\mu} - \left( q \cdot r \, p^{\mu} + q \cdot p \, r^{\mu} \right) \Big) , \nonumber\\
\varepsilon^{\mu}_{Y} &=& \mathcal{N}_{Y}~ \epsilon^{\mu}_{p\, q\, r} \, .
\end{eqnarray}
With a light-like $q^{\mu}$, the normalization factors read 
$\mathcal{N}_{X} = \mathcal{N}_{Y} = 1/\sqrt{2 p \cdot q \, p \cdot r \, q \cdot r}$. 
By the same procedure, a similar formula can be  derived also for the case with a non-lightlike reference vector $r^{\mu}$.
Eq.~(\ref{EQ:TranspolMBRGeneric}) resembles the Voronov polarization vectors~\cite{Voronov:1973kga,DeCausmaecker:1981wzb}, apart from the previously discussed promotion of open Lorentz indices formally to D dimensions, the appearance of a term proportional to $p^{\mu}$ in $\varepsilon^{\mu}_{X}$ as well as the treatment of $\epsilon^{\mu}_{p\, q\, r}$ to be discussed below.
With all three Lorentz vectors $\big\{ p^{\mu}, r^{\mu}, q^{\mu} \big\}$ being light-like with a positive temporal component, 
the dot products under the square root in the normalization factors are always non-negative.
As long as $q^{\mu}$ is linearly independent of $\big\{ p^{\mu}, r^{\mu} \big\}$, $\varepsilon^{\mu}_{X}$ will not become identically zero.
Geometrically it is not hard to see that the direction of $\varepsilon^{\mu}_{X}$ in eq.~(\ref{EQ:TranspolMBRGeneric}) 
is given by $q^{\mu}$ after subtracting from the latter all components that can be linearly composed out
of $\big\{ p^{\mu}, r^{\mu} \big\}$.
It is straightforward to check that the two vectors defined in eq.~(\ref{EQ:TranspolMBRGeneric}) satisfy all 
requirements for being representations of the two physical polarization states of a massless gauge boson 
with momentum $p$ and a reference vector $r$.
In addition, their subscripts $X,Y$ indicate that the usual helicity states (or circular polarization states) of this massless
(incoming) gauge boson can be obtained by $\varepsilon^{\mu}_{\pm} = \frac{1}{\sqrt{2}}\, \left(\varepsilon^{\mu}_{X} \, \pm\, i \varepsilon^{\mu}_{Y} \right)$.
One can go to the center-of-mass frame of $p+r$ to get an illustrative geometric picture of these polarization vectors.
In particular, with the term proportional to $p^{\mu}$ included in eq.~(\ref{EQ:TranspolMBRGeneric}), 
the normalization factor $\mathcal{N}_{X}$ in this reference frame has only 
spatial components transverse to $\vec{p}$ that are not vanishing.
~\\

The formula eq.~(\ref{EQ:TranspolMBRGeneric}) works for any massless gauge boson in an arbitrary multiple-parton 
scattering process in massless QFT.
However, as mentioned already, when there are at least 5 external particles involved (non-trivially) in the scattering,
e.g.,~a $2\to 3$ process, one has the opportunity to eliminate the explicit appearance of the Levi-Civita tensor 
$\epsilon^{\mu\nu\rho\sigma}$ from external projectors (i.e.,~in $\epsilon_Y$), by applying the trick used in defining 
the van Neerven-Vermaseren basis~\cite{vanNeerven:1983vr}.
To be more specific, let us consider a $2\to 3$ scattering process where the four linearly independent four-momenta 
are denoted by $p_1, p_2,p_3,p_4$.
A single power of $\epsilon^{\mu\nu\rho\sigma}$ in an external polarization projector can be rewritten as 
\begin{eqnarray} \label{EQ:NVtrick}
\epsilon^{\mu\nu\rho\sigma} &=& 
\frac{\epsilon_{p_1 p_2 p_3 p_4}}{\epsilon_{p_1 p_2 p_3 p_4}} 
~\epsilon^{\mu\nu\rho\sigma} \nonumber\\
&=& \Delta ~  
\Big(
p_1^{\rho } p_2^{\nu } p_3^{\mu } p_4^{\sigma }-p_1^{\nu } p_2^{\rho } p_3^{\mu } p_4^{\sigma }-p_1^{\rho } p_2^{\mu } p_3^{\nu } p_4^{\sigma }+p_1^{\mu } p_2^{\rho } p_3^{\nu } p_4^{\sigma }+p_1^{\nu } p_2^{\mu } p_3^{\rho } p_4^{\sigma }-p_1^{\mu } p_2^{\nu} p_3^{\rho } p_4^{\sigma }
\nonumber\\&&~~~~~
-p_1^{\rho } p_2^{\nu } p_3^{\sigma } p_4^{\mu }+p_1^{\nu } p_2^{\rho } p_3^{\sigma } p_4^{\mu}+p_1^{\rho } p_2^{\sigma } p_3^{\nu } p_4^{\mu }-p_1^{\sigma } p_2^{\rho } p_3^{\nu } p_4^{\mu }-p_1^{\nu } p_2^{\sigma }
p_3^{\rho } p_4^{\mu }+p_1^{\sigma } p_2^{\nu } p_3^{\rho } p_4^{\mu }
\nonumber\\&&~~~~~
+p_1^{\rho } p_2^{\mu } p_3^{\sigma } p_4^{\nu }-p_1^{\mu }p_2^{\rho } p_3^{\sigma } p_4^{\nu }-p_1^{\rho } p_2^{\sigma } p_3^{\mu } p_4^{\nu }+p_1^{\sigma } p_2^{\rho } p_3^{\mu } p_4^{\nu
   }+p_1^{\mu } p_2^{\sigma } p_3^{\rho } p_4^{\nu }-p_1^{\sigma } p_2^{\mu } p_3^{\rho } p_4^{\nu }
\nonumber\\&&~~~~~
-p_1^{\nu } p_2^{\mu } p_3^{\sigma } p_4^{\rho }+p_1^{\mu } p_2^{\nu } p_3^{\sigma } p_4^{\rho }+p_1^{\nu } p_2^{\sigma } p_3^{\mu } p_4^{\rho}-p_1^{\sigma } p_2^{\nu } p_3^{\mu } p_4^{\rho }-p_1^{\mu } p_2^{\sigma } 
p_3^{\nu } p_4^{\rho }+p_1^{\sigma } p_2^{\mu } p_3^{\nu } p_4^{\rho } \Big),
\nonumber\\
\end{eqnarray}
where the normalization factor $\Delta \equiv \frac{1}{\epsilon_{p_1 p_2 p_3 p_4}}$ can be conveniently pulled out and
grouped together with other normalization factors of external projectors (and used consistently throughout the whole calculation). 
The treatment of the Levi-Civita tensor in eq.~(\ref{EQ:NVtrick}) complies with the two rules listed in section~\ref{SEC:prescription:MBR}. 
Eq.~(\ref{EQ:NVtrick}) can be formally regarded as the momentum basis representation of the Levi-Civita tensor.
This is made feasible because one now has at hand a set of linearly independent four-momenta forming a complete basis 
of the 4-dimensional Minkowski space-time.
This, of course, holds also for other Lorentz tensors with a different rank (See Appendix~\ref{append:vNVbasis}), 
in particular the space-time metric tensor in 4 dimensions.
One can obtain this decomposition either by performing the Gram-Schmidt orthogonalization procedure, like in section~\ref{SEC:prescription:MBR}, or one can even directly read off the decomposition by making use of the van Neerven-Vermaseren basis~\cite{vanNeerven:1983vr}. 
With eq.~(\ref{EQ:NVtrick}), no Levi-Civita tensor appears in external polarization projectors for $2 \to 3$ gluon-scattering amplitudes any more, up to a global normalization factor, and hence it is manifest that the form of external projectors can be unambiguously constructed.
At this point, it is worthy to mention that FORM~\cite{Vermaseren:2000nd} has a built-in (pseudo) Levi-Civita tensor where one has $\epsilon\_(p_1, p_2, p_3, p_4) = -i\epsilon_{p_1 p_2 p_3 p_4} = \frac{1}{4} \mathrm{Tr}[\gamma_5\, \slashed{p}_1\, \slashed{p}_2 \,\slashed{p}_3\, \slashed{p}_4]$. 
The imaginary unit $i$ in the linear superposition formula connecting the linearly polarized 
amplitudes and helicity (i.e.,~circularly polarized) amplitudes can thus always be absorbed into 
quantities defined by a consistent usage of FORM's (pseudo) Levi-Civita tensor.

Thus when there are at least 5 external particles involved, one could rewrite the $\epsilon^{\mu}_{p\, q\, r}$ in eq.~(\ref{EQ:TranspolMBRGeneric}) for $\varepsilon^{\mu}_{Y}$ by making use of the momentum basis representation of the Levi-Civita tensor eq.~(\ref{EQ:NVtrick}).
Afterwards, all open Lorentz indices carried by the momenta therein are promoted to be D-dimensional just as done in previous sections.
Just as discussed in section~\ref{SEC:prescription:MBR} and~\ref{SEC:prescription:NTS} for $2 \to 2$ scatterings, the polarization projectors for a $2 \rightarrow 3$ scattering amplitude among gauge bosons will be given conveniently by the tensor products of momentum basis representations of all external gauge bosons' polarization vectors as determined in eq.~(\ref{EQ:TranspolMBRGeneric}) with the aid of eq.~(\ref{EQ:NVtrick}).
Note that in this way, both $\varepsilon^{\mu}_{X}$ and $\varepsilon^{\mu}_{Y}$ are given explicitly in terms of linear combinations of just external momenta, and thus there is no more explicit appearance of the space-time metric tensor $g_{\mu\nu}$ in the linear polarization projectors.\footnote{Since there is no more need to contract pairs of Levi-Civita tensors in this treatment, the polarization projector for the scattering amplitude with $N$$\geq$5 gauge bosons as a whole remains strictly a factorized product of the individual polarization vectors.}
In consequence, the helicity amplitudes reconstructed from these projections automatically comply with those defined in the HV scheme.

With both $\varepsilon^{\mu}_{X}$ and $\varepsilon^{\mu}_{Y}$ given explicitly and solely in terms of gauge bosons' momenta, one could then trim the projectors by the virtue of Ward identities in a local gauge theory.
To be more specific about this, one can drop in $\varepsilon^{\mu}_{X}$ and $\varepsilon^{\mu}_{Y}$ terms proportional to the momentum of the corresponding gauge boson (e.g.~gluon).
This kind of trimming on linear-polarization projectors is allowed because the orthogonality between each linear-polarization vector and the corresponding gauge boson's momentum is not affected by this.
Note that the contraction rule associated with this trimmed version of linear-polarization projectors is still simply the space-time metric tensor $g_{\mu\nu}$.
If one would use the physical polarization sum rule for each external gauge boson in the contraction between external projectors and the amplitude, one can drop even more terms appearing in the projectors.
However, this does not necessarily reduce the complexity of the computation at all, because dressing all external gauge bosons by their polarization sums is a very costly action in multiple-parton scatterings.

\subsection{Comments} 
\label{SEC:prescription:comments}

\subsubsection*{The complex phase factors}

The specific linear combinations in eq.~(\ref{EQ:LP2HLmassless}) imply a definite choice of phase conventions for the
vector boson helicity states and hence for the helicity amplitudes composed out of the original projected amplitudes.
We know that, in principle, the phase conventions for the helicity states of different external particles in a scattering amplitude can be set differently and independently of each other, without altering any genuine physics.
For an intermediate on-shell particle, e.g.,~an intermediate particle produced approximately on-shell and subsequently decays, its phase convention needs to be used consistently throughout the computations of the two ``on-shell factorized'' parts of the complete amplitude where this particle state and its complex conjugate appear respectively.

Although the definite phase convention in use is not relevant in many practical applications, there is still the question of how one can quickly determine the appropriate complex phase factors appearing in the linear combinations like eq.~(\ref{EQ:LP2HLmassless}) needed to transform into helicity states defined with a particular phase convention, e.g.~, the Jacob-Wick phase convention~\cite{Jacob:1959at}, especially without knowing the definite geometric interpretations of the original linear polarization projectors used. 
This question can be most easily resolved by appealing to the form-factor decomposition perspective of the projections made using the linear-polarization projectors, which was alluded to at the end of section~\ref{SEC:prescription:NTS}.
From the point of view of a form-factor decomposition, the set of linear-polarization projectors represent precisely the Lorentz tensor decomposition basis in use, and the projected linearly polarized amplitudes (after dividing out the normalization factors, if not accounted for in the projectors) are the corresponding form-factor coefficients, at least at the level of the finite remainders in the four-dimensional limit.
(This is a consequence of the orthogonality among these projectors by construction, although the linear completeness is only ensured in the four-dimensional limit in general.)
Just like how one evaluates helicity amplitudes from a given form-factor decomposition representation of the amplitude, the expectation values of the chosen Lorentz-tensor decomposition structures, which in our case are the tensor products of the linear-polarization states, over the targeted helicity states provide exactly the answer to the question of appropriate complex phases required in linear combinations like eq.~(\ref{EQ:LP2HLmassless}). 
On the other hand, the knowledge of these so-projected quantities as linearly polarized amplitudes offers us a convenient short-cut to derive the linear combinations needed to transform into the chosen helicity basis.
From this discussion, it is thus clear that with these ``linear-polarization form factors'' at hand, one can also easily reproduce polarized amplitudes defined in other helicity conventions or polarization basis, just like how helicity amplitudes are computed with the usual form-factor decomposition representation of an amplitude.
~\\

Of course, instead of projectors for polarized amplitudes in the linear-polarization basis, one could also choose to use directly 
those corresponding to the helicity basis which can be linearly composed from the former, as should be clear from the discussions above.
However, there may not be much advantage in doing so at least in the initial stage of projecting out the raw unreduced amplitudes where all (scalar) 
loop integrals therein are denoted just symbolically in terms of some bookkeeping notations.
On the other hand, if one knows that the final results of the amplitudes in question are simpler in the helicity basis 
than in other polarization bases, it is then expected that it should be advantageous 
to directly reconstruct the final explicit results in the helicity basis.
In particular, this means that substitution of the table of IBP relations (even with the analytic results of all 
master integrals involved) as well as simplifications of their rational coefficients could be performed for amplitudes in the helicity basis, 
linearly composed out of the original projections.
To this end, it is helpful to note the following points.
The imaginary unit $i$ in the linear superposition formula eq.~(\ref{EQ:LP2HLmassless}) connecting the linearly polarized amplitudes and helicity (circularly polarized) amplitudes can always be absorbed into quantities defined by a consistent usage of FORM's (pseudo) Levi-Civita tensor.
The possibly remaining square roots in the products of normalization factors associated with linear-polarization state vectors, e.g.,~eq.~(\ref{EQ:norms_massive}) and eq.(\ref{EQ:TranspolMBRmassiveGenericNFs}), could be eliminated by a suitable re-parameterization of the external kinematical variables.

\subsubsection*{$1\to 2$ decay.} 

For a $1\to 2$ decay amplitude, the conventional Lorentz tensor decomposition and projection method can be carried out quite 
simply (due to the limited number of basis structures and scales). 
For instance, for the fermion gauge interaction vertex a general form-factor decomposition can be found in the literature, e.g.,~in~\cite{Hollik:1998vz}. 
Here we briefly comment on how one can compute polarized $1\to 2$ decay amplitudes if one wants to use the above prescription.

The computation requires the introduction of an intermediate auxiliary reference vector, denoted by $\hat{r}^{\mu}$, 
which will be formally treated on the same footing as an external four-momentum.
The reference-vector $\hat{r}^{\mu}$ may be associated with the polarization vector of 
the decaying particle (in which case it has a physical meaning), or chosen to be 
an auxiliary coordinate-frame dependent vector merely for intermediate usage. 
The important point we would like to emphasize here is that 
the definition of $\hat{r}^{\mu}$ can be achieved by simply specifying the values of 
a complete set of quadratic Lorentz invariant products between $\hat{r}^{\mu}$ 
and the two linearly independent external momenta, which we denote by $p_1$ and $p_2$.  
For instance, the normalized space-like $\hat{r}^{\mu}$ can be implicitly specified by 
\begin{eqnarray} \label{EQ:defRV}
\hat{r} \cdot p_1 = 0~, ~~ \hat{r} \cdot p_2 = 0~,~~ \hat{r} \cdot \hat{r} = -1, 
\end{eqnarray}
which guarantees that it lies in the plane transverse to $p_1$ and $p_2$. 
This set of assignments \eqref{EQ:defRV} is sufficient to algebraically manipulate
$\hat{r}$ in the computation of polarized $1\to 2$ decay amplitudes. 
There is no need for its explicit component-wise specification in a definite coordinate system. 
With the aid of the thus-defined $\hat{r}$, all procedures outlined above for the 
$2 \to 2$ scattering processes, discussed in section~\ref{SEC:prescription:MBR}, can be repeated here. 
To be a bit more specific, in this case the set of three linearly independent four-vectors 
$\{ p_1, p_2, \hat{r}\}$ will take over the roles that were played by the three linearly independent 
external momenta $\{ p_1, p_2, p_3\}$ in the $2 \to 2$ scattering processes.
In fact the $\hat{r}$ defined in eq.~(\ref{EQ:defRV}) fulfills the same set of conditions
that $\varepsilon_{X}$ satisfies in eq.~(\ref{EQ:XpolEQs}). 
Moreover, it never appears in Feynman propagators\footnote{This means that the sectors of loop 
integrals appearing in the projected amplitudes will not be enlarged by the introduction 
of this external reference-vector $\hat{r}$.}, and the Lorentz invariants appearing in 
the resulting projections are still just those made out of $p_1$ and $p_2$ (as the right-hand side 
of eq.~(\ref{EQ:defRV}) are all constants).
In the end the physical decay rates are independent of the choice of this auxiliary vector $\hat{r}$. 
In the case of a scalar decaying into a pair of fermions, the introduction of such an auxiliary vector 
can be avoided because the helicity polarization vector of a massive fermion, eq.~(\ref{EQ:PLpolMBR}), 
makes no reference at all to any transverse direction w.r.t. its momentum.

\section{Unitarity of the Prescription}
\label{SEC:unitarity}

The potential RS dependence of amplitudes is intimately connected to the
structure of their UV and IR singularities. Fortunately, in QCD they obey a 
factorized form at the amplitude level~\cite{Sen:1982bt,Collins:1989bt,Catani:1998bh,Sterman:2002qn,Dixon:2008gr,Aybat:2006wq,Gardi:2009qi,Gardi:2009zv,Becher:2009cu,Becher:2009kw,Becher:2009qa,Feige:2014wja}. 
The final result for a physical quantity, for instance a cross section, 
is of course finite and must not depend on the RS used.

The usage of the polarization projectors defined in the previous sections 
yields helicity amplitudes that differ in general from those defined in many existing dimensional regularization 
variants, in particular the CDR. In this section, we argue that our prescription 
of external state vectors will however lead to the same RS-independent finite remainders as for instance
in CDR, and can therefore be used in a hybrid way with CDR to achieve a maximal convenience 
owing to the amplitude-level factorization of UV and IR singularities in QCD amplitudes. 

\subsection{Pole subtracted amplitudes}
\label{SEC:unitarity:PSA}

We recall that in the D-dimensional Lorentz decomposition representation 
of a scattering amplitude, the Lorentz-invariant form factors encode all dependence 
on dimensionally regularized loop integrals and are independent of the external polarization vectors.
Once the (renormalized) loop amplitudes are available in such a tensor decomposed form, 
with all (singular) Lorentz-invariant form factors computed in D dimensions, 
then merely changing the RS for the external particles' state vectors, consistently both for the loop amplitudes 
and the corresponding IR subtraction terms, should not alter the finite remainders resulting from 
subtracting all poles and subsequently taking the 4-dimensional limit.\footnote{The equivalence between CDR and HV 
in leading to the same RS-independent finite remainders with the identical set of renormalization constants 
and anomalous dimensions~\cite{Kilgore:2012tb,Broggio:2015dga} can be appreciated this way, 
and the same arguments apply here as well.}  
Because in the form-factor representation of an amplitude the loop-integral dependent part 
is separated from the part depending on the external states, it is thus unambiguous to implement 
whatever non-CDR convention for external state vectors in the computation of singular amplitudes.
The crucial question for our purpose is 
whether our non-CDR prescription for external state vectors can still be unambiguously and 
directly applied in the computation of amplitudes without performing the form-factor decomposition first.

In our prescription all open Lorentz indices of the polarization projectors 
defined in section~\ref{SEC:Prescription} are set to be D-dimensional and no dimensional 
splitting is ever introduced, just like in CDR. Thus, commutation between Lorentz 
index contraction and loop integration is preserved within our prescription. 
This means that applying our polarization projectors directly to the original 
Feynman-diagrammatic representation of a loop amplitude should lead to the same 
polarized amplitudes as those that are obtained by applying these projectors 
to the D-dimensional form-factor decomposition representation of that amplitude.
No matter whether or not  evanescent Lorentz structures appear explicitly 
or implicitly in the form-factor decomposition of the  loop amplitude, 
they are taken into account exactly as they appear in the original Feynman-diagrammatic 
representation of this amplitude.
From this perspective we could already expect to end up with the same 
(4-dimensional) finite remainder as the one obtained from a computation purely within CDR.~\\

Below we demonstrate this crucial point more clearly via providing an alternative formulation of 
finite remainders introduced in the proposed prescription, which also helps to clarify a few points 
alluded in the preceding section.
Let us consider the finite remainders of amplitudes in CDR as defined by the celebrated amplitude-level 
factorization formula. 
Singularities in the dimensionally regularized QCD amplitudes are known to 
factorize~\cite{Sen:1982bt,Collins:1989bt,Catani:1998bh,Sterman:2002qn,Dixon:2008gr,Aybat:2006wq,Gardi:2009qi,Gardi:2009zv,Becher:2009cu,Becher:2009kw,Becher:2009qa,Feige:2014wja}.
For our purpose, we can sketch this factorization property 
of a bare QCD scattering amplitude $\hat{\mathcal{A}}(\epsilon)$ among several resolved 
external particles (with fixed external kinematics) schematically as follows:
\begin{eqnarray}\label{EQ:AmpPoleFactorization}
\hat{\mathcal{A}}(\epsilon) = \hat{\mathcal{Z}}_{\mathrm{IR}}(\epsilon)~ 
\mathcal{Z}_{\mathrm{UV}}(\epsilon)~ \hat{\mathcal{F}}(\epsilon) \, ,
\end{eqnarray}
where\footnote{The need of mass renormalizations in the case of massive quarks is understood.}
we have suppressed the dependence of the quantities on external kinematics and masses
as well as on auxiliary dimensional scales except the dimensional regulator $\epsilon$, 
(for a detailed exposition, see e.g.~\cite{Catani:1998bh,Feige:2014wja,Broggio:2015dga,Magnea:2018ebr} 
and references therein).
The bare amplitude $\hat{\mathcal{A}}(\epsilon)$ and the finite pole-subtracted amplitude 
$\hat{\mathcal{F}}(\epsilon)$ should be viewed as vectors in the color space of the external particles, 
and the multiplicative singular IR-factor $\hat{\mathcal{Z}}_{\mathrm{IR}}(\epsilon)$ is a matrix. 
The RS-dependent singular factors $\mathcal{Z}_{\mathrm{UV}}(\epsilon)$ and $\hat{\mathcal{Z}}_{\mathrm{IR}}(\epsilon)$ encode all UV and IR pole-singularities of $\hat{\mathcal{A}}(\epsilon)$. 
What is essential for our discussion below is that these singular factors are independent of the detailed kinematic configuration, in particular the polarization states, of the external resolved particles.
By the very meaning of pole factorization in eq.~(\ref{EQ:AmpPoleFactorization}), $\hat{\mathcal{F}}(\epsilon)$ is regular in $\epsilon$ and has a finite 4-dimensional limit, $\hat{\mathcal{F}}(\epsilon = 0)$. 
We call this quantity the (4-dimensional) \textit{finite remainder} of $\hat{\mathcal{A}}(\epsilon)$ defined by subtracting all poles minimally by the multiplicative factors as sketched in eq.~(\ref{EQ:AmpPoleFactorization}).

We may summarize this by the following expression for the finite remainder 
$\hat{\mathcal{F}}_{4} \equiv \hat{\mathcal{F}}(\epsilon = 0)$, namely 
\begin{eqnarray}\label{EQ:AmpsFiniteRemainder}
\hat{\mathcal{F}}_{4} = 
\Big(
\hat{\mathcal{Z}}^{-1}_{\mathrm{IR};\mathrm{CDR}}(\epsilon)~ 
\mathcal{Z}^{-1}_{\mathrm{UV};\mathrm{CDR}}(\epsilon)~\hat{\mathcal{A}}_{\mathrm{CDR}}(\epsilon) 
\Big)_{\epsilon = 0} \, ,
\end{eqnarray}
where we added the subscript ``CDR'' to all singular RS-dependent quantities given in CDR.  
For the point to be demonstrated here, the concrete expressions of these singular multiplicative factors 
taken from CDR are irrelevant. 
The claim is that replacing all CDR-regularized external states of the fixed-angle bare scattering
amplitude $\hat{\mathcal{A}}_{\mathrm{CDR}}(\epsilon)$  by their respective counterparts 
given in terms of momentum basis representations defined in section~\ref{SEC:Prescription} 
will still result in the same finite remainder $\hat{\mathcal{F}}_{4}$, 
where all poles have been subtracted in a minimal way by the same untouched 
$\hat{\mathcal{Z}}^{-1}_{\mathrm{IR};\mathrm{CDR}}(\epsilon) ~ \mathcal{Z}^{-1}_{\mathrm{UV};\mathrm{CDR}}(\epsilon)$, 
without appealing to the Lorentz tensor decomposition representation of $\hat{\mathcal{A}}_{\mathrm{CDR}}(\epsilon)$.

In order to facilitate the discussion, let us exhibit the dependence of 
$\hat{\mathcal{A}}_{\mathrm{CDR}}(\epsilon)$ on the CDR-regularized polarization state 
$\bar{\varepsilon}_{\bar{\lambda}}(p_i,r_i)$ of a \textit{representative} external 
massless gauge boson with momentum $p_i$ and reference vector $r_i$. 
Because the bare scattering amplitude $\hat{\mathcal{A}}_{\mathrm{CDR}}$ is linear in 
$\bar{\varepsilon}_{\bar{\lambda}}(p_i,r_i)$,  
we write 
\begin{equation} \label{eq:ACDReps}
\hat{\mathcal{A}}_{\mathrm{CDR}} 
\big(\epsilon; \bar{\varepsilon}_{\bar{\lambda}}(p_i,r_i) \big)
= g_{\mu \nu}~
\big(\hat{\mathcal{A}}_{\mathrm{CDR}}\big)^{\mu} ~  
\bar{\varepsilon}^{~\nu}_{\bar{\lambda}}(p_i,r_i) \, ,
\end{equation}
where we have introduced a compact notation $\big(\hat{\mathcal{A}}_{\mathrm{CDR}}\big)^{\mu}$.
For the pole-subracted amplitude we have 
\begin{eqnarray}\label{EQ:AmpsFiniteRemainderCDR}
\hat{\mathcal{F}}_{\mathrm{CDR}}\big( \epsilon; \bar{\varepsilon}_{\bar{\lambda}}(p_i,r_i) \big) &\equiv& 
\hat{\mathcal{Z}}^{-1}_{\mathrm{IR};\mathrm{CDR}}\left(\epsilon\right)~ 
\mathcal{Z}^{-1}_{\mathrm{UV};\mathrm{CDR}}\left(\epsilon\right)~
\hat{\mathcal{A}}_{\mathrm{CDR}}
\big(\epsilon; \bar{\varepsilon}_{\bar{\lambda}}(p_i,r_i) \big) \nonumber\\
&=& \hat{\mathcal{F}}_{4}\big(\varepsilon_{\lambda}(p_i,r_i) \big) + \mathcal{O}(\epsilon) \, ,
\end{eqnarray} 
whose limit at $\epsilon = 0$ is precisely the finite remainder $\hat{\mathcal{F}}_{4}$ 
in eq.~(\ref{EQ:AmpsFiniteRemainder}) with 4-dimensional external polarization  vector
$\varepsilon_{\lambda}(p_i,r_i)$. 
Now we multiply this regular finite quantity by a generalized D-dependent 
Lorentz-invariant norm-orthogonal factor $\Delta_{\bar{\lambda} \lambda}$ defined by
\begin{eqnarray}\label{EQ:generalizedDelta}
\Delta_{\bar{\lambda} \lambda} 
&\equiv& - \bar{\varepsilon}^{~*}_{\bar{\lambda}}(p_i,r_i) \cdot 
\varepsilon^{\text{MBR}}_{\lambda}(p_i,r_i) \nonumber\\
&=& \delta_{\bar{\lambda} \lambda} + \mathcal{O}(\epsilon). 
\end{eqnarray}
Here $\varepsilon^{\text{MBR}}_{\lambda}$ refers to a polarization vector for a massless
gauge boson of our prescription\footnote{The acronym ``MBR'' denotes
momentum basis representation.} of section~\ref{SEC:Prescription}, and the dot product in  
\eqref{EQ:generalizedDelta} refers to the D-dimensional Minkowski scalar product. 
We recall that  the polarization index $\bar{\lambda}$ labels the $D$$-$$2$ polarization
states of CDR while in our prescription the index $\lambda$ of $\varepsilon^{\text{MBR}}_{\lambda}$
takes only two values for massless gauge bosons (and three for massive ones), 
which are $\pm$ in the helicity basis.   
The 4-dimensional limits of these simple Lorentz-invariant contractions 
$\Delta_{\bar{\lambda} \lambda}$ are the norm-orthogonal factors 
(i.e.~the Kronecker deltas) among different 4-dimensional physical 
polarization/helicity states.

Next we consider the sum of products 
\begin{equation} \label{EQ:AmpsFiniteRemainderMBRstartpoint}
\sum_{\bar{\lambda} = \pm,~ D-4}
\hat{\mathcal{F}}_{\mathrm{CDR}}\big( \epsilon; \bar{\varepsilon}_{\bar{\lambda}}(p_i,r_i) \big)
~\Delta_{\bar{\lambda} \lambda}.
\end{equation} 
As exhibited in eqs.~(\ref{EQ:AmpsFiniteRemainderCDR}) and \eqref{EQ:generalizedDelta}, both 
$\hat{\mathcal{F}}_{\mathrm{CDR}}\big( \epsilon; \bar{\varepsilon}_{\bar{\lambda}}(p_i,r_i) \big)$ 
and $\Delta_{\bar{\lambda} \lambda}$ are regular in $\epsilon$. 
Thus they can be expanded in powers of $\epsilon$, and their four-dimensional limits can be taken separately before being 
multiplied together and subsequently summed over polarizations. 
Proceeding in this way, we first insert the $\epsilon$-expanded expressions of these two factors given
above, and the resulting quantity is precisely the finite remainder  
$\hat{\mathcal{F}}_{4}\big(\varepsilon_{\lambda}(p_i,r_i) \big)$ of eq.~(\ref{EQ:AmpsFiniteRemainder}):
\begin{equation} \label{EQ:AmpsFiniteRemainderMBRleft}
\sum_{\bar{\lambda} = \pm,~ D-4} 
\hat{\mathcal{F}}_{\mathrm{CDR}}\big( \epsilon; \bar{\varepsilon}_{\bar{\lambda}}(p_i,r_i) \big)
~\Delta_{\bar{\lambda} \lambda}
= \hat{\mathcal{F}}_{4}\big(\varepsilon_{\lambda}(p_i,r_i) \big) + \mathcal{O}(\epsilon). 
\end{equation}

On the other hand, we can first perform the polarization sum in \eqref{EQ:AmpsFiniteRemainderMBRstartpoint}
in D dimensions and take the 4-dimensional limit afterwards. 
Proceeding this way, we have 
\begin{eqnarray}\label{EQ:AmpsFiniteRemainderMBRright1}
&&\sum_{\bar{\lambda} = \pm,~ D-4}
\hat{\mathcal{F}}_{\mathrm{CDR}}\big( \epsilon; \bar{\varepsilon}_{\bar{\lambda}}(p_i,r_i) \big)
~\Delta_{\bar{\lambda} \lambda}  \nonumber\\
&=& 
 - \sum_{\bar{\lambda} = \pm,~ D-4} 
\hat{\mathcal{Z}}^{-1}_{\mathrm{IR};\mathrm{CDR}}\left(\epsilon\right)~ 
\mathcal{Z}^{-1}_{\mathrm{UV};\mathrm{CDR}}\left(\epsilon\right)~
\hat{\mathcal{A}}_{\mathrm{CDR}} 
\big(\epsilon;~ \bar{\varepsilon}_{\bar{\lambda}}(p_i,r_i) \big)~ 
\bar{\varepsilon}^{~*}_{\bar{\lambda}}(p_i,r_i) \cdot 
\varepsilon^{\text{MBR}}_{\lambda}(p_i,r_i) \nonumber\\
&=& 
- \hat{\mathcal{Z}}^{-1}_{\mathrm{IR};\mathrm{CDR}}\left(\epsilon\right)~ 
\mathcal{Z}^{-1}_{\mathrm{UV};\mathrm{CDR}}\left(\epsilon\right)~
\sum_{\bar{\lambda} = \pm,~ D-4} 
\big(\hat{\mathcal{A}}_{\mathrm{CDR}}\big)_{\mu} ~ 
\bar{\varepsilon}_{\bar{\lambda}}^{~\mu}(p_i,r_i) ~
\bar{\varepsilon}^{~*\nu}_{\bar{\lambda}}(p_i,r_i) ~
\varepsilon^{\text{MBR}}_{\lambda,~\nu} (p_i,r_i) \, ,\nonumber\\
\end{eqnarray} 
where we have used the fact that $\hat{\mathcal{A}}_{\mathrm{CDR}} 
\big(\epsilon;~ \bar{\varepsilon}_{\bar{\lambda}}(p_i,r_i) \big)$ is linear 
in the external polarization vector $\bar{\varepsilon}_{\bar{\lambda}}(p_i,r_i)$.
Now we employ eq.~(\ref{EQ:polsumCDRPhys}) for summing over the $D$$-$$2$ polarization states 
of the CDR-regularized external gauge boson\footnote{Note that here we should sum over 
physical polarizations only, especially in the case of gluons, which ensures that unphysical components 
such as scalar and longitudinal polarizations are absent from the outset. 
With this choice there is no need to incorporate diagrams involving ghost fields 
in the external states (when there are multiple external non-Abelian gauge bosons).}
and obtain
\begin{eqnarray}\label{EQ:AmpsFiniteRemainderMBRright2}
&& -\sum_{\bar{\lambda} = \pm,~ D-4} 
\big(\hat{\mathcal{A}}_{\mathrm{CDR}}\big)_{\mu} ~ 
\bar{\varepsilon}_{\bar{\lambda}}^{~\mu}(p_i,r_i) ~
\bar{\varepsilon}^{~*\nu}_{\bar{\lambda}}(p_i,r_i) ~
\varepsilon^{\text{MBR}}_{\lambda,~\nu} (p_i,r_i) \nonumber\\
&=& 
\big(\hat{\mathcal{A}}_{\mathrm{CDR}}\big)_{\mu} 
~\Big(
g^{\mu\nu} - \frac{p_i^{\mu} r_i^{\nu} + r_i^{\mu} p_i^{\nu}}{p_i \cdot r_i}
\Big)~ 
\varepsilon^{\text{MBR}}_{\lambda,~\nu} (p_i,r_i) \nonumber\\
&=&  \hat{\mathcal{A}}_{\mathrm{CDR}} 
\big(\epsilon;~ \varepsilon^{\text{MBR}}_{\lambda}(p_i,r_i)\big)\, ,
\end{eqnarray} 
where we have used the orthogonality of $\varepsilon^{\text{MBR}}_{\lambda}(p_i,r_i)$ 
w.r.t. the particle's momentum $p_i$ and its reference vector $r_i$ in D dimensions, 
which $\varepsilon^{\text{MBR}}_{\lambda}(p_i,r_i)$  has to satisfy by construction.
Inserting  eq.~(\ref{EQ:AmpsFiniteRemainderMBRright2}) back into 
eq.~(\ref{EQ:AmpsFiniteRemainderMBRright1}) we end up with 
\begin{eqnarray}\label{EQ:AmpsFiniteRemainderMBRright3}
\sum_{\bar{\lambda} = \pm,~ D-4}
\hat{\mathcal{F}}_{\mathrm{CDR}}\big( \epsilon;~ \bar{\varepsilon}_{\bar{\lambda}}(p_i,r_i) \big)
~\Delta_{\bar{\lambda} \lambda} 
&=& 
\hat{\mathcal{F}}_{\mathrm{CDR}}\big( \epsilon;~ \varepsilon^{\text{MBR}}_{\lambda}(p_i,r_i) \big) \, ,
\end{eqnarray}  
whose left-hand side has, according to  eq.~(\ref{EQ:AmpsFiniteRemainderMBRleft}), 
a 4-dimensional limit that is equal to the  finite remainder 
$\hat{\mathcal{F}}_{4}\Big(\varepsilon_{\lambda}(p_i,r_i) \Big)$ given in eq.~(\ref{EQ:AmpsFiniteRemainder}).
Notice that eq.~(\ref{EQ:AmpsFiniteRemainderMBRright3}) is an identity holding to all orders in $\epsilon$.
The right-hand side of \eqref{EQ:AmpsFiniteRemainderMBRright3}, more explicitly, 
\begin{equation} \label{eq:rhsFcdr}
\hat{\mathcal{F}}_{\mathrm{CDR}}\big( \epsilon;~ \varepsilon^{\text{MBR}}_{\lambda}(p_i,r_i) \big) 
= \hat{\mathcal{Z}}^{-1}_{\mathrm{IR};\mathrm{CDR}}\left(\epsilon\right)~ 
\mathcal{Z}^{-1}_{\mathrm{UV};\mathrm{CDR}}\left(\epsilon\right)~
\hat{\mathcal{A}}_{\mathrm{CDR}} 
\big(\epsilon;~ \varepsilon^{\text{MBR}}_{\lambda}(p_i,r_i)\big)
\end{equation}
is precisely the quantity suggested by our prescription.
In order to avoid confusion we emphasize that the subscript ``CDR'' on $\hat{\mathcal{F}}_{\mathrm{CDR}}$
at the right-hand side of \eqref{EQ:AmpsFiniteRemainderMBRright3}, and on $\hat{\mathcal{F}}_{\mathrm{CDR}}$
and $\hat{\mathcal{A}}_{\mathrm{CDR}}$ in eq.~\eqref{eq:rhsFcdr} means that these are the respective 
CDR expressions with the exception that the CDR polarization vector of the external gluon with momentum $p_i$
is replaced by the polarization vector of our hybrid MBR prescription. If there are more gluons in the external
state then the procedure outlined by eqs.~\eqref{eq:ACDReps} - \eqref{eq:rhsFcdr} can be iterated.

What the above reformulations show is that, to all orders in $\epsilon$, the 
$\hat{\mathcal{F}}_{\mathrm{CDR}}\big(\epsilon;~\varepsilon^{\text{MBR}}_{\lambda}(p_i,r_i) \big)$ can be formally viewed as an unpolarized interference between $\hat{\mathcal{F}}_{\mathrm{CDR}}\big( \epsilon;~ \bar{\varepsilon}_{\bar{\lambda}}(p_i,r_i) \big)$ and the Lorentz-invariant generalized norm-orthogonal factor defined in eq.~(\ref{EQ:generalizedDelta}), using physical polarization sum rules for all CDR external states. 
The unpolarized Landau density matrices of external gauge bosons reduce to the unique space-time metric tensor by virtue of the built-in orthogonality between $\varepsilon^{\text{MBR}}_{\lambda}(p_i,r_i)$ and $p_i,r_i$.~\\

An analogous reformulation can be made for external fermions in the scattering amplitude.
In fact, for each open fermion line such a reformulation is more straightforward than 
in the above gauge boson case, because there is no redundancy in the spinor representation of the Lorentz algebra, 
and the number of the polarization/helicity states of a fermion is two both in CDR and in our prescription.
The unpolarized Landau density matrix of an external fermion is the well-known projection operator  
onto the space of on-shell Dirac-spinors. 
After performing a similar reformulation of an open fermion line in the scattering amplitude, 
denoted by $\langle \psi_A|\,\hat{\mathrm{M}}\,|\psi_B \rangle$ as in eq.~(\ref{EQ:SFLtrace1}),
we end up with the following replacement:
\begin{eqnarray} \label{EQ:fermionlinerewritten}
\langle \psi^{\mathrm{CDR}}_A|\,\hat{\mathrm{M}}\,|\psi^{\mathrm{CDR}}_B \rangle 
~\longrightarrow~ 
\mathrm{Tr} \Big{[} 
\,\hat{\mathrm{M}}\,
\frac{\hat{P}_{on}\left(p_B,m_B\right)}{2 \lambda_B\, m_B}
\Big(
~|\psi^{\mathrm{MBR}}_B \rangle \langle \psi^{\mathrm{MBR}}_A|~
\Big)
\frac{\hat{P}_{on}\left(p_A,m_A\right)}{2 \lambda_A\, m_A}
\Big{]}\,,
\end{eqnarray}
where $\hat{P}_{on}(p,m)=(\slashed{p} \pm m)$ denotes the aforementioned on-shell projection operator 
for a $u$- respectively $v$-type Dirac spinor with momentum $p$ and mass $m$,
and $|\psi^{\mathrm{MBR}}_B \rangle \langle \psi^{\mathrm{MBR}}_A|$
is exactly the matrix  \eqref{EQ:TPextsps1} that was further discussed in 
eqs.~\eqref{EQ:LDMofDSP} - (\ref{EQ:TPextsps2}).
The appearance of ${1}/{(2 \lambda_A\, m_A)}$ and ${1}/{(2 \lambda_B\, m_B)}$ in eq.~(\ref{EQ:fermionlinerewritten})
is due to the conventional choice of normalization factors of on-shell Dirac spinors.
Here the factors $\lambda_A, \lambda_B = 1 ~(-1)$ when the fermion $A$ respectively
$B$ is associated with a $u$-type ($v$-type) spinor.

Quantities that are sandwiched \textit{between} the pair of on-shell projection operators, 
$\hat{P}_{on}(p_A,m_A)$ and $\hat{P}_{on}(p_B,m_B)$, associated with the two external spinors of 
the open fermion line, can be manipulated and simplified according to the 4-dimensional 
Lorentz/Dirac-algebra. We just have to agree on one definite form that will be
taken as its canonical form (out of all the forms that are equivalent in 4 dimensions)
and used unambiguously in D-dimensional algebraic computations. 
This pair of on-shell projection operators sets the domain where matrices related to 
external fermions' states, namely $|\psi^{\mathrm{MBR}}_B \rangle \langle \psi^{\mathrm{MBR}}_A|$,
can be manipulated and moved around using just 4 dimensional Lorentz/Dirac-algebra.
While, in general and to be on the safe side, moving any of these matrices beyond this range must be done in accordance with the 
D-dimensional Lorentz/Dirac-algebra in order not to introduce artificial terms by mistake. 
For instance, the object $\gamma_{\mu} \gamma_{\nu} \gamma_{\rho} S_{\sigma} \epsilon^{\mu\nu\rho\sigma}$ 
commutes with $\slashed{P}$ in 4 dimensions because of the orthogonality condition $S \cdot P = 0$. 
However, this is no longer true w.r.t.~the D-dimensional algebra (with a non-anticommuting $\gamma_5$), 
and there is thus a non-vanishing evanescent commutator resulting from interchanging the product order between the two.
In section~\ref{SEC:examples:eeQQ} we will briefly comment on this subtle point again 
in context of a practical 1-loop example.

Finally, in order to bring the external projector in eq.~(\ref{EQ:fermionlinerewritten}) into a form analogous to 
eq.~(\ref{EQ:SFLtrace1}) with the tensor product of external spinors given by eq.~(\ref{EQ:TPextsps2}),
the following defining property of the on-shell projection operators, valid for $p^2=m^2$ in D dimensions, 
can be used:
\begin{eqnarray}\label{EQ:onshellprojectorID}
\hat{P}_{on}(p,m) \frac{\hat{P}_{on}(p,m)}{2 \lambda_f\, m} = \hat{P}_{on}(p,m) \, ,
\end{eqnarray}
where $\lambda_f =\pm 1$ depending on whether  ${\hat P}_{on}$ is associated with a $u$-type or $v$-type spinor.
Notice also that such an identity has a continuous limit at $m \rightarrow 0$, 
despite the superficial appearance of the singular $1/m$ factor 
which does prevent setting $m=0$ directly in eq.(\ref{EQ:onshellprojectorID}).
Such an alternative perspective thus helps to explain the choice made  
in eq.~\eqref{EQ:TPextsps2} where the polarization projection operators, 
especially those with Dirac matrices contracted with the Levi-Civita tensor,
were placed \textit{inside} the on-shell projection operators.~\\

We thus achieved what we aimed at in this subsection.
We found an alternative formulation of pole-subtracted finite amplitudes  
which helps to prove the following claim:
despite the fact that usage of the polarization projectors defined in and manipulated according to section~\ref{SEC:Prescription} 
results in (bare) helicity amplitudes different from those in CDR (or HV), 
replacing all CDR-regularized external polarization states of $\hat{\mathcal{A}}_{\mathrm{CDR}}(\epsilon)$ in eq.~(\ref{EQ:AmpPoleFactorization}) 
by their counterparts given in terms of momentum basis representations constructed in section~\ref{SEC:Prescription} still results in the  same 
RS-independent finite remainder, where all poles are chosen to be subtracted by the same factorized (singular) coefficients given in CDR, 
without appealing to Lorentz tensor decomposition representations of $\hat{\mathcal{A}}_{\mathrm{CDR}}(\epsilon)$.
The validity of this statement is not confined to one-loop or next-to-leading order (NLO) corrections to a Born-level scattering 
amplitude, but holds as long as the amplitude-level factorization formula sketched in eq.~(\ref{EQ:AmpPoleFactorization}) holds in CDR.
Since the 4-dimensional limit of the properly defined IR-subtracted finite remainder of a renormalized virtual amplitude should remain the same in different variants of unitary dimensional regularization schemes, the aforementioned equivalence carries out automatically to other unitary dimensional schemes as well, in particular the HV scheme.

\subsection{Finite remainders in an IR subtraction framework}
\label{SEC:unitarity:FRIR}

In this subsection, we move on and analyze finite remainders defined in an IR-subtraction method that are obtained with our hybrid MBR prescription for external polarization vectors.
We will then show that this hybrid CDR-compatible prescription is unitary as defined in the sense of refs.~\cite{vanDamme:1984ig,Catani:1996pk}.~\\

In practice the finite RS-independent physical observables at NLO and beyond are usually computed 
as combinations of separate, in general UV and/or IR divergent contributions living in different 
partonic phase spaces. (UV renormalization is understood in what follows.)
To render individual contributions from each partonic phase space IR-finite and RS-independent, 
one can add and subtract properly defined auxiliary IR-subtraction terms. 
The introduction of these auxiliary terms are designed to ensure the cancellation of all intermediate 
IR-divergences of amplitudes in each partonic phase space, while on the other hand  
they leave no trace in the final properly combined physical observables. 
This is the idea of IR-subtraction methods~\cite{Ellis:1980wv,Kunszt:1992tn}, 
which are nowadays available in many different versions (e.g.,~\cite{Frixione:1995ms,Catani:1996vz,Nagy:2003qn,Kosower:1997zr,GehrmannDeRidder:2005cm,Czakon:2010td,Czakon:2014oma,Caola:2017dug,Magnea:2018hab,Herzog:2018ily,Catani:2007vq,Somogyi:2006da}).

Let us now sketch an IR-subtraction method by only being explicit about aspects that are relevant for 
showing that our hybrid MBR prescription of external states is unitary.

Assume that the Born-level scattering amplitude $\mathcal{A}_n$ lives in a n-particle 
phase space, and we consider an IR-safe observable defined by the measurement function $F_J$. 
The leading-order (LO) observable $\sigma_{\rm LO}$ is given by 
\begin{eqnarray}\label{EQ:LOamp}
\sigma_{\rm LO} & = & \int_{d\Phi_n} |\mathcal{A}_n|^2 ~ \mathit{F}_{\mathit{J}}^{~(n)},
\end{eqnarray}
where we suppressed all prefactors related to spin averaging for the initial state 
and the incident flux.
The NLO QCD correction $\sigma_{\rm NLO}$ consists of real radiations $\int_{d\Phi_{n+1}} d\sigma_{\rm NLO}^{\mathcal{R}}$
in the (n+1)-particle phase space and the (renormalized) virtual corrections 
$\int_{d\Phi_{n}} d\sigma_{\rm NLO}^{\mathcal{V}}$ in the n-particle phase space. 
To render individual contributions in each of these two partonic phase spaces finite, 
one adds and subtracts an appropriate IR-subtraction term $d\sigma^{\mathcal{S}}$. 
Subsequently $\sigma_{\rm NLO}$ can then be rewritten in an IR subtraction method as follows:\footnote{
For the sake of simplicity, we suppressed here an initial-state collinear subtraction 
term related to the (re)definition of parton-distribution functions, 
which does not add any additional conceptual complexity to what we want to show.}    
\begin{eqnarray}\label{EQ:NLOamp}
\sigma_{\rm NLO} & = & \int_{d\Phi_{n+1}} d\sigma_{\rm NLO}^{\mathcal{R}} + 
                      \int_{d\Phi_{n}} d\sigma_{\rm NLO}^{\mathcal{V}} \nonumber\\ 
                  & = & \int_{d\Phi_{n+1}} |\mathcal{A}_{n+1}^{\mathcal{R}}|^2 ~ 
                        \mathit{F}_{\mathit{J}}^{~(n+1)} 
                      +  \Big(   
                         \int_{d\Phi_{n+1}} d\sigma^{\mathcal{S}} ~\mathit{F}_{\mathit{J}}^{~(n)}
                         - 
                         \int_{d\Phi_{n+1}} d\sigma^{\mathcal{S}} ~\mathit{F}_{\mathit{J}}^{~(n)}
                           \Big) \nonumber\\
                       &+& \int_{d\Phi_{n}} 2\mathrm{Re}\left[\mathcal{A}_{n}^{*} \mathcal{A}_{n}^{\mathcal{V}}\right] ~ 
                         \mathit{F}_{\mathit{J}}^{~(n)} \nonumber\\
                   & = & \int_{d\Phi_{n+1}} \Big[ \left( 
                        |\mathcal{A}_{n+1}^{\mathcal{R}}|^2 ~ 
                        \mathit{F}_{\mathit{J}}^{~(n+1)}\right)_{\epsilon = 0} 
                         -  
                       \left(  d\sigma^{\mathcal{S}} ~ \mathit{F}_{\mathit{J}}^{~(n)}
                       \right)_{ \epsilon = 0}  \Big] \nonumber\\
                      &+& \int_{d\Phi_{n}}  \left[ 2~ \mathrm{Re}\left[\mathcal{A}_{n}^{*} \mathcal{A}_{n}^{\mathcal{V}}\right]  
                       + \int_{1} d\sigma^{\mathcal{S}} 
                        \right]_{ \epsilon = 0} \mathit{F}_{\mathit{J}}^{~(n)}\,.
\end{eqnarray}
By construction, the subtraction term $d\sigma^{\mathcal{S}}$ should have the same local 
IR-singular behavior as the squared real-radiation matrix $|\mathcal{A}_{n+1}^{\mathcal{R}}|^2$ 
everywhere in the (n+1)-particle phase space (subject to the constraint implied by $\mathit{F}_{\mathit{J}}$). 
Consequently, the resulting subtracted phase-space integrand 
$\big[ \big(|\mathcal{A}_{n+1}^{\mathcal{R}}|^2 ~ \mathit{F}_{\mathit{J}}^{~(n+1)}\big)_{\epsilon = 0}  - \big(  d\sigma^{\mathcal{S}} ~\mathit{F}_{\mathit{J}}^{~(n)} \big)_{ \epsilon = 0}  \big]$ 
can be numerically evaluated and integrated over the phase space in 4 dimensions, 
as indicated by $\epsilon = 0$.
Notice that  $\mathit{F}_{\mathit{J}}^{~(n)}$  associated with 
$d\sigma^{\mathcal{S}}$ is the same as for the virtual corrections living in the n-particle phase space.
The integration of $d\sigma^{\mathcal{S}}$ over the unresolved phase space 
has to be done in D dimensions with the IR unresolved partonic d.o.f. regularized in the same way as 
those in the virtual correction $2~\mathrm{Re}\left[\mathcal{A}_{n}^{*} \mathcal{A}_{n}^{\mathcal{V}}\right]$,
following from the unitarity constraint. 
The resulting IR singularities that appear as poles in $\epsilon$ must cancel  
those appearing in $2~\mathrm{Re}\left[\mathcal{A}_{n}^{*} \mathcal{A}_{n}^{\mathcal{V}}\right]$, 
which renders the  quantity in the second square bracket of 
the last line of eq.~(\ref{EQ:NLOamp}) finite in 4 dimensions as well.

In order that eq.~(\ref{EQ:NLOamp}) is useful in practice, one must be able to perform  
the D-dimensional integration $\int_{1} d\sigma^{\mathcal{S}}$,
either analytically or numerically. 
Thanks to the IR factorization, $d\sigma^{\mathcal{S}}$ and likewise its integrated counterpart 
$\int_{1} d\sigma^{\mathcal{S}}$ can be constructed, schematically,
as a convoluted product of certain universal (process-independent) multiplicative coefficient 
and the (process-specific) squared Born amplitude  
$|\mathcal{A}_n|^2$:
\begin{eqnarray}\label{EQ:IRsubtractionterm}
d\sigma^{\mathcal{S}} &=& \left( d \hat{I}_{RS} \right) \otimes |\mathcal{A}_n|^2~,\nonumber\\
\int_{1} d\sigma^{\mathcal{S}} &=& \hat{I}_{RS} \otimes |\mathcal{A}_n|^2.
\end{eqnarray}
The factor $\hat{I}_{RS}$ plays a similar role as the multiplicative factors 
$\hat{\mathcal{Z}}_{\mathrm{IR}}(\epsilon)$ in eq.~(\ref{EQ:AmpPoleFactorization}).
At NLO it encodes all IR pole-singularities and is to be viewed 
as an operator in the color space of the external particles.

In fact each variant of an IR-subtraction method can be seen as providing a concrete constructive
prescription for the integral representations of the factorized IR-subtraction coefficients, 
like the factor $\hat{I}_{RS}$, that contain all the explicit pole-singularities of the 
loop amplitudes (after multiplication with certain relevant process-dependent hard-scattering amplitudes). 
The crucial point relevant for the following discussion is that 
these integral representations are based on the amplitude-level IR factorization, 
and are manifestly independent of the polarization states of external particles which 
appear in the (remaining) hard-scattering matrix elements.\footnote{The dependence of factorized collinear singularities 
on the polarization of a parent parton in the real-radiation diagrams drops once one sums over the polarizations of all other particles 
and also integrates over all unresolved degrees of freedom in the collinear limit, 
notably the transverse plane of the radiated partons 
(which essentially eliminates any preference in the transverse direction).}

All quantities in eq.~(\ref{EQ:NLOamp}) that contain explicit IR-divergences, i.e.
poles in $\epsilon$, contain RS-dependent pieces in their truncated 
Laurent series to order $\epsilon^0$, especially the integrated $\hat{I}_{RS}$. 
At NLO, this concerns only $\int_{d\Phi_{n}} d\sigma_{\rm NLO}^{\mathcal{V}}$ and 
$\int_{1} d\sigma^{\mathcal{S}} = \hat{I}_{RS} \otimes |\mathcal{A}_n|^2$
that live in the same n-particle phase space. 
By appealing to an IR-subtraction method the unitarity constraint, originally imposed 
between the calculations of $\int_{d\Phi_{n+1}} d\sigma_{\rm NLO}^{\mathcal{R}}$ and 
$\int_{d\Phi_{n}} d\sigma_{\rm NLO}^{\mathcal{V}}$ is translated into the following ``locally distributed'' 
version: 
we just need to make sure that contributions associated with the same partonic phase space 
are computed consistently with a unitarity-respecting prescription, while 
pole-subtracted 4-dimensional remainders living in different partonic phase spaces 
can be computed independently of each other (using different methods).
Thus, as argued in ref.~\cite{Catani:1996pk}, IR subtraction methods offer a 
convenient way to isolate and investigate the RS-dependence of individual 
singular pieces and subsequently ensure the unitarity of regularization prescriptions 
used in the calculation.~\\

With the above sketch of essential aspects of an IR-subtraction framework ready, we can discuss 
how each of the two square brackets in the last line of eq.~(\ref{EQ:NLOamp}) should be evaluated 
with our proposed prescription in order to ensure a correct NLO observable $\sigma_{\rm NLO}$.

First, the subtraction of implicit IR-singularities in $d\sigma_{\rm NLO}^{\mathcal{R}}$, 
i.e.~terms in the first square bracket of the last line of eq.~(\ref{EQ:NLOamp}), 
is to be done at the integrand level of phase-space integrals. This results in a subtracted 
real-radiation contribution that is numerically integrable in 4 dimensions. 
In the 4-dimensional limit ($\epsilon = 0$) the external polarization states 
defined by the momentum basis representations given in section~\ref{SEC:Prescription}, 
all coincide with their respective standard 4-dimensional expressions. 
Therefore the RS-independence of the finite remainders of real-radiation contributions 
associated with the 4-dimensional (n+1)-particle phase space is manifest 
as dimensional regularization can be avoided from the outset. 
Thus we just have to make sure that in this hybrid prescription, 
the integral-level subtraction of the explicit $\epsilon$-pole singularities 
in $2~\mathrm{Re}\left[\mathcal{A}_{n}^{*} \mathcal{A}_{n}^{\mathcal{V}}\right]$, 
i.e.~the second square bracket of the last line of eq.~(\ref{EQ:NLOamp}),  
is also done in a unitarity-respecting way so as to lead to the 
correct RS-independent finite remainder in the n-particle phase space.

To this end, we can proceed in two ways. 
We could devise a proof analogous to the previous subsection, but now applied to the finite remainder 
$\left[ 2\mathrm{Re}\left[\mathcal{A}_{n}^{*} \mathcal{A}_{n}^{\mathcal{V}}\right] + \hat{I}_{RS} \otimes |\mathcal{A}_n|^2 \right]_{\epsilon = 0}$, 
where the integrated factor $\hat{I}_{RS}$ plays a similar role
as the perturbatively-expanded multiplicative factor $\hat{\mathcal{Z}}_{\mathrm{IR}}(\epsilon)$ in eq.~(\ref{EQ:AmpPoleFactorization}). 
Alternatively, we argue in this subsection that the unitarization recipe of ref.~\cite{Catani:1996pk} 
is indeed respected by our hybrid prescription. 
We examine this now one by one. 
\begin{enumerate}
\item 
The external partons in the Born-level hard-scattering matrix element $\mathcal{A}_n$ 
of the factorized IR-subtraction term $\hat{I}_{RS} \otimes |\mathcal{A}_n|^2$ have to be 
treated in the same way as for the external partons in the virtual loop amplitude $\mathcal{A}_{n}^{\mathcal{V}}$ 
(of the same external kinematic configuration).

This is guaranteed by applying the same set of polarization projectors defined in 
section~\ref{SEC:Prescription} consistently to $\mathcal{A}_n$ at LO and $\mathcal{A}^{\mathcal{V}}_n$ 
at NLO in the same partonic phase space, computed respectively to the required powers in $\epsilon$.

\item 
The parent parton and its (soft and collinear) daughter partons involved in the 
integral representation of the factorized process-independent (singular) coefficient function 
$\hat{I}_{RS}$ have to be treated like the corresponding partons inside the
loop integrals of $\mathcal{A}_{n}^{\mathcal{V}}$.

This is guaranteed by performing integrals involving IR-unresolved d.o.f.
consistently regularized with CDR. In particular, the phase-space integrals 
in $\hat{I}_{RS}$ are done in D dimensions like D-dimensional loop integrals 
subject to Cutkosky cuts. 
\end{enumerate}

Concerning the first point, as long as there is an unambiguous and consistent way of 
directly applying such a non-CDR regularization convention of external states 
in the computation of the virtual loop amplitude $\mathcal{A}_{n}^{\mathcal{V}}$ 
(without appealing to its Lorentz tensor decomposition representation), 
then the demonstration is completed. 
Similar as in section~\ref{SEC:unitarity:PSA}, this point is guaranteed in our
projection prescription by the fact that all open Lorentz indices 
of the polarization projectors defined in section~\ref{SEC:Prescription} are taken to be
a D-dimensional and no dimensional splitting is ever introduced, just like in CDR.

Thus we have argued that our hybrid prescription can be conveniently 
used in a NLO IR subtraction framework to correctly obtain all RS-independent 
finite remainders needed for computing physical observables, 
with the (process-independent) integrated IR-subtraction coefficients directly 
taken from CDR. 
In other words, we have argued that our hybrid CDR-compatible prescription is unitary.~\\

Although beyond the scope of this article, it is possible, by analogy to the NLO case, 
to ensure unitarity of the prescription at NNLO and beyond, owing to the following generic features 
of an IR subtraction method (on which the above NLO discussions essentially rely).
\begin{itemize}
\item 
In a typical IR subtraction framework, all \textit{explicit} IR-singularities 
in loop amplitudes, manifested as poles in $\epsilon$, are always subtracted 
by IR subtraction terms whose constructions are based on amplitude-level singularity factorization 
formulae, and the factorized IR-subtraction coefficients are independent of all 
external polarization states;
\item  
Any potential \textit{implicit} IR singularity of the ($\epsilon$-pole-free) finite remainders 
will always be further subtracted at the integrand level of phase-space integrals 
over the external kinematics, and will be directly evaluated in 4 dimensions without employing 
dimensional regularization.
\end{itemize}
Thus concerning the 4-dimensional integrand level subtractions of implicit IR-singularities
in those finite remainders, their $\epsilon$-suppressed terms are never needed 
because the phase-space integration over the external kinematics is done (numerically) in 4 dimensions.
We leave a detailed exposition of this issue at NNLO for a future publication.

\section{A few examples}
\label{SEC:examples}

The polarization projectors constructed in section~\ref{SEC:Prescription} are independent of the loop order of virtual amplitudes, regardless of possible evanescent Lorentz structures that may be generated in D dimensions. 
To illustrate its usage without being overwhelmed by irrelevant complications, we consider two prototype examples, virtual 2-loop corrections to $gg \rightarrow gg$ in massless QCD and $e^+ e^- \rightarrow Q \bar{Q}$ at order $\alpha_s$, in order to show that the finite remainders obtained are indeed RS-independent as discussed in the preceding sections. 
We will comment along the way points worthy of attention.

\subsection{$gg \rightarrow gg$}
\label{SEC:examples:gggg}

We consider the scattering process among 4 gluons in massless QCD:
\begin{equation} \label{EQ:gggg}
g_1(p_1)~+~ g_2(p_2) \to g_3(p_3)~+~g_4(p_4)\,.
\end{equation}
The Mandelstam variables are given in eq.~(\ref{EQ:kinematicinvariants}).
The corresponding scattering amplitude perturbatively expanded up to 2-loop order reads 
\begin{eqnarray} \label{EQ:ggggAMP}
\Big|\mathcal{A}_{gggg} \Big\rangle = \Big|\mathcal{A}^{[0]}_{gggg} \Big\rangle + \Big|\mathcal{A}^{[1]}_{gggg} \Big\rangle + \Big|\mathcal{A}^{[2]}_{gggg} \Big\rangle 
+ \mathcal{O}(\alpha^{4}_s) \, , 
\end{eqnarray}
which can be viewed as a vector in the color space of the external gluons. 
The virtual corrections to the 4-gluon scattering amplitude to 2-loop order were computed in refs.~\cite{Bern:1991aq,Schubert:2001he,Binoth:2006hk,Bern:2000dn,Glover:2001af,Bern:2002tk,Ahmed:2019qtg}.
For representing color structures of multi-gluon scattering amplitudes, like eq.~(\ref{EQ:ggggAMP}), it is very  convenient to perform a color decomposition using the choice of basis of refs.~\cite{Berends:1987cv,Mangano:1987xk,Mangano:1987kp,Mangano:1988kk,Bern:1990ux}. 
It is well known that the amplitude eq.~(\ref{EQ:ggggAMP}) at the tree level can be decomposed into color-ordered partial amplitudes, multiplied by associated single color traces (over all noncylic permutations of fundamental color generators).
Decomposition of color structures of the 4-gluon scattering amplitude at higher orders in QCD can be done in a similar way but with an extended color basis including products of two color traces\footnote{This can be easily understood by combing the statement about tree-level color decomposition and the Fierz identities of SU(N) color algebra.}.

We decompose the amplitude eq.~(\ref{EQ:ggggAMP}) as follows:
\begin{eqnarray}\label{EQ:colordecomposition}
\Big|\mathcal{A}^{[0]}_{gggg} \Big\rangle = \sum_{i=1}^{6} \mathcal{A}^{[0,i]}_{gggg}\, |c_i \rangle\,,\,
\Big|\mathcal{A}^{[1]}_{gggg} \Big\rangle = \sum_{i=1}^{9} \mathcal{A}^{[1,i]}_{gggg}\, |c_i \rangle\,,\,
\Big|\mathcal{A}^{[2]}_{gggg} \Big\rangle = \sum_{i=1}^{9} \mathcal{A}^{[2,i]}_{gggg}\, |c_i \rangle\,,
\end{eqnarray}
using the following basis of 9 color structures, 
\begin{eqnarray}\label{EQ:colorbasis}
|c_1\rangle &=& \mathrm{Tr}\Big[T_1~ T_2~ T_3~ T_4 \Big]~,~~
|c_2\rangle = \mathrm{Tr}\Big[T_1~ T_2~ T_4~ T_3 \Big] ~,~~
|c_3\rangle = \mathrm{Tr}\Big[T_1~ T_3~ T_4~ T_2 \Big] \, , \nonumber\\
|c_4\rangle &=& \mathrm{Tr}\Big[T_1~ T_3~ T_2~ T_4 \Big] ~,~~
|c_5\rangle = \mathrm{Tr}\Big[T_1~ T_4~ T_3~ T_2 \Big] ~,~~
|c_6\rangle = \mathrm{Tr}\Big[T_1~ T_4~ T_2~ T_3 \Big] \, ,\nonumber\\
|c_7\rangle &=& \mathrm{Tr}\Big[T_1~ T_2\Big] \mathrm{Tr}\Big[T_3~ T_4 \Big]~,~
|c_8\rangle = \mathrm{Tr}\Big[T_1~ T_3\Big] \mathrm{Tr}\Big[T_2~ T_4 \Big] ~,~\nonumber\\
|c_9\rangle &=& \mathrm{Tr}\Big[T_1~ T_4\Big] \mathrm{Tr}\Big[T_2~ T_3 \Big] \,.
\end{eqnarray}
The subscripts of these color generators label the associated gluons while their color indices are suppressed.
These 9 color structures are linearly independent, as can be checked by computing its Gram matrix. 
The tree-level amplitude $\big|\mathcal{A}^{[0]}_{gggg} \big\rangle$ involves only the first 6 non-cylic single color traces given in eq.~\eqref{EQ:colorbasis}, which can be further reduced to 4 structures by reflection symmetries. 
The color structures $|c_7\rangle,~ |c_8\rangle,~|c_9\rangle$ are needed in addition to represent the loop amplitudes $\big|\mathcal{A}^{[1]}_{gggg} \big\rangle$ and $\big|\mathcal{A}^{[2]}_{gggg} \big\rangle$. 
If the Bose symmetry among the external gluons are explicitly taken into account, the linear basis of the color space for the 4-gluon scattering amplitude has only 6 elements, which we choose to be $\Big\{ |c_1\rangle + |c_5\rangle\,, |c_2\rangle+|c_3\rangle\,, |c_4\rangle+|c_6\rangle\,, |c_7\rangle\,, |c_8\rangle\,, |c_9\rangle \Big\}$.

Each of the color decomposition coefficients $\mathcal{A}^{[l,i]}_{gggg}$ (with $l = 1,2$) in eq.~(\ref{EQ:colordecomposition}) is a function of external kinematics and polarization state vectors, to which we now apply the polarization projectors prescribed in section~\ref{SEC:Prescription}.
We extract polarized amplitudes in the linear polarization basis for all four external gluons (cf.~section~\ref{SEC:prescription:MBR:2to2}), from which helicity amplitudes can be easily obtained. 
Because the reaction (\ref{EQ:gggg}) is parity-invariant the scattering amplitude does not contain terms involving 
$\gamma_5$ or an odd number of Levi-Civita tensors. 
We thus need to consider only the following 8 linear polarization projectors, which are even in $\varepsilon_Y$, respectively in the number of Levi-Civita tensors:
\begin{eqnarray}\label{EQ:LPgggg8}
&&\varepsilon^{\mu_1}_X \varepsilon^{\mu_2}_X \varepsilon^{\mu_3}_T \varepsilon^{\mu_4}_T ~,~ 
\varepsilon^{\mu_1}_X \varepsilon^{\mu_2}_X \varepsilon^{\mu_3}_Y \varepsilon^{\mu_4}_Y ~,~ 
\varepsilon^{\mu_1}_X \varepsilon^{\mu_2}_Y \varepsilon^{\mu_3}_T \varepsilon^{\mu_4}_Y ~,~ 
\varepsilon^{\mu_1}_X \varepsilon^{\mu_2}_Y \varepsilon^{\mu_3}_Y \varepsilon^{\mu_4}_T ~,~ \nonumber\\
&&\varepsilon^{\mu_1}_Y \varepsilon^{\mu_2}_X \varepsilon^{\mu_3}_T \varepsilon^{\mu_4}_Y ~,~
\varepsilon^{\mu_1}_Y \varepsilon^{\mu_2}_X \varepsilon^{\mu_3}_Y \varepsilon^{\mu_4}_T ~,~
\varepsilon^{\mu_1}_Y \varepsilon^{\mu_2}_Y \varepsilon^{\mu_3}_T \varepsilon^{\mu_4}_T ~,~
\varepsilon^{\mu_1}_Y \varepsilon^{\mu_2}_Y \varepsilon^{\mu_3}_Y \varepsilon^{\mu_4}_Y.
\end{eqnarray}
For the sake of simplicity of notation, the arguments of these polarization vectors are suppressed 
while their subscripts at the open Lorentz indices indicate the associated gluons.

The number of linear polarization projectors in eq.~(\ref{EQ:LPgggg8}) equals the number of independent helicity amplitudes, taking into account the parity symmetry of the scattering amplitude. 
We do not consider additional relations among the linear polarized amplitudes arising from Bose symmetry, which involve kinematic crossings.
The 8 linear polarization projectors in eq.~(\ref{EQ:LPgggg8}) are sufficient for any parity-even scattering amplitude among four external massless bosons to any loop order, irrespective of any possible (evanescent) Lorentz structures therein.\footnote{In case a $2\to 2$ scattering amplitude involves parity-violating couplings, 8 linear polarization projectors containing an odd number of $\varepsilon_Y$ (or Levi-Civita tensors) can be used in addition.} 
We insert in (\ref{EQ:LPgggg8}) the expressions (\ref{EQ:XpolMBR}), \eqref{EQ:TpolMBR}, and \eqref{EQ:YpolMBR} for the polarization vectors. 
Let us emphasize again that, in order to avoid possible ambiguities in the definition and application 
of these external projectors, all pairs of Levi-Civita tensors in eq.~(\ref{EQ:LPgggg8})
are replaced according to the contraction rule eq.~(\ref{EQ:LeviCivitaContRule}) 
\textit{before} being used in the projection, especially the projector $\varepsilon^{\mu_1}_Y \varepsilon^{\mu_2}_Y \varepsilon^{\mu_3}_Y \varepsilon^{\mu_4}_Y$ with 4 Levi-Civita tensors.  
Then the projectors (\ref{EQ:LPgggg8}) are expressed solely in terms of external momenta and space-time metric tensors, which have an unambiguous extension in D dimensions.\footnote{As a consequence of this operation, for those polarization projectors with multiple $\varepsilon_Y$, new non-factorized forms arise. Because of this, one may not be able to single out dot products with individual polarization vectors and rewrite them, in contrast to the computations done in FDH using the spinor-helicity representations of polarization vectors.}
After pulling out the normalization factors as prescribed in section~(\ref{SEC:Prescription}), 
the resulting tensor projectors (which have only a polynomial dependence on external momenta and kinematics) 
will be applied to the color stripped amplitudes $\mathcal{A}^{[l,i]}_{gggg}$.
We use the convention to set the variable $D=4$ in the projectors (\ref{EQ:LPgggg8}), in particular in the normalization factors that are pulled out. 
Of course, this convention is used both for the bare virtual amplitudes and the associated UV and/or IR subtraction terms (where amplitudes at lower loop orders occur).
Then the normalization factors pulled out from the respective projectors \eqref{EQ:LPgggg8} are equal 
in this case and given by $1 / (s^2 t^2 (s + t)^2)$.

The linear polarized amplitudes projected out by applying eq.~(\ref{EQ:LPgggg8}) to $\mathcal{A}^{[1,i]}_{gggg}$ and $\mathcal{A}^{[2,i]}_{gggg}$ contain both UV and IR singularities, manifested as poles in $\epsilon$. 
We are only interested in the finite remainders defined by subtracting all these singularities in accordance with a certain convention.
For our purpose, there is no need to stick to a specific IR-subtraction scheme. 
All we need to know is a factorization formula providing us with a set of terms that capture all singularities in 
$\mathcal{A}^{[l,i]}_{gggg}$ (with the process-independent singular coefficients obtained in CDR as explained in previous sections). 
We choose to define the finite remainders of the virtual amplitude $\big|\mathcal{A}_{gggg} \big\rangle$ in accordance with the IR factorization formulae in refs.~\cite{Catani:1998bh,Sterman:2002qn,Aybat:2006wq,Dixon:2008gr,Becher:2009cu}, conveniently denoted by 
\begin{eqnarray}\label{EQ:ggggUVIRcounterterms}
\Big| \mathcal{A}^{[\text{fin}]}_{gggg} \Big\rangle &=& \hat{\mathcal{Z}}_{\text{IR}}\big(\alpha_s,\epsilon, \left\{p_i\right\}\big) \, \Big| \mathcal{A}_{gggg}\big(\alpha^B_s \rightarrow Z_{\alpha_s} \, \alpha_s \big) \Big\rangle \, ,
\end{eqnarray}
where $\alpha^B_s$ is the bare QCD coupling, subsequently renormalized in the $\overline{\mathrm{MS}}$ scheme (with the renormalization scale $\mu = 1$), and $\left\{p_i\right\}$ denotes collectively the external momenta in eq.~(\ref{EQ:gggg}). 
The UV divergence of the on-shell 4-gluon amplitude $\big|\mathcal{A}_{gggg} \big\rangle$ in massless QCD are removed by the renormalization of the QCD coupling $\alpha_s$, which we need to 2-loop order.  
Unlike the UV divergence, the IR factorization or subtraction coefficients $\hat{\mathcal{Z}}_{\text{IR}}\big(\alpha_s,\epsilon, \left\{p_i\right\}\big)$ needed for $\big|\mathcal{A}_{gggg} \big\rangle$ is not proportional to a unit matrix in the color space: it is a $6\times 6$ dimensional  matrix of (kinematic-dependent) IR singular factors
in this color space, given explicitly in ref.~\cite{Ahmed:2019qtg}, which we use.
The crucial point relevant here is that both $Z_{\alpha_s}$ and $\hat{\mathcal{Z}}_{\text{IR}}\big(\alpha_s,\epsilon, \left\{p_i\right\}\big)$, which capture the UV and IR divergences (regularized as poles in $\epsilon$) in the virtual amplitude with fixed legs, are independent of the polarization states of the external particles.
We emphasize again that we use the expressions of these universal factors as defined in the CDR scheme.

Regarding the technical aspect of the computation, we obtain the unreduced symbolic form of the projected 4-gluon amplitudes using an extension of the program GoSam~\cite{Cullen:2011ac,Cullen:2014yla,Jones:2016bci} at 1-loop and 2-loop order. 
In particular, all the Lorentz and Dirac algebra involved in the projection are carried out using FORM~\cite{Vermaseren:2000nd}.
The list of unreduced loop integrals appearing is then extracted and fed to Kira~\cite{Maierhoefer:2017hyi,Maierhofer:2018gpa} to obtain a table of IBP rules. 
Insertion of the IBP table and subsequent simplification of rational coefficients in front of master integrals are performed with an in-house routine based on a parallelized usage of Mathematica and fermat~\cite{fermat}. 
Analytic expressions of the 1- and 2-loop master integrals involved, sufficiently expanded in $\epsilon$ to get the 2-loop finite remainders of the 4-gluon amplitudes, are taken from ref.~\cite{Ahmed:2019qtg}.~\footnote{Private communication of Taushif Ahmed.}

With this computational set-up, we get the analytic results for all 8 non-vanishing finite remainders of the interferences between $\big|\mathcal{A}^{[2]}_{gggg} \big\rangle$, $\big|\mathcal{A}^{[1]}_{gggg} \big\rangle$ and $\big|\mathcal{A}^{[0]}_{gggg} \big\rangle$ in linear polarization basis.\footnote{Up to 1-loop order, the projections and computations of the 4-gluon amplitudes are cross-checked with an alternative set-up using QGRAF~\cite{Nogueira:1991ex}, FORM~\cite{Vermaseren:2000nd} and Package-X~\cite{Patel:2015tea}.}
The finite remainder of the unpolarized interference in 4 dimensions is obtained by summing over these 8 quantities.
On the other hand, one can compute this finite remainder within CDR using a polarization sum formula like \eqref {EQ:polsumCDRPhys} for each of the 4 external gluons. 
We have checked analytically that both ways lead to the same finite remainders, while the unsubtracted bare results differ starting from the sub-leading power in $\epsilon$ due to the usage of our hybrid dimensional regularization scheme.

In addition, we have composed the helicity amplitudes with the aid of the constant transformation matrix from the linearly polarized amplitudes projected out using eq.~(\ref{EQ:LPgggg8}).
We confirm that for all helicity amplitudes we have obtained the same finite remainders analytically as those computed in ref.~\cite{Ahmed:2019qtg} where the helicity amplitudes in HV scheme are computed conventionally by first obtaining the Lorentz tensor decomposition representation of the 4-gluon amplitudes using the form-factor projectors and then evaluating contractions between Lorentz structures and external polarization vectors in 4 dimensions.

\subsection{$e^+ e^- \rightarrow Q \bar{Q}$}
\label{SEC:examples:eeQQ}

Next we consider quark-pair production in $e^+e^-$ collisions:
\begin{equation} \label{EQ:eeQQ}
e^-(p_1) ~+~ e^+(p_2) \to Z^* \to Q(p_3)~+~{\bar Q}(p_4)~,
\end{equation}
mediated by a Z-boson where $Q$ denotes a massive quark with mass 
$m$, i.e., $p_3^2 = p_4^2 = m^2$, and the electron (positron) is taken to be massless. 
The corresponding bare scattering amplitude perturbatively expanded to NLO in QCD reads
\begin{eqnarray}\label{EQ:eeQQAMP} 
\Big|\mathcal{A}_{eeQQ} \Big\rangle &=&  
\mathcal{A}^{[\text{tree}]}_{eeQQ}(1_{e^-}, 2_{e^+}, 3_{Q}, 4_{\bar{Q}}) ~\delta_{i_3 i_4} \nonumber\\
&+&\frac{\alpha_s}{4 \pi} \bar{C}( \epsilon ) 
\mathcal{A}^{[\text{1-loop}]}_{eeQQ}(1_{e^-}, 2_{e^+}, 3_{Q}, 4_{\bar{Q}}) ~2\mathrm{C}_F \delta_{i_3 i_4}
+ \mathcal{O}(\alpha_s^2) \, ,
\end{eqnarray} 
where $i_3$ ($i_4$) denotes the color index of the heavy quark (antiquark), 
$\mathrm{C}_F=({N_c^2-1})/({2N_c})$, and $\bar{C}(\epsilon) \equiv 
( 4 \pi )^\epsilon e^{-\epsilon \gamma_E }$ with $\gamma_E = 0.57721 \ldots $ 
denoting the Euler--Mascheroni constant.
In eq.~(\ref{EQ:eeQQAMP}) we introduced symbolic labels $i_X$ in order to encode the dependence on  
the momentum $p_i$ and helicity $\lambda_i$ of an external particle $i$ of type $X$.
These 1-loop QCD corrections were first computed in ref.~\cite{Jersak:1981sp}.

Because we work to the lowest order in electroweak couplings, the UV renormalization counterterms 
can be introduced by the following replacement of the bare coupling vertex of the Z boson and the heavy quark:
\begin{eqnarray} \label{EQ:eeQQUVcounterterms}
\Big( v_Q \gamma^{\mu} + a_Q \gamma^{\mu} \gamma_5 \Big) 
\rightarrow 
Z^{[1]}_{\psi,OS}(\epsilon,\alpha_s)
\Big( v_Q \gamma^{\mu} + Z^{ns}_5(\alpha_s) ~a_Q  
\frac{-i}{3!} \epsilon^{\mu \nu \rho \sigma} 
\gamma_{\nu} \gamma_{\rho} \gamma_{\sigma} \Big) \, .
\end{eqnarray}
Here $v_Q$ and $a_Q$ denote the vector and axial vector couplings of $Q$,
\begin{equation*}
 Z^{[1]}_{\psi,OS}(\epsilon,\alpha_s) = 
-\frac{\alpha_s}{4\pi} (4\pi)^\epsilon ~ \Gamma(1+\epsilon)\frac{1}{\epsilon} 
\left(\frac{\mu_{DR}^2}{m^2}\right)^\epsilon \mathrm{C}_F  \frac{(3-2\epsilon)}{(1-2\epsilon)} + \mathcal{O}(\alpha_s^2) \, ,
\end{equation*}
and we use Larin's prescription~\cite{Larin:1991tj,Larin:1993tq} for the non-singlet axial vector current 
which involves $Z^{ns}_5(\alpha_s) = 1 +  \frac{\alpha_s}{4\pi} \left( -4 \mathrm{C}_F \right)+ \mathcal{O}(\alpha_s^2)$.

For subtracting the IR singularities of the renormalized 1-loop amplitude $\mathcal{A}^{[\text{1-loop,R}]}_{eeQQ}$, 
we use the antenna subtraction method \cite{Kosower:1997zr,GehrmannDeRidder:2005cm}.
The antenna subtraction term needed here reads \cite{GehrmannDeRidder:2009fz}: 
\begin{eqnarray}\label{EQ:eeQQIRcounterterms} 
\big|\mathcal{A}^{[\text{IR}]}_{eeQQ} \big\rangle &=&  
\frac{\alpha_s}{4 \pi} \bar{C}( \epsilon ) 
\mathcal{A}^0_3\big(\epsilon, \frac{\mu_{DR}^2}{s}; y \big)
\mathcal{A}^{[\text{tree}]}_{eeQQ}(1_{e^-}, 2_{e^+}, 3_{Q}, 4_{\bar{Q}}) ~2\mathrm{C}_F \delta_{i_3 i_4}
+ \mathcal{O}(\alpha_s^2), 
\end{eqnarray} 
where $y=\frac{1-\beta}{1+\beta}~,~ \beta=\sqrt{1-4m^2/s}$, and $\mathcal{A}^0_3\big(\epsilon, \frac{\mu_{DR}^2}{s}; y \big)$ denotes the integrated three-parton tree-level massive quark-antiquark antenna function given in refs.~\cite{GehrmannDeRidder:2009fz,Abelof:2011jv}. 

Because we take the leptons to be massless, there are only 8 non-vanishing helicity amplitudes
which, in the absence of parity symmetry\footnote{In the Standard Model the 1-loop scattering
amplitude of (\ref{EQ:eeQQ}) still respects the combined symmetry of parity and charge conjugation,
which relates the helicity amplitude with helicity configuration $+- ++$ to $+- --$, and similarly $-+ ++$ to $-+ --$.},
differ from each other. 
We now consider the extraction of polarized amplitudes in the helicity basis both at the tree level and the 1-loop level.
Following the discussion of section~\ref{SEC:prescription:NTS}, 
we choose to attach an auxiliary spinor inner product 
\begin{equation}\label{EQ:flHLnormfactors}
\mathcal{N}_{\lambda_{e} \lambda_{Q} \lambda_{\bar{Q}}} = 
\bar{u}(p_1, \lambda_{e}) \slashed{p}_3 v(p_2, -\lambda_{e}) \otimes  
\bar{v}(p_4, \lambda_{\bar{Q}}) \slashed{p}_1 u(p_3, \lambda_{Q}) 
\end{equation}
to each helicity amplitude characterized by $\lambda_{e},~ \lambda_{Q},~\lambda_{\bar{Q}}$. 
This factor is to be removed by numerical division  at the end of the computation in 4 dimensions. 
Pulling off $\mathcal{N}^{~ -1}_{\lambda_{e} \lambda_{Q} \lambda_{\bar{Q}}}$  
from each helicity amplitude, the polarization projections can be most conveniently performed, in analogy to eq.~\eqref{EQ:SFLtrace1}, using 
the following 8 regrouped projectors according to eqs.~(\ref{EQ:TPextsps2}), \eqref{EQ:TPextsps2massless}:
\begin{eqnarray} \label{EQ:LPeeQQ8}
\hat{\mathrm{P}}_1 &=&  \Big(\slashed{p}_1  \slashed{p}_3 \slashed{p}_2 \Big) \otimes
\Big( \left(\slashed{p}_4 - m \right) \slashed{p}_1 \left(\slashed{p}_3 + m \right) \Big)~,~\nonumber\\
 \hat{\mathrm{P}}_2 &=&  \Big(\slashed{p}_1  \slashed{p}_3 \slashed{p}_2 \Big) \otimes
\Big( \left(\slashed{p}_4 - m \right) \left(\frac{-i}{3!} \epsilon_{\gamma \gamma \gamma S_{\bar{Q}}} \right) \slashed{p}_1 \left(\slashed{p}_3 + m \right) \Big)~,~\nonumber\\
\hat{\mathrm{P}}_3 &=&  \Big(\slashed{p}_1  \slashed{p}_3 \slashed{p}_2 \Big) \otimes
\Big( \left(\slashed{p}_4 - m \right) \slashed{p}_1 \left(\frac{-i}{3!} \epsilon_{\gamma \gamma \gamma S_{Q}} \right) \left(\slashed{p}_3 + m \right) \Big)~,~\nonumber\\
 \hat{\mathrm{P}}_4 &=&  \Big(\slashed{p}_1  \slashed{p}_3 \slashed{p}_2 \Big) \otimes
\Big( \left(\slashed{p}_4 - m \right) \slashed{S}_{\bar{Q}} \slashed{p}_1 \slashed{S}_{Q}  \left(\slashed{p}_3 + m \right) \Big)~,~\nonumber\\
\hat{\mathrm{P}}_5 &=&  \Big(\slashed{p}_1 \frac{i}{3!}\epsilon_{\gamma \gamma \gamma p_3}  \slashed{p}_2 \Big) \otimes
\Big( \left(\slashed{p}_4 - m \right) \slashed{p}_1 \left(\slashed{p}_3 + m \right) \Big)~,~\nonumber\\
 \hat{\mathrm{P}}_6 &=&  \Big(\slashed{p}_1 \frac{i}{3!}\epsilon_{\gamma \gamma \gamma p_3}  \slashed{p}_2 \Big) \otimes
\Big( \left(\slashed{p}_4 - m \right) \left(\frac{-i}{3!} \epsilon_{\gamma \gamma \gamma S_{\bar{Q}}} \right) \slashed{p}_1 \left(\slashed{p}_3 + m \right) \Big)~,~\nonumber\\
\hat{\mathrm{P}}_7 &=&  \Big(\slashed{p}_1 \frac{i}{3!}\epsilon_{\gamma \gamma \gamma p_3}  \slashed{p}_2 \Big) \otimes
\Big( \left(\slashed{p}_4 - m \right) \slashed{p}_1 \left(\frac{-i}{3!} \epsilon_{\gamma \gamma \gamma S_{Q}} \right) \left(\slashed{p}_3 + m \right) \Big)~,~\nonumber\\
 \hat{\mathrm{P}}_8 &=&  \Big(\slashed{p}_1 \frac{i}{3!}\epsilon_{\gamma \gamma \gamma p_3} \slashed{p}_2 \Big) \otimes
\Big( \left(\slashed{p}_4 - m \right) \slashed{S}_{\bar{Q}} \slashed{p}_1 \slashed{S}_{Q}  \left(\slashed{p}_3 + m \right) \Big)~,
\end{eqnarray}
where the momentum basis representations of the two helicity polarization vectors  
$S_{Q}^{\mu}$ and $S_{\bar{Q}}^{\mu}$, in analogy to  eq.~(\ref{EQ:PLpolMBR}), 
will be inserted during the computation\footnote{This insertion can 
conveniently be done after having performed the Dirac traces and having used 
$p_3 \cdot S_{Q}=p_4 \cdot S_{\bar{Q}}=0$ and $S_{Q} \cdot S_{Q} = S_{\bar{Q}} \cdot S_{\bar{Q}} = -1$.}
so that eventually the resulting projections are functions of the external momenta only.
Of course, the manipulation of Dirac matrices associated with two disconnected fermion lines 
(separated by $\otimes$ in eq.~(\ref{EQ:LPeeQQ8})) can be performed independently and should not be confused. 
Notice that the set of polarization projectors in eq.~(\ref{EQ:LPeeQQ8}) is also sufficient for computing virtual 
amplitudes that involve box contributions, for instance $q(p_1)~\bar{q}(p_2) \to Q(p_3)~{\bar Q}(p_4)$ in QCD, 
irrespective of any possible evanescent 
Lorentz structure that can be generated at high loop orders in D dimensions. 
In case $q(p_1)~\bar{q}(p_2) \to Q(p_3)~{\bar Q}(p_4)$ is parity invariant, 
which is the case if one considers only QCD interactions, 
then $\hat{\mathrm{P}}_2~,\hat{\mathrm{P}}_3~,\hat{\mathrm{P}}_5~,\hat{\mathrm{P}}_8$ 
can be safely discarded and only 4 projectors are needed.

In the simple example considered here, where the amplitude \eqref{EQ:eeQQAMP} involves only 3-point
vertex functions, there is not much technical advantage in using eq.~(\ref{EQ:LPeeQQ8}) instead of  
the conventional form-factor decomposition. If one nevertheless chooses to use the projectors 
(\ref{EQ:LPeeQQ8}) for computing helicity amplitudes including QCD corrections, 
one can compute the trace \eqref{EQ:SFLtrace1} of the string of Dirac matrices along the lepton line,
both for the renormalized amplitude and the IR subtraction term \eqref{EQ:eeQQIRcounterterms},
in 4 dimensions, because the lepton line receives no QCD correction and remains purely tree level. 
In this case we can replace $\frac{i}{3!}\epsilon_{\gamma \gamma \gamma p_3}$ 
in eq.~(\ref{EQ:LPeeQQ8}) by $\slashed{p}_3 \gamma_5$.

Helicity amplitudes can be assembled by linear combinations of the 
projections made with (\ref{EQ:LPeeQQ8}), and the linear combination coefficients can be 
read off from eqs.~(\ref{EQ:TPextsps2}), \eqref{EQ:TPextsps2massless}. 
It is convenient to perform such a transformation at a later stage of the 
 computation where explicit analytic results have been inserted. 
The explicit form of the overall normalization factor given in 
eq.~(\ref{EQ:flHLnormfactors}) is usually needed only at the level  
of squared amplitudes (or interferences).
The squared modulus of  $\mathcal{N}_{\lambda_{e} \lambda_{Q} \lambda_{\bar{Q}}}$ is
\begin{eqnarray}
\Big| \mathcal{N}_{\lambda_{e} \lambda_{Q} \lambda_{\bar{Q}}} \Big|^2 &=& 
-\frac{m^4 - 2 m^2 t + t (s + t)}{2} \Big(
\lambda_{Q} \lambda_{\bar{Q}} ~\Big(
2(m^2 - t)  
(p_1 \cdot S_{Q} ~ p_3 \cdot S_{\bar{Q}} - p_1 \cdot S_{\bar{Q}} ~ p_4 \cdot S_{Q}) \nonumber\\
&&~~~~ + 
2 s ~(p_1 \cdot S_{\bar{Q}} ~ p_4 \cdot S_{Q} - p_1 \cdot S_{Q} ~ p_1 \cdot S_{\bar{Q}} ) 
\Big) \nonumber\\
&&~~ + (m^2 - t) (m^2 - s - t) \left(-1 + \lambda_{Q} \lambda_{\bar{Q}} ~ S_{Q} \cdot S_{\bar{Q}} \right)
\Big)\nonumber\\
&=&
\frac{1}{2} (m^2 - t) (m^2 - s - t) (m^4 - 2 m^2 t + t (s + t)) 
- \frac{\lambda_{Q} \lambda_{\bar{Q}}}{2 (s - 4 m^2)} \nonumber\\
&& (m^4 - 2 m^2 t + t (s + t)) (4 m^6 + s t (s + t) - m^4 (3 s + 8 t) + m^2 (s^2 + 2 s t + 4 t^2))\, \nonumber\\
\end{eqnarray}
where in the last line we have inserted momentum basis representations of 
$S_{Q}^{\mu}$ and $S_{\bar{Q}}^{\mu}$ that are given in analogy to eq.~(\ref{EQ:PLpolMBR}). 
In case the normalization factors are to be included at the amplitude level, 
we can use for their computation either the concrete 4-dimensional representations
of spinors and Dirac matrices, as listed for instance in ref.~\cite{Murayama:1992gi}, 
or employ the 4-dimensional spinor-helicity representation of 
these objects~\cite{Gunion:1985vca,Kleiss:1985yh,Xu:1986xb,Kleiss:1986qc,Dittmaier:1998nn,Schwinn:2005pi,Arkani-Hamed:2017jhn}.

With the ingredients just outlined we computed the finite remainders of the interferences between the  
tree-level and 1-loop helicity amplitudes, multiplied, for convenience, with the inverse square of the Z-boson propagator: 
\begin{equation}\label{EQ:tr1lohel}
\left(s-m^2_Z\right)^2 \times 2~ {\rm Re}\Big[\mathcal{A}^{[\text{tree}]*}_{eeQQ}(1_{e^-}, 2_{e^+}, 3_{Q}, 4_{\bar{Q}})~
 \mathcal{A}^{[\text{1-loop}]}_{eeQQ}(1_{e^-}, 2_{e^+}, 3_{Q}, 4_{\bar{Q}}) \Big] \, .
\end{equation}
We calculated \eqref{EQ:tr1lohel} analytically using FORM~\cite{Vermaseren:2000nd} and the involved loop integrals 
with Package-X~\cite{Patel:2015tea}. Table~\ref{TAB:numbersFinInf}  contains the finite remainders of \eqref{EQ:tr1lohel}
for all helicity configurations evaluated at the test point $m=17.3$ GeV, $s= 10^6$ $({\rm GeV})^2$, $t = -90$ $({\rm GeV})^2$.
($v_e$ and $a_e$ denote the vector and axial vector couplings of electron.)

\vspace{2mm}
\begin{table}[tbh!]
\begin{center}
\begin{tabular}{|c|l|}
\hline
Helicities  &  Finite remainders of the interferences \eqref{EQ:tr1lohel}  in units of $({\rm GeV})^2$ \\
\hline 
$+-,++$ & $-1.4211829*10^6 ~a_e^2 v_Q^2 - 2.8423658*10^6 ~a_e v_e v_Q^2-1.4211829*10^6 ~v_e^2 v_Q^2$ \\
\hline
\multirow{3}{*}{$+-,+-$} & $~~ 2.4731876*10^4 ~a_e^2 a_Q^2 + 4.9463752*10^4 ~a_e a_Q^2 v_e + 2.4731876*10^4 ~a_Q^2 v_e^2 $ \\
                         & $~+ 4.9178930*10^4 ~a_e^2 a_Q v_Q + 9.8357861*10^4 ~a_e a_Q v_e v_Q + 4.9178930*10^4 ~a_Q v_e^2 v_Q $ \\ 
                         & $~+ 2.4446875*10^4 ~a_e^2 v_Q^2 + 4.8893750*10^4 ~a_e v_e v_Q^2 + 2.4446875*10^4 ~v_e^2 v_Q^2$ \\
\hline
\multirow{3}{*}{$+-,-+$} & $~~ 3.0551961 *10^{12}~a_e^2 a_Q^2 + 6.1103923 *10^{12}~a_e a_Q^2 v_e + 3.0551961 *10^{12}~a_Q^2 v_e^2 $ \\
                         & $~+ 6.0752075 *10^{12}~a_e^2 a_Q v_Q -1.2150415 * 10^{13}~a_e a_Q v_e v_Q -6.0752075 *10^{12}~a_Q v_e^2 v_Q $ \\ 
                         & $~+ 3.0199891 *10^{12}~a_e^2 v_Q^2 + 6.0399783 *10^{12}~a_e v_e  v_Q^2 + 3.0199891 *10^{12}~v_e^2 v_Q^2$ \\
\hline
$+-,--$ & $-1.4211829*10^6 ~a_e^2 v_Q^2 - 2.8423658*10^6 ~a_e v_e v_Q^2-1.4211829*10^6 ~v_e^2 v_Q^2$ \\
\hline
$-+,++$ & $-1.4211829*10^6 ~a_e^2 v_Q^2 + 2.8423658*10^6 ~a_e v_e v_Q^2-1.4211829*10^6 ~v_e^2 v_Q^2$ \\
\hline
\multirow{3}{*}{$-+,+-$} & $~~ 3.0551961 *10^{12}~a_e^2 a_Q^2 - 6.1103923 *10^{12}~a_e a_Q^2 v_e + 3.0551961 *10^{12}~a_Q^2 v_e^2 $ \\
                         & $~+ 6.0752075 *10^{12}~a_e^2 a_Q v_Q -1.2150415 * 10^{13}~a_e a_Q v_e v_Q +6.0752075 *10^{12}~a_Q v_e^2 v_Q $ \\ 
                         & $~+ 3.0199891 *10^{12}~a_e^2 v_Q^2 - 6.0399783 *10^{12}~a_e v_e  v_Q^2 + 3.0199891 *10^{12}~v_e^2 v_Q^2$ \\
\hline
\multirow{3}{*}{$-+,-+$} & $~~ 2.4731876*10^4 ~a_e^2 a_Q^2 - 4.9463752*10^4 ~a_e a_Q^2 v_e + 2.4731876*10^4 ~a_Q^2 v_e^2 $ \\
                         & $~+ 4.9178930*10^4 ~a_e^2 a_Q v_Q + 9.8357861*10^4 ~a_e a_Q v_e v_Q - 4.9178930*10^4 ~a_Q v_e^2 v_Q $ \\ 
                         & $~+ 2.4446875*10^4 ~a_e^2 v_Q^2 - 4.8893750*10^4 ~a_e v_e v_Q^2 + 2.4446875*10^4 ~v_e^2 v_Q^2$ \\
\hline
$-+,--$ & $-1.4211829*10^6 ~a_e^2 v_Q^2 + 2.8423658*10^6 ~a_e v_e v_Q^2-1.4211829*10^6 ~v_e^2 v_Q^2$ \\
\hline
\end{tabular}
\caption{\label{TAB:numbersFinInf} 
Numerical values of the finite remainders of the interferences \eqref{EQ:tr1lohel}
at the test point $m=17.3$ GeV, $s= 10^6$ $({\rm GeV})^2$, $t = -90$ $({\rm GeV})^2$.}
\end{center}
\end{table} 
The interferences were computed to about 30 significant digits while only the
first 8 significant digits are shown in table~\ref{TAB:numbersFinInf}
for simplicity. (There is no rounding in the shown digits.) 
 CP invariance dictates that the helicity configurations 
$+- ++$ and $+- --$ yield identical expressions, and likewise $-+ ++$ and $-+ --$.
The large differences between the values of these helicity amplitudes are  
due to the particular kinematic point considered: it corresponds to 
a high-energy (small mass) limit of the scattering amplitude 
in the near-forward scattering region.

We computed also the finite remainder of the unpolarized interferences \eqref{EQ:tr1lohel} within CDR at the same kinematic point with the renormalized virtual amplitudes from refs.~\cite{Bernreuther:2004ih,Bernreuther:2004th} 
available in a form-factor decomposed form.
For this unpolarized interference we obtain
\begin{equation*}
6.1103923*10^{12} ~\big(a_e^2 a_Q^2 + v_e^2 a_Q^2 \big)
~-~ 2.4300829*10^{13} ~a_e v_e a_Q v_Q ~+~ 
6.0399727*10^{12} ~\big(a_e^2 v_Q^2 + v_e^2 v_Q^2 \big),
\end{equation*}
which precisely reproduces the sum of all helicity configurations listed in table~\ref{TAB:numbersFinInf}.
~\\

Let us comment on a point that was already alluded to in section~\ref{SEC:prescription:NTS} and discussed in section~\ref{SEC:unitarity:PSA}. 
It concerns the  placing of Dirac matrices between pairs of on-shell projection operators.
Moving the matrix $\big(\frac{-i}{3!} \epsilon_{\gamma \gamma \gamma S_{Q}} \big)$ around 
in the external projectors in eq.~\eqref{EQ:LPeeQQ8} in accordance with the 4-dimensional algebra 
between the pair of on-shell projection operators, 
$\big(\slashed{p}_4 - m \big)$ and $\big(\slashed{p}_3 + m \big)$,
always leads to the same finite remainders documented in table~\ref{TAB:numbersFinInf}. 
Yet, as expected, these different choices result in different bare (unsubtracted) virtual amplitudes.
Once we decide to move $\big(\frac{-i}{3!} \epsilon_{\gamma \gamma \gamma S_{Q}} \big)$ 
beyond $\big(\slashed{p}_4 - m \big)$ or $\big(\slashed{p}_3 + m \big)$,
this operation has to be made in accordance with the D-dimensional algebra in order to end up 
with the same finite remainders (with the same IR subtraction coefficients). 
For instance, the commutator between $\big(\frac{-i}{3!} \epsilon_{\gamma \gamma \gamma S_{Q}} \big)$
and $\slashed{p}_3$, which vanishes in 4 dimensions because of
$p_3 \cdot S_{Q} = 0$, must not be omitted.
~\\

We conclude this subsection with a remark on a subtle point concerning the specification of a definite contraction order among multiple Levi-Civita tensors, in order to reach an unambiguous canonical form for a projector as well as for the resulting projection in D dimensions. 
As discussed in ref.~\cite{Moch:2015usa}, the contraction of four Levi-Civita tensors can lead to different expressions in D dimensions depending on the choice of pairings, which are not algebraically identical due to the lack of a Schouten identity.
This issue is of no concern for the amplitude of eq.~(\ref{EQ:eeQQAMP}), especially if we do the trace over the lepton line using 4-dimensional Dirac algebra before dealing with the heavy quark line. 
Nevertheless, in more general situations to which our projector prescriptions also apply,
one clear and safe choice would be to pair Levi-Civita tensors from inner vertices (of the same fermion line) in the contraction~\cite{Moch:2015usa}, leaving all other Levi-Civita tensors appearing 
in the external projectors in a different category that are to be manipulated among themselves (in 4 dimensions). 
Once a definite choice of pairing and ordering of Levi-Civita tensors in the contraction is made, it should be consistently applied in the computations of all terms that contribute to a (renormalized and subtracted) helicity amplitude.
Let us stress again that the prescription for the \textit{external} projectors proposed here is not tied to applying a non-anticommuting $\gamma_5$ prescription to the axial currents or other $\gamma_5$-related objects \textit{inside} the amplitudes (stripped off external states). 
Any appropriate $\gamma_5$ prescription, such as those featuring an anticommuting $\gamma_5$ to some extent~\cite{Chanowitz:1979zu,Buras:1989xd,Kreimer:1989ke,Korner:1991sx,Zerf:2019ynn}, can of course be used as long as its application to the amplitudes in question is carefully implemented.
In particular, for an open fermion line to which an external projector with Dirac matrices, like those in eqs.~(\ref{EQ:TPextsps1},\ref{EQ:LDMofDSP}), is applied, if several $\gamma_5$ from non-singlet axial-current vertices and/or pseudoscalar vertices are present on the same line, including possibly the one from the external projector, one can resort to a fully anti-commuting $\gamma_5$ and use the rule $\gamma_5^2 = 1$ in $D$ dimensions~\cite{Chanowitz:1979zu} to reduce them. 
Furthermore, if the total number of $\gamma_5$ on this open fermion line is odd, one can choose to move the remaining single $\gamma_5$ after the above anticommuting manipulation into the external projector and placed in accordance with the prescription formulated in section~\ref{SEC:prescription:NTS} and the related comments given in section~\ref{SEC:unitarity:PSA}.
This shall lead to the same final result one would get with a thorough implementation of Larin's 
prescription of non-singlet axial vector vertices and pseudoscalar vertices~\cite{Larin:1993tq,Moch:2015usa}, albeit it is computationally more convenient.

\section{Conclusions}
\label{SEC:conclusion}

The aim of this article was to formulate a prescription for obtaining polarized 
dimensionally regularized amplitudes and to provide a recipe for constructing simple 
and general polarized amplitude projectors in D dimensions, which circumvents  
the conventional Lorentz tensor decomposition, and difficulties associated with it, in a manifestly CDR-compatible way.
The polarization projectors devised in this article are based on the momentum basis representations of external state vectors, 
and all their open Lorentz indices are taken to be D-dimensional. 
This avoids dimensional splitting when applied to loop amplitudes. 
The momentum basis representations of external gauge bosons' polarization vectors as well as polarization vectors of massive fermions were discussed in detail in the first half of this article. 
In particular, the way of dealing with massive external polarized fermions, 
i.e.,~by inserting momentum basis representations of their polarization vectors appearing in Landau density matrices, 
has not been been discussed before in the literature.
Subtleties related to the proper arrangement of pieces in the respective projectors in D-dimensional computations are discussed for the first time in this article.
It is also worth pointing out that this treatment is fully compatible with Larin's prescription of $\gamma_5$ in $D$ dimensions, 
and hence it is convenient to use when there are axial (or pseudoscalar) couplings involved in the loop amplitude in dimensional regularization.
It is, however, worth emphasizing that the prescription for the external projectors proposed here is not tied to applying a 
non-anticommuting $\gamma_5$ prescription to the axial currents or other $\gamma_5$-related objects inside an amplitude.

As shown in section~\ref{SEC:Prescription}, it is quite straightforward to construct these projectors,
and their structures depend only on the masses and spins of the external particles. 
The construction procedure requires almost no knowledge of the Lorentz structures present in the loop amplitude, 
nor whether or not they are linearly independent of each other (in D dimensions).
In particular, there is no need to trim any unphysical Lorentz structure off the original Feynman-diagrammatic 
representation of the amplitude before applying these external projectors. 
The number and forms of these projectors are truly independent of the loop order of the virtual amplitude 
as well as of possible evanescent Lorentz structures that may be generated in D dimensions. 
In fact, the number of these projections needed are equal to the number of independent helicity amplitudes in 4 dimensions.
Constraints from symmetry properties such as parity symmetry can be accounted for in a simple way in terms of this set of projectors.

From the point of view of the projection method as recapped in section~\ref{SEC:projectionmethod:recap} and~\ref{SEC:prescription:comments}, the set of projectors prescribed in this article may be loosely viewed as a special choice of Lorentz decomposition basis structures which by construction are orthogonal to each other. 
This perspective is also very useful in showing how one can easily reproduce polarized amplitudes defined in other helicity conventions or polarization bases, starting from the original projections with the proposed projectors.
Furthermore, each of these decomposition structures is directly related to a physical quantity, a linearly polarized amplitude up to a normalization factor, and thus patterns of (explicit and/or implicit) singularities therein are protected by physical conditions observed by these physical quantities.
In this way the issues related to the conventional form-factor decomposition as discussed in section~\ref{SEC:projectionmethod:comments} are avoided.~\\

The usage of these D-dimensional polarized amplitude projectors results in helicity amplitudes 
which are eventually expressed solely in terms of Lorentz invariants made out of external momenta. 
The resulting (bare) helicity amplitudes (and the incoherent sum of their squared moduli) are, 
however, different from those defined in many existing dimensional regularization schemes, in particular CDR.
Despite being different from CDR, owing to the amplitude-level factorization of UV and IR singularities 
(which are independent of polarization states of external particles),
combined with the commutation between D-dimensional Lorentz-index contraction and loop integration, 
our prescription for external states can be used in a hybrid way with CDR to obtain the same 
finite remainders of loop amplitudes as in CDR, 
without having to re-calculate the (process-independent) pole-subtraction coefficients. 
This was demonstrated in section~\ref{SEC:unitarity:PSA} in a formal way for minimally pole-subtracted amplitudes
where a few subtle points related to manipulating fermions are discussed along the way. 
The validity of our argumentation is not confined to one-loop corrections to Born amplitudes, 
but persists as long as the amplitude-level factorization formulas hold in CDR, as sketched in eq.~(\ref{EQ:AmpPoleFactorization}). 
 
Subsequently, the same issue was discussed in section~\ref{SEC:unitarity:FRIR} for finite remainders 
defined in an IR subtraction framework, where we argued that the unitarization recipe of ref.~\cite{Catani:1996pk} 
is properly respected by our method. 
Thus we have shown that our hybrid CDR-compatible prescription is unitary.
We emphasize again that in order to unambiguously and consistently apply our prescription for external states to the calculation of loop amplitudes in D dimensions, there is no need to appeal to their Lorentz tensor decomposition representations.

In order to illustrate the usage of our hybrid prescription in practical applications, we discussed in section~\ref{SEC:examples} 
the construction of polarization projectors for $e^+ e^- \rightarrow Q \bar{Q}$ and $gg \rightarrow gg$, 
and computed their RS-independent finite remainders respectively to 1-loop and 2-loop order in QCD.  
While the arguments presented in section~\ref{SEC:unitarity:FRIR}, as well as the examples of section~\ref{SEC:examples}, 
mainly focus on NLO computations, it is possible to ensure unitarity of the prescription at NNLO in QCD and beyond, with the aid of an IR-subtraction method as briefly commented on at the end of the section~\ref{SEC:unitarity:FRIR}. 
This is, however, beyond the scope of the current article, and we leave a detailed exposition of this in a future publication.~\\

Given the impressive list of calculations of unpolarized observables done using CDR, we hope that, 
with this add-on, the resulting hybrid CDR-compatible prescription formulated in this article offers 
a convenient and efficient set-up for computing physical observables associated with polarization effects 
for phenomenologically interesting processes in perturbative QCD.
~\\

\textit{Note added}: While this work was under reversion, there appeared refs.~\cite{Ahmed:2019udm,Peraro:2019cjj,Peraro:2020sfm} aimed to address some of the issues related to the evanescent tensor structures in the conventional form factor decomposition formalism, highlighting the advantage of removing evanescent tensor structures in a scattering amplitude.

\section*{Acknowledgments}

The author is grateful to W.~Bernreuther, M.~Czakon, and G.~Heinrich for discussions and comments on the manuscript.
The author would like to thank T.~Ahmed for pleasant communication regarding the cross-check of the finite remainders of helicity amplitudes computed in ref.~\cite{Ahmed:2019qtg}, and R.~Poncelet for discussions regarding application of this approach on 5-point scattering amplitudes.
The author also wishes to thank S.~Jahn, S.~Jones and M.~Kerner for helpful discussions and feedback on the draft, 
and T.~Ahmed, M.~Capozi, H.~Luo, J.~Schlenk, Z.G.~Si, Y.~Zhang for reading the manuscript. 


\appendix

\section{An explicit formula for linear polarization states of a (massive) gauge boson}
\label{append:ffvp}

We have seen in section~\ref{SEC:prescription:MBR} and \ref{SEC:prescription:NTS}  
that three linearly independent external momenta are sufficient to build momentum basis representations of external polarization vectors, regardless of their masses, and the concrete decomposition coefficients depend on the particular kinematics.

In section~\ref{SEC:prescription:beyond4P}, we have provided a compact formula for linear polarization states of a massless gauge boson that can be conveniently used in any multiple-parton scattering process in massless QCD with a flexible choice of the (lightlike) reference vectors as well as the additional auxiliary vectors.
For constructing momentum basis representations of polarization vectors for final-state vector bosons in general, it is also convenient to take a group of three linearly independent external momenta of which two are always chosen to be the momenta of the initial-state (massless) particles and the third one is the particular final-state particle in question.
Using this approach, we document here the momentum basis representations of linear polarization vectors introduced in section~\ref{SEC:prescription:MBR}, but without specializing the concrete external kinematic configuration.
We consider a generic configuration with two massless initial state particles with momenta $p_1$ and $p_2$, 
applicable to most of the phenomenologically interesting high-energy scattering processes, 
while the mass of the particular final state particle, with momentum $p_3$ is left unspecified. 
These three external momenta are assumed to be linearly independent. 
No specification is made of the kinematics of the other particles in the final state.

For the kinematic invariants required here are
\begin{eqnarray}
s_{12} = 2~ p_1 \cdot p_2 ~,~~  s_{13} = 2~ p_1 \cdot p_3 ~,~~  
s_{23} = 2~ p_2 \cdot p_3 ~,~~ m^2 = p_3 \cdot p_3 ~,
\end{eqnarray} 
which are assumed to be independent of each other. 
Repeating the construction made in section~\ref{SEC:prescription:MBR}, 
we obtain for this generic kinematic setting:
\begin{eqnarray} \label{EQ:TranspolMBRmassiveGeneric}
\varepsilon^{\mu}_{X} &=& \mathcal{N}_{X} 
\Big((-s_{23})~ p^{\mu}_1 + (-s_{13})~ p^{\mu}_2 + s_{12}~ p^{\mu}_3 \Big) , \nonumber\\
\varepsilon^{\mu}_{T} &=&  \mathcal{N}_{T} 
\Big((-s_{23} (s_{13} + s_{23}) + 2 m^2 s_{12} )~ p^{\mu}_1 + (s_{13} (s_{13} + s_{23}) - 2 m^2 s_{12})~ p^{\mu}_2 + (s_{12} (-s_{13} + s_{23}))~ p^{\mu}_3 \Big), \nonumber\\
\varepsilon^{\mu}_{Y} &=& \mathcal{N}_{Y}~ 2 \epsilon^{\mu}_{p_1 p_2 p_3}, \nonumber\\
\varepsilon^{\mu}_{L3}&=& \mathcal{N}_{L3} 
\Big(-2 m^2~ \left( p^{\mu}_1 + p^{\mu}_2 \right) + (s_{13} + s_{23}) ~ p^{\mu}_3 \Big) \, ,
\end{eqnarray}
with the respective normalization factors 
\begin{eqnarray} \label{EQ:TranspolMBRmassiveGenericNFs}
\mathcal{N}_{X}^{~-2} &=& s_{12} \left(s_{13} s_{23} - m^2 s_{12}\right) \, ,\nonumber\\
\mathcal{N}_{T}^{~-2} &=& s_{12} \left( s_{13} s_{23} (s_{13} + s_{23})^2 - m^2 s_{12} (s_{13}^2 + 6 s_{13} s_{23} + s_{23}^2) + 4 m^4 s_{12}^2 \right) \, , \nonumber\\
\mathcal{N}_{Y}^{~-2} &=& s_{12} \left(s_{13} s_{23} - m^2 s_{12}\right) \, , \nonumber\\
\mathcal{N}_{L3}^{~-2} &=& m^2 \left((s_{13} + s_{23})^2 - 4 m^2 s_{12}\right) \, .
\end{eqnarray}
The comments on polarization vectors and normalization factors made in section~\ref{SEC:prescription:beyond4P} apply here as well.
In particular, when there are no less than 3 particles in the final state of the scattering, one could rewrite the $\epsilon^{\mu}_{p_1 p_2 p_3}$ in eq.~(\ref{EQ:TranspolMBRmassiveGeneric}) for $\varepsilon^{\mu}_{Y}$ by making use of the momentum basis representation of the Levi-Civita tensor eq.~(\ref{EQ:NVtrick}).
If the target particle with momentum $p_3$ is a light-like gauge boson, then there will be no  longitudinal polarization mode, 
and the transverse polarization vectors given above amount to taking the ``beam-axis'' vector $p_1 + p_2$ as the reference vector.

\section{Conventional form-factor projectors for $N (\geq 5)$ vector-boson scattering from the van Neerven-Vermaseren basis} 
\label{append:vNVbasis}

Although this article is mainly concerned with a constructive prescription for projectors that directly project out polarized amplitudes in a new hybrid CDR-compatible scheme, it is still interesting to see how the van Neerven-Vermaseren basis~\cite{vanNeerven:1983vr} allows us to read off conventional form-factor projectors for scattering amplitudes among $N \geq 5$ vector bosons, straightforwardly at almost zero computational cost.

The single most important quantity in the construction of the van Neerven-Vermaseren basis is the generalized Kronecker 
delta $\delta^{\mu_1 \cdots \mu_n}_{\nu_1 \cdots \nu_n}$, which can be written as the determinant of $n \times n$ space-time metric tensors:
\begin{eqnarray} \label{EQ:GKD}
\delta^{\mu _{1} \dots \mu _{n}}_{\nu _{1}\dots \nu_{n}}
=
{\begin{vmatrix}
g_{\nu _{1}}^{~\mu_{1}}&\cdots &g_{\nu_{n}}^{~\mu _{1}}\\\vdots &\ddots &\vdots \\
g_{\nu _{1}}^{~\mu_{n}}&\cdots &g_{\nu_{n}}^{~\mu _{n}}
\end{vmatrix}}\,.
\end{eqnarray}
In the case of $n=4$, the dimension of the Minkowski space, one has 
$\delta^{\mu_1 \cdots \mu_4}_{\nu_1 \cdots \nu_4} = \epsilon^{\mu_1 \mu_2 \mu_3 \mu_4} \, \epsilon_{\nu_1 \nu_2 \nu_3 \nu_4}$, i.e.,
 the contraction given in eq.~(\ref{EQ:LeviCivitaContRule}).
Note that if one takes eq.(\ref{EQ:GKD}) as the definition of the symbol $\delta^{\mu_1 \cdots \mu_n}_{\nu_1 \cdots \nu_n}$, it then has,
unlike the Levi-Civita tensor, a straightforward extension to D dimensions, because the r.h.s. consists of only the space-time metric tensors.
To simplify the discussion of form-factor projectors below, let us confine ourselves to  the scattering among exactly N=5 vector bosons where 
the 4 linearly independent external momenta are denoted by $\{p_1, p_2, p_3, p_4\}$.\footnote{On the boundaries of the phase space, 
the number of linearly independent momenta is known to become smaller.}
The close relation between the generalized Kronecker delta and the Gram matrix of $\{p_1, p_2, p_3, p_4\}$ makes this object 
very useful for constructing the \textit{dual} vectors, or the van Neerven-Vermaseren basis, 
$\{\mathcal{P}_1, \mathcal{P}_2, \mathcal{P}_3, \mathcal{P}_4\}$ of the linear space spanned by $\{p_1, p_2, p_3, p_4\}$.
To be specific, this vector basis is given by
\begin{eqnarray} \label{EQ:vNVbasis}
\mathcal{P}^{\mu}_1 &= \frac{\delta^{\,\mu \, p_2\, p_3\, p_4}_{\,p_1\, p_2\, p_3\, p_4}}{\delta^{\,p_1\, p_2\, p_3\, p_4}_{\,p_1\, p_2\, p_3\, p_4}}\, , \, 
\mathcal{P}^{\mu}_2 = \frac{\delta^{\, p_1 \,\mu \, p_3\, p_4}_{\,p_1\, p_2\, p_3\, p_4}}{\delta^{\,p_1\, p_2\, p_3\, p_4}_{\,p_1\, p_2\, p_3\, p_4}}\, , 
\mathcal{P}^{\mu}_3 = \frac{\delta^{\, p_1\, p_2\,\mu \, p_4}_{\,p_1\, p_2\, p_3\, p_4}}{\delta^{\,p_1\, p_2\, p_3\, p_4}_{\,p_1\, p_2\, p_3\, p_4}}\, , \, 
\mathcal{P}^{\mu}_4 = \frac{\delta^{\, p_1 \, p_2\, p_3\,\mu }_{\,p_1\, p_2\, p_3\, p_4}}{\delta^{\,p_1\, p_2\, p_3\, p_4}_{\,p_1\, p_2\, p_3\, p_4}}\, ,
\end{eqnarray}
where a compact notation for the generalized Kronecker delta contracted with momenta has been used, namely $\delta^{\,p\,\mu_2 \dots  \mu _{n}}_{\,q\, \nu_2 \dots \nu_{n}} \equiv p_{\mu_1} \, q^{\nu_1}\, \delta^{\,\mu_{1} \dots \mu_{n}}_{\,\nu _{1}\dots \nu_{n}}$.
One recognizes the common denominator $\delta^{\,p_1\, p_2\, p_3\, p_4}_{\,p_1\, p_2\, p_3\, p_4}$ as the Gram determinant of the list of independent momenta $\{p_1, p_2, p_3, p_4\}$.
It is straightforward to see that $\mathcal{P}_i \cdot p_j = \delta_{ij}$ for $i,j \in \{1,2,3,4\}$.
Consequently, for a rank-$5$ Lorentz tensor $M^{\mu_1 \cdots \mu_5}$ that can be linearly decomposed in terms of a set of Lorentz structures formed by tensor products of 5 momenta from $\{p_1, p_2, p_3, p_4\}$, 
\begin{eqnarray} \label{EQ:rankNamp}
M^{\mu_1 \mu_2 \mu_3 \mu_4 \mu_5} = \sum_{i_n \in \{1,2,3,4\}}  C_{i_1 i_2 i_3 i_4 i_5} \, p_{i_1}^{\mu_1} \, p_{i_2}^{\mu_2} \, p_{i_3}^{\mu_3} \, p_{i_4}^{\mu_4} \, p_{i_5}^{\mu_5} \, ,
\end{eqnarray}
the projectors for the linear decomposition coefficients $ C_{i_1 i_2 i_3 i_4 i_5}$ can be composed simply by tensor products of the van Neerven-Vermaseren basis $\{\mathcal{P}_1, \mathcal{P}_2, \mathcal{P}_3, \mathcal{P}_4\}$: 
\begin{eqnarray} \label{EQ:vNVbasisProjectors}
\mathcal{P}_{i_1 i_2 i_3 i_4 i_5} = \mathcal{P}_{i_1}^{\mu_1} \, \mathcal{P}_{i_2}^{\mu_2} \, \mathcal{P}_{i_3}^{\mu_3} \, \mathcal{P}_{i_4}^{\mu_4} \, \mathcal{P}_{i_5}^{\mu_5}\,.
\end{eqnarray}
In this way, the form-factor projectors for 5-vector-boson scattering amplitudes are obtained bypassing completely the explicit
procedure of building up and inverting the Gram matrix of a large set of Lorentz structures (such as those discussed in section~\ref{SEC:projectionmethod:recap}), at almost zero computational cost.
See ref.~\cite{Peraro:2019cjj} for a detailed discussion of how one can determine the form factor projectors for 5 gluon scattering amplitudes in the HV scheme alternatively via solving the linear equations involved with finite-field methods.

The identity operator of the linear space spanned by the momenta basis $\{p_1, \cdots, p_n\}$ (with $n \leq 4$) can also be easily 
composed as~\cite{vanNeerven:1983vr,Ellis:2011cr}
\begin{eqnarray} \label{EQ:identityoperator}
\hat{\mathrm{I}}^{\,\mu \nu}_n = \sum_{i=1}^{n} p_i^{\mu}\, \mathcal{P}_i^{\nu} \,,
\end{eqnarray} 
owing to $\mathcal{P}_i \cdot p_j = \delta_{ij}$.
In the case of $n=4$, eq.~(\ref{EQ:identityoperator}) provides the momentum basis representation of the space-time metric tensor $g_{\mu\nu}$ in the four-dimensional Minkowski space.\footnote{Projection operators to the complementary subspace that is orthogonal to the subspace spanned by the given vector basis (with $n < 4$) can also be composed by subtracting the identity operator eq.~(\ref{EQ:identityoperator}) of the subspace spanned by $\{p_1, \cdots, p_n\}$ from the underlying $g^{\mu\nu}$. This can also be conveniently achieved by making use of $\delta^{\mu _{1} \dots \mu _{n}}_{\nu _{1}\dots \nu_{n}}$.} 
In eq.~(\ref{EQ:NVtrick}), the momentum basis representation of the rank-4 Levi-Civita tensor was given. 
From the discussions of section~\ref{SEC:prescription:beyond4P} and the points made in ref~\cite{Ahmed:2019udm}, it should be clear that the Lorentz tensor structures needed for scattering amplitudes among $N \geq 5$ vector bosons, regardless of whether or not the interactions are parity-even, can all be expressed in terms products of the 4 linearly independent external momenta.
Indeed, the projectors given in eq.~(\ref{EQ:vNVbasisProjectors}) hold in general: the tensor amplitude $M^{\mu_1 \cdots \mu_5}$ can contain, apart from the structures given in eq.~(\ref{EQ:rankNamp}), terms involving the space-time metric tensor and the Levi-Civita tensor.
The possible appearances of any additional momentum, the space-time metric tensor, and also the Levi-Civita tensor
in the original $M^{\mu_1 \mu_2 \mu_3 \mu_4 \mu_5}$, given directly by Feynman diagrams, are effectively seen by the projectors $\mathcal{P}_{i_1 i_2 i_3 i_4 i_5}$ in eq.~(\ref{EQ:vNVbasisProjectors}) as merely intermediate short-hand notations of composite objects made out of $p_{i_1}^{\mu_1} \, p_{i_2}^{\mu_2} \, p_{i_3}^{\mu_3} \, p_{i_4}^{\mu_4} \, p_{i_5}^{\mu_5}$, because they are all linearly dependent on the later, clearly shown by eq.~(\ref{EQ:NVtrick}) and eq.~(\ref{EQ:identityoperator}).  
Furthermore, one only needs to project out the form-factor coefficients in front of the set of Lorentz tensor structures that survive and contribute under the chosen reference vectors.
One important and nice feature about the form-factor projectors in eq.(\ref{EQ:vNVbasisProjectors}) is that their contraction with the tensor amplitude $M^{\mu_1 \cdots \mu_5}$ can be done with the spacetime-metric tensor $g_{\mu\nu}$, i.e.~no need to insert the physical polarization sum rules of all $5$ external gauge bosons. 
Note that there are no explicit appearances of the space-time metric tensor in the form-factor projectors
given in eq.~(\ref{EQ:vNVbasisProjectors}), but only the external momenta.
Consequently, the helicity amplitudes reconstructed from form factors
projected out by this type of projectors are automatically those of the HV scheme.

Under the condition that one would first dress the multiple-parton scattering amplitudes by the physical polarization sums for each external gauge boson before being contracted with the external projectors in eq.(\ref{EQ:vNVbasisProjectors}), these projectors can be dramatically reduced by dropping terms that are nullified by these physical polarization sums.
However, this does not necessarily reduce the complexity of the computation, because dressing all external gauge bosons by their polarization sums is a very costly action in multiple-parton scatterings.

\bibliography{HelicityProjectors} 
\bibliographystyle{utphysM}

\end{document}